\documentclass[aps,prd,twocolumn,floatfix,amsmath,amssymb,superscriptaddress,notitlepage,10pt]{revtex4-1}

\usepackage{xcolor}
\usepackage{graphicx}
\usepackage{dcolumn}
\usepackage{bm}
\usepackage{amsmath}
\usepackage{color}
\usepackage{xcolor}
\usepackage{physics}
\usepackage{lipsum}
\usepackage{hyperref}

\def\avrg#1{\left\langle #1 \right\rangle}

\newcommand{\simgt}{\lower.5ex\hbox{$\; \buildrel > \over \sim \;$}}
\newcommand{\simlt}{\lower.5ex\hbox{$\; \buildrel < \over \sim \;$}}

\newcommand{\mnras}{Mon.~Not.~Roy.~Astron.~Soc.}
\newcommand{\physrep}{Phys.~Rep.}
\newcommand{\jcap}{JCAP}
\newcommand{\pasj}{PASJ}
\newcommand{\apjl}{\apj~Lett.}
\newcommand{\apjs}{\apj~Suppl.}
\newcommand{\aap}{Astronomy \& Astrophysics}
\newcommand{\aj}{Astron.~J.}

\newcommand{\bk}{{\bf k}}

\newcommand{\hiMpc}{h^{-1}{\rm Mpc}}

\newcommand{\wgg}{w_{\rm p}}
\newcommand{\dSigma}{\Delta\!\Sigma}

\begin{document}


\title{Cosmological inference from emulator based halo model I: Validation tests with HSC and SDSS mock catalogs}

\author{Hironao~Miyatake}
\email{hironao.miyatake@iar.nagoya-u.ac.jp}
\affiliation{Institute for Advanced Research, Nagoya University, Nagoya 464-8601, Japan}
\affiliation{Division of Particle and Astrophysical Science, Graduate School of Science, Nagoya University, Nagoya 464-8602, Japan}
\affiliation{Kavli Institute for the Physics and Mathematics of the Universe (WPI), The University of Tokyo Institutes for Advanced Study (UTIAS), The University of Tokyo, Chiba 277-8583, Japan}

\author{Yosuke~Kobayashi}
\affiliation{Kavli Institute for the Physics and Mathematics of the Universe
(WPI), The University of Tokyo Institutes for Advanced Study (UTIAS),
The University of Tokyo, Chiba 277-8583, Japan}
\affiliation{Physics Department, The University of Tokyo, Bunkyo, Tokyo 113-0031, Japan}

\author{Masahiro~Takada}
\email{masahiro.takada@ipmu.jp}
\affiliation{Kavli Institute for the Physics and Mathematics of the Universe (WPI), The University of Tokyo Institutes for Advanced Study (UTIAS), The University of Tokyo, Chiba 277-8583, Japan}

\author{Takahiro~Nishimichi}
\affiliation{Center for Gravitational Physics, Yukawa Institute for Theoretical Physics, Kyoto University, Kyoto 606-8502, Japan}
\affiliation{Kavli Institute for the Physics and Mathematics of the Universe
(WPI), The University of Tokyo Institutes for Advanced Study (UTIAS),
The University of Tokyo, Chiba 277-8583, Japan}

\author{Masato~Shirasaki}
\affiliation{National Astronomical Observatory of Japan, Mitaka, Tokyo 181-8588, Japan}
\affiliation{The Institute of Statistical Mathematics,
Tachikawa, Tokyo 190-8562, Japan}

\author{Sunao~Sugiyama}
\affiliation{Kavli Institute for the Physics and Mathematics of the Universe (WPI), The University of Tokyo Institutes for Advanced Study (UTIAS), The University of Tokyo, Chiba 277-8583, Japan}
\affiliation{Physics Department, The University of Tokyo, Bunkyo, Tokyo 113-0031, Japan}

\author{Ryuichi~Takahashi}
\affiliation{Faculty of Science and Technology, Hirosaki University, 3 Bunkyo-cho, Hirosaki, Aomori 036-8561, Japan}

\author{Ken~Osato}
\affiliation{Institut d'Astrophysique de Paris, Sorbonne Universit\'{e}, CNRS, UMR 7095, 98bis boulevard Arago, 75014 Paris, France}

\author{Surhud~More}
\affiliation{The Inter-University Centre for Astronomy and Astrophysics, Post bag 4, Ganeshkhind, Pune 411007, India}
\affiliation{Kavli Institute for the Physics and Mathematics of the Universe
(WPI), The University of Tokyo Institutes for Advanced Study (UTIAS),
The University of Tokyo, Chiba 277-8583, Japan}

\author{Youngsoo~Park}
\affiliation{Kavli Institute for the Physics and Mathematics of the Universe
(WPI), The University of Tokyo Institutes for Advanced Study (UTIAS),
The University of Tokyo, Chiba 277-8583, Japan}

\date{\today}

\begin{abstract}
We present validation tests of emulator-based halo model method for cosmological parameter 
inference,
assuming hypothetical measurements of the projected correlation function of galaxies, 
$\wgg(R)$, and the galaxy-galaxy weak lensing, $\dSigma(R)$, from the spectroscopic SDSS galaxies and the Hyper Suprime-Cam Year1  (HSC-Y1)  galaxies.
To do this, we use \textsc{Dark Emulator} developed in Nishimichi {\it et al.}
based on an ensemble of $N$-body simulations, which is an emulation package enabling a fast, accurate computation of 
halo clustering quantities (halo mass function, halo auto-correlation and halo-matter cross-correlation) for flat-geometry cold dark matter cosmologies. 
Adopting the halo occupation distribution,
 the emulator allows us to obtain model predictions of $\dSigma$ and $\wgg$ for the SDSS-like galaxies at a few CPU seconds 
for an input set of parameters. We present performance and validation of the method by carrying out Markov Chain Monte Carlo analyses
of the mock signals measured from a variety of mock catalogs that mimic the SDSS and HSC-Y1 galaxies. We show that the halo model method can recover the underlying true cosmological parameters to within the 68\% credible interval, except for the mocks including the assembly bias effect (although we consider the unrealistically-large 
amplitude of assembly bias effect).  
Even for the assembly bias mock, we demonstrate that the  cosmological parameters can be 
recovered {\it if} the analysis is restricted to scales 
$R\gtrsim 10~h^{-1}{\rm Mpc}$
(i.e., if the information on the average mass of halos hosting SDSS galaxies inherent in 
the 1-halo term of $\dSigma$ is not included).
We also show that, by using a single population of source galaxies to infer the relative strengths of $\dSigma$ for multiple lens samples at different redshifts, the joint probes method allows for self-calibration of photometric redshift errors and multiplicative shear bias. Thus we conclude that the emulator-based halo model method can be safely applied to the HSC-Y1 dataset, achieving a precision of $\sigma(S_8)\simeq 0.04$ after marginalization over nuisance parameters such as the halo-galaxy connection parameters and 
the photo-$z$ error parameter, and our method is 
complementary to methods based on perturbation theory.
\end{abstract}

\maketitle

\section{introduction}
\label{sec:introduction}

Wide-area imaging galaxy surveys offer exciting opportunities to address the fundamental questions in cosmology such as the nature of 
dark matter and the origin of cosmic acceleration
\citep{Weinbergetal:13}. The current-generation imaging surveys such as the Subaru Hyper Suprime-Cam 
\footnote{\url{https://hsc.mtk.nao.ac.jp/ssp/}}
\citep[HSC][]{HSCoverview:17,2019PASJ...71...43H,2020PASJ...72...16H}, the Dark Energy Survey \footnote{\url{https://www.darkenergysurvey.org}}
\citep[DES][]{2018PhRvD..98d3526A}, and the Kilo-Degree Survey \citep[KiDS][]{2020arXiv200715632H} 
have succeeded in using accurate measurements of weak gravitational lensing effects to obtain tight constraints on cosmological parameters. Interestingly, the cosmological model inferred from these large-scale structure probes show 
the so-called $\sigma_8$- or $S_8$-tension \citep[e.g.][]{2019PASJ...71...43H,2020MNRAS.tmp.2035P} compared with cosmological models inferred from
the {\it Planck} cosmic microwave background (CMB)
measurement \citep{2020A&A...641A...6P}, perhaps indicating a signature beyond the standard cosmological model, i.e. the flat-geometry $\Lambda$CDM model \citep[e.g.][]{2020MNRAS.tmp.2035P}. Upcoming galaxy surveys such as the Subaru Prime Focus Spectrograph \footnote{\url{https://pfs.ipmu.jp/index.html}} \citep{Takadaetal:14}, the Dark Energy 
Spectrograph Instrument \footnote{\url{https://www.desi.lbl.gov}}, the VRO Legacy Survey of Space and Time \footnote{\url{https://www.lsst.org}}, 
the ESA Euclid \footnote{\url{https://www.euclid-ec.org}} and the NASA Roman Space Telescope \footnote{\url{https://roman.gsfc.nasa.gov}} \cite{WFIRST:15} are expected to deliver a decisive conclusion on the possible tension and also revolutionize our understanding of the Universe. 

Main challenges of large-scale structure probes lie in uncertainty in galaxy bias, which refers to unknown relation between the distributions of matter and galaxies in large-scale structure \citep{2014Natur.509..177V,2018MNRAS.475..676S}. 
Since physical processes inherent in formation and evolution of galaxies are still difficult to accurately model from the first principles, we 
need both observational and theoretical approaches to tackle the galaxy bias uncertainty in order for us to obtain ``unbiased'' and ``precise'' estimation of the underlying cosmological parameters from large-scale structure observables.

For observational approach, 
joint probes cosmology offers a promising way to mitigate 
the impact of galaxy bias uncertainty on cosmology inference \citep{2005PhRvD..71j3515S,OguriTakada:11,2013MNRAS.430..767C,Mandelbaumetal:13,Miyatakeetal:15,2015ApJ...806....2M,2018PhRvD..98d3526A,2020arXiv200715632H,2020arXiv200806873S}. In particular, 
galaxy-galaxy weak lensing, obtained by cross-correlating positions of foreground (lens) galaxies with shapes of background galaxies, can be used to infer the average mass distribution around lens galaxies. 
A combination of
galaxy-galaxy weak lensing with auto-correlation function of the same sample of galaxies as the lens galaxies can be used to observationally disentangle the galaxy bias and the correlation function of the underlying matter distribution from the measured clustering signal
of galaxies. 

On theory side, there are mainly two empirical approaches to tackle the galaxy bias uncertainty. First one is a model based on perturbation theory (PT) of large-scale structure \citep{Bernardeauetal:02,Desjacquesetal:16}. As long as only the large-scale information of clustering observables in the linear or quasi-nonlinear regime is used and nuisance parameters to model galaxy bias are introduced, such a PT-based method is expected to serve as an ``accurate'' theoretical template of galaxy clustering \citep{Kuraseetal:17,2018MNRAS.480.4614M,2020arXiv200308277N,2020arXiv200806873S}.
Accuracy here refers to the fact that a PT model 
can reproduce the observed clustering correlation of galaxies by varying the bias parameters, down to a certain scale still in the quasi-nonlinear regime, 
 where PT is valid. 
An advantage of this method is that the model can be used for any type of galaxies, because PT is formulated based on properties of gravity and primordial fluctuations \citep{1980lssu.book.....P} and the free bias parameters absorb large-scale clustering
properties of galaxies irrespective of galaxy types. 
A price to pay is that the method breaks down at scales below 
a certain nonlinear scale,
and cannot be used to extract cosmological information from the small-scale clustering signals, which generally carry higher signal-to-noise ratios than in 
the large scales. 

An alternative theoretical method is the halo model approach \citep{Seljak:00,PeacockSmith:00,MaFry:00,Scoccimarroetal:01}. Halos are places where galaxies likely form, and clustering properties of halos are relatively well understood, on both analytical approach and 
$N$-body simulations \citep[][]{CooraySheth:02}. Then an empirical model such as the halo occupation distribution (HOD) method \cite{1998ApJ...494....1J,Zhengetal:05} can be used to connect halos to galaxies. An advantage of this method is that it would allow one to use the small-scale information in cosmology inference, thereby yielding tighter constraints on cosmological parameters. However, a danger is that, if the model is not sufficiently accurate 
nor flexible enough to capture the complicated galaxy-scale physics, the method might lead to a significant bias in cosmological parameters, more than the statistical credible interval. A worst-case scenario is that one might claim a wrong cosmology, e.g. a time-varying dark energy model, from a given dataset due to  the inaccurate theoretical templates. 

Hence, the purpose of this paper is to assess performance and limitation of the halo model method for cosmology inference. To do this, 
we use the 
\textsc{Dark Emulator} developed in \citet{2018arXiv181109504N}, which enables a fast, accurate computation of halo clustering quantities
(halo mass function, halo auto-correlation function and halo-matter cross-correlation) for an input set of cosmological parameters within flat-geometry $w$CDM framework with adiabatic Gaussian initial conditions. 
The Dark Emulator is particularly useful for our halo model approach, as it enables accurate predictions for the clustering quantities well into the non-linear regime and thereby allows us to make robust use of the cosmological information from small scales.
We combine \textsc{Dark Emulator} with the HOD model to make model predictions of 
the projected correlation functions of galaxies,  $\wgg(R)$ and
the galaxy-galaxy weak lensing, $\dSigma(R)$, that mimic those measured from the spectroscopic SDSS DR11 galaxies \citep{Alametal:16}
and the HSC Year1 (HSC-Y1) galaxies \citep{HSCDR1_shear:17}. More precisely, we consider mock galaxies of LOWZ and CMASS galaxies in the redshift range 
$0.15\lesssim z\lesssim 0.7$ for the spectroscopic galaxies in the $\wgg$ measurement and for the lens sample in $\dSigma$, and then consider mock galaxies of the deep HSC-Y1 data for the background galaxy sample in $\dSigma$. 
We use a variety of mock catalogs for the SDSS galaxies to assess performance and limitation of the halo model method for cosmology inference, including a mock 
where we implement an extreme version of galaxy assembly bias
\cite{2019arXiv190708515K}. Here we quantify the performance by studying whether the halo model method can recover the cosmological parameters employed in the mock catalogs, after marginalization over the halo-galaxy connection parameters -- hereafter we will often refer to this exercise as {\it cosmology challenges}.
Moreover, following \citet{OguriTakada:11},
we assess the ability of our method to perform self-calibration of photometric redshift errors and multiplicative bias error for the joint probes cosmology.
Those errors are among the most important, systematic errors for weak lensing cosmology, and we show below that the method allows for an accurate self-calibration of the errors, i.e. a robust estimation of cosmological parameters mitigating the impact of the systematic errors. This paper gives a validation of the halo model methodology for cosmology inference that we are planning to carry out with the actual SDSS and HSC datasets (Miyatake et al. in preparation). The results of this paper are also compared to those of the companion paper, \citet{2020arXiv200806873S}, 
which used a perturbation theory inspired method and was validated against the same mock catalogs.

The structure of this paper is as follows. In Section~\ref{sec:theory} we briefly review the halo model for $\wgg$ and $\dSigma$ and discuss 
the methodology of how cosmological parameters can be estimated from the clustering observables. In Section~\ref{sec:simulations} we describe details of $N$-body simulations, the halo catalogs, the mock catalogs of SDSS and HSC galaxies, and \textsc{Dark Emulator}.
In Section~\ref{sec:methodology} we describe our strategy for assessing performance and limitation of the halo model method, i.e. cosmology challenges done from the comparison of the halo model with the mock signals. In Section~\ref{sec:results}, we show the main results of this paper. 
Section~\ref{sec:conclusion} is devoted to conclusion and discussion. 
Throughout this paper, unless otherwise stated,  
we employ the flat-geometry {\it Planck} cosmology as a target cosmology for cosmology challenges 
and as a fiducial cosmology when we compute cosmological dependences of observables. 
The model is characterized by $\Omega_{\rm m}=0.3156$ (the present-day matter density parameter), $h = 0.672$ and $\sigma_8 = 0.831$, respectively, and we
adopt the units of $c=1$ for the speed of light. 

\section{Theory}
\label{sec:theory}

In this paper we focus on two observables that are obtained from imaging and spectroscopic data of galaxies in
the overlapping regions of the sky. One is the galaxy-galaxy weak lensing that can be measured by stacking 
shapes of background galaxies around a sample of foreground lensing galaxies.
Here we assume that the background galaxy sample is taken from the Subaru HSC images, while the foreground lensing galaxies are from the SDSS spectroscopic 
galaxies. 
The other is the projected correlation function of the spectroscopic galaxies that are from the same population of 
galaxies used as lens (foreground) galaxies
in the galaxy-galaxy weak lensing. 

\subsection{Observables: galaxy-galaxy weak lensing and projected auto-correlation function}
\label{sec:observables}

Galaxy-galaxy weak lensing 
\cite{Mandelbaumetal:06,Miyatakeetal:15}
probes the {\it averaged} excess surface mass density profile, $\dSigma$, around the lensing (foreground)
galaxies that is given in terms of the surface mass density profile, $\Sigma_{\rm gm}(R)$, as
\begin{align}
\dSigma(R;z_{\rm l})&=\avrg{\Sigma_{\rm gm}}\!(<R)-\Sigma_{\rm gm}\!(R)\nonumber\\
&=\Sigma_{\rm crit}(z_{\rm l},z_{\rm s})\left.\gamma_+(R)\right|_{R=\chi_{\rm l}\Delta\theta},
\label{eq:dSigma_def}
\end{align}
where $\gamma_+$ is the average tangential shear of background galaxies in the circular annulus of projected centric radius $R$ from the foreground 
galaxies, and $\chi_{\rm l}$ is the comoving angular diameter distance to each foreground 
galaxy. Note that background galaxy shapes 
are averaged over all the pairs of foreground-background galaxies in the same projected separation $R$, not the angular separation $\Delta\theta$, even when 
the lensing galaxies have a redshift distribution. 
$\avrg{\Sigma_{\rm gm}}\!(<R)$ is the average surface mass density within 
a circular aperture of radius $R$, defined as $\avrg{\Sigma_{\rm gm}}\!(<R)\equiv 
1/(\pi R^2)\int^R_0\!2\pi R'\mathrm{d}R'~\Sigma_{\rm gm}(R')$.
$\Sigma_{\rm crit}$ is the critical surface mass density that describes a lensing efficiency for pairs of foreground/background galaxies as a function of their redshifts, and is defined as 
\begin{align}
\Sigma_{\rm crit}\!(z_{\rm l},z_{\rm s})\equiv \frac{\chi_{\rm s}(z_{\rm s})}{4\pi G\chi_{\rm ls}(z_{\rm l},z_{\rm s})\chi_{\rm l}(z_{\rm l})(1+z_{\rm l})}, 
\label{eq:Sigma_cr}
\end{align}
where $\chi_{\rm s}$ and $\chi_{\rm ls}$ are the comoving angular diameter distances to each source galaxy and between source and lens galaxies in each pair, respectively. 
The factor $(1+z_{\rm l})$ is due to 
our use of the comoving coordinates. 
In practice we need to 
take into account the redshift distributions of lens and source galaxies that are straightforward to include, e.g. following the method 
in Ref.~\citep{Miyatakeetal:15}.

The surface mass density profile around lensing galaxies, $\Sigma_{\rm gm}(R)$, 
is given in terms of the three-dimensional 
galaxy-matter cross-correlation function as
\begin{align}
\label{eq:Sigma}
\Sigma_{\rm gm}(R; z_{\rm l})&=\bar{\rho}_{\rm m0}\int_{-\infty}^{\infty}\!\!\mathrm{d}\pi
~\xi_{\rm gm}\!\left(\sqrt{\pi^2+R^2}; z_{\rm l}\right)
\end{align}
where $\bar{\rho}_{\rm m0}$ is the mean mass density today, $\xi_{\rm gm}(r)$ is the three-dimensional cross-correlation function 
between the galaxies and matter, and $\pi$ is the 
separation along the line-of-sight direction, and $R$ is the projected separation perpendicular to the line-of-sight direction. 
Here we ignored the contribution from the background mean mass density because it is not relevant for weak lensing observables \citep[compared to Eq.~3 in Ref.][]{2020arXiv200806873S}. 
Fourier-transformed counterpart of $\xi_{\rm gm}(r)$ is the galaxy-matter cross power spectrum, 
$P_{\rm gm}(k)$,
defined as
\begin{align}
\xi_{\rm gm}(r;z_{\rm l})\equiv \int_0^\infty\frac{k^2\mathrm{d}k}{2\pi^2}~P_{\rm gm}(k;z_l)j_0(kr),
\label{eq:xigm_def}
\end{align}
where $j_0(x)$ is the zero-th order spherical Bessel function. Hereafter we  omit the redshift of the lensing galaxies, $z_{\rm l}$, in the argument of the correlation functions for notational simplicity. Throughout this paper we assume a distant observer approximation for computations of projected correlation functions. In other words, we ignore the effects of curved sky for definitions of separations along or perpendicular to the line-of-sight direction. 
The excess surface mass density profile 
is given in terms of the galaxy-matter cross power spectrum as
\begin{align}
\Delta\!\Sigma_{\rm gm}(R)&\equiv \avrg{\Sigma_{\rm gm}}\!(<R)-\Sigma_{\rm gm}(R)\nonumber\\
&=\bar{\rho}_{\rm m0}\int_0^\infty\!\frac{k^2\mathrm{d}k}{2\pi^2}P_{\rm gm}(k)J_2(kR), 
\label{eq:dsigma}
\end{align}
where $J_2(x)$ is the 2nd-order Bessel function. 

To obtain the theoretical template of $\dSigma(R)$ for a given cosmological model within the flat $\Lambda$CDM framework, we use \textsc{Dark Emulator}, developed in 
Ref.~\cite{2018arXiv181109504N}, that enables accurate and fast computations of halo statistical quantities; the halo mass function, the halo-matter 
cross-correlation function and the halo-auto correlation function as a function of redshift, halo masses, and separations for an input cosmological model. 
As demonstrated in Ref.~\cite{2018arXiv181109504N}, the emulator outputs the predictions to better than a few percent in the fractional amplitude
over scales of separations we are interested in. 

However, the emulator does not take into account the effects of baryonic physics, and this is a limitation we should keep in mind.
Nevertheless we do not think this limitation causes a catastrophic failure of our approach due to the following reasons. 
The baryonic physics is local in the sense that it affects the matter distribution at scales smaller than a maximum scale, denoted as $R_\ast$. 
For example, even if we consider a violent effect of the AGN feedbacks, it would affect the mass distribution around halos up to a few Mpc at maximum. 
As nicely discussed in Refs.~\cite{2015JCAP...12..049S,2019JCAP...03..020S}, we can safely consider that the baryonic effect causes a ``redistribution'' of matter  at $r\lesssim R_\ast$ around galaxies, and does not alter the mass distribution at $r\gtrsim R_\ast$ ($r$ is the 
three-dimensional
radius from the center of galaxy). In other words, even in the presence of the baryonic physics, the mass conservation at $r\lesssim R_\ast$ holds. 
Keeping  this in mind, we can say that \textsc{Dark Emulator}, as designed,  
can accurately model the matter distribution at $r\gtrsim R_\ast$ around the host halos of galaxies, and also satisfies 
the mass conservation to within  $r\simeq R_\ast$, even though \textsc{Dark Emulator} ceases to accurately predict the mass profile at $r\lesssim R_\ast$. 

Based on the above consideration, we can rewrite the lensing profile (Eq.~\ref{eq:dsigma}) as
\begin{align}
\dSigma(R)&=
\avrg{\Sigma_{\rm gm}}\!(<R)-\Sigma_{\rm gm}\!(R)\nonumber\\
&\hspace{-2em}=\frac{M(<R_\ast)}{\pi R^2}+\frac{2}{R^2}\int_{R_\ast}^R\!R'\mathrm{d}R'~\Sigma_{\rm gm}\!(R')-\Sigma_{\rm gm}(R).
\label{eq:dSigma_1halo}
\end{align}
In the second line on the r.h.s., we used the relation $M(<R_\ast)\equiv \int_0^{R_\ast}\!2\pi R'\mathrm{d}R'~\Sigma_{\rm gm}(R')$,
where $M(<R_\ast)$ is the mass interior of a circular aperture of radius $R_\ast$.
The second term is the contribution over the range of $[R_\ast,R]$ to the average mass density. 
As long as we focus on the lensing profiles 
at $R\ge R_\ast$, \textsc{Dark Emulator} can accurately predict the lensing profile, including the interior mass at the aperture of $R\simeq R_\ast$.
On the other hand, 
we need to introduce 
various nuisance parameters to model the lensing ``profile'' at $R\le R_\ast$, 
if the information is included, and then marginalize over the parameters when estimating the interior mass 
$M(<R_\ast)$ and performing cosmology inference.

Another clustering observable we use is the projected auto-correlation function for spectroscopic galaxy sample that is the same sample used as lens (foreground) galaxies in the galaxy-galaxy weak lensing measurement. The projected correlation function is defined by a line-of-sight projection of the three-dimensional 
auto-correlation function of galaxies, $\xi_{\rm gg}(r)$, as 
\begin{align}
w_{\rm p}(R)\equiv 2\int_0^{\pi_{\rm max}}\mathrm{d}\Pi~ \xi_{\rm gg}\!\left(\sqrt{R^2+\Pi^2}\right),
\label{eq:wp_def}
\end{align} 
where $\pi_{\rm max}$ is the length of the line-of-sight projection. Note that the projected correlation function has 
the unit
of $[h^{-1}{\rm Mpc}]$. Throughout this paper, unless explicitly stated,
we employ $\pi_{\rm max}=100~\hiMpc$ as our default choice. Also notice that $\xi_{\rm gg}(r)$ is given in terms of the auto-power spectrum of the galaxy number density field, $P_{\rm gg}$, as 
\begin{align}
\xi_{\rm gg}(r)=\int_0^\infty\!\!k^2\mathrm{d}k/(2\pi^2)~P_{\rm gg}(k)j_0(kr).
\label{eq:xigg_def}
\end{align}
The projected correlation function is not sensitive to the redshift-space distortion (RSD) effect due to peculiar velocities of galaxies, if a sufficiently large projection length ($\pi_{\rm max}$) is taken \citep[see Fig.~6 in Ref.][]{2013MNRAS.430..725V}.
The RSD effect itself is a useful cosmological probe, but its use requires an accurate modeling \citep{2019arXiv190708515K,2020arXiv200506122K},
which is not straightforward. Hence, the projected correlation function makes it somewhat easier to compare with theory in a cosmological analysis. In the following, we ignore the RSD effect in most cases of our cosmology challenges,
but will separately discuss the impact of 
the RSD effect in parameter estimation.

\subsection{Theoretical template: \textsc{Dark Emulator} implementation of halo model}
\label{sec:halo_model}

In this section we describe details of the halo model implementation to make model predictions for the observables $\dSigma(R)$ and 
$\wgg(R)$. As can be found around Eqs.~(\ref{eq:dsigma}) and (\ref{eq:wp_def}), we need 
the real-space cross-power spectrum of galaxies and matter $P_{\rm gm}(k;z)$, and
the real-space auto-power spectrum of galaxies,
$P_{\rm gg}(k;z)$,
as a function of cosmological models to compute 
the observables. 

\subsubsection{Halo Occupation Distribution}
\label{subsub:hod}

In the halo model we assume that all matter is associated with halos, and the correlation function of matter is given by the contributions 
from pairs of matter in the same halo and those in two different halos, which are referred to as the 1- and 2-halo 
terms,  respectively.
To connect halos to galaxies in a given sample, 
we employ the halo occupation distribution 
\citep[HOD][]{1998ApJ...494....1J,PeacockSmith:00,Scoccimarroetal:01}.
The HOD model gives 
the mean number of central and satellite galaxies in halos of mass $M$ as
\begin{align}
\avrg{N}\!(M)=\avrg{N_{\rm c}}\!(M)+\avrg{N_{\rm s}}\!(M),
\label{eq:N_HOD}
\end{align}
where $\avrg{\hspace{1em}}\!(M)$ denotes the average of a quantity for halos of mass $M$. 

We employ the mean HOD for central galaxies, given as
\begin{align}
\avrg{N_{\rm c}}\!(M)=\frac{1}{2}\left[1+{\rm erf}\left(\frac{\log M-\log M_{\rm min}}{\sigma_{\log M}}\right)\right],
\label{eq:Nc}
\end{align}
where ${\rm erf}(x)$ is the error function and $M_{\rm min}$ and $\sigma_{\log M}$ are model parameters. 

For the mean HOD of satellite galaxies, we employ the following form: 
\begin{align}
\avrg{N_{\rm s}}\!(M)\equiv \avrg{N_{\rm c}}\!(M)\lambda_{\rm s}(M)=\avrg{N_{\rm c}}\!(M)\left(\frac{M-\kappa M_{\rm min}}{M_1}\right)^{\alpha},
\end{align}
where $\kappa,
M_1$ and $\alpha$ are model parameters, and we have introduced the notation $\lambda_{\rm s}(M)=[(M-\kappa M_{\rm min})/M_1]^{\alpha}$. 
For our default prescription we assume that satellite galaxies reside only in a halo that already hosts a central galaxy. This means $N_{\rm c}=1$ for such halos. Then we assume that the number 
of satellite galaxies
in halos of mass $M$ follows the Poisson distribution with mean $\lambda_{\rm s}(M)$; ${\rm Prob}(N_{\rm s})=(\lambda_{\rm s})^{N_{\rm s}}\exp(-\lambda_{\rm s})/N_{\rm s}!$ for halos that host a central galaxy inside, 
and ${\rm Prob}(N_{\rm s})=\delta^K_{N_{\rm s},0}$ for halos that do not host a central galaxy, where $\delta^K_{ij}$ is the Kronecker delta function.
Under these assumptions, the mean number of galaxy pairs living in the same halo with mass $M$, which is relevant for the 1-halo term, can be computed as
\begin{align}
\avrg{N(N-1)}&=\avrg{N_{\rm c}}
\left.\avrg{N(N-1)}\right|_{N_{\rm c}=1}\nonumber\\
&\hspace{3em}+
(1-\avrg{N_{\rm c}})
\left.\avrg{N(N-1)}\right|_{N_{\rm c}=0}\nonumber\\
&=\avrg{N_{\rm c}}\left[\avrg{(N_{\rm c}+N_{\rm s})(N_{\rm c}+N_{\rm s}-1)}\right]_{N_{\rm c}=1}\nonumber\\
&=\avrg{N_{\rm c}}\left[\avrg{N_{\rm s}^2}+\avrg{N_{\rm s}}\right]\nonumber\\
&=\avrg{N_{\rm c}}\left[2\lambda_{\rm s}+\lambda_{\rm s}^2\right].
\end{align}
where we have used the 
fact
$N=0$ for halos which have no central galaxy for our default prescription.

We have 5 model parameters, $\{M_{\rm min},\sigma_{\log M},\kappa, M_1, \alpha\}$, 
to characterize the central and satellite HODs in total for each galaxy sample 
for a given cosmological model.

Once the HOD model is given, the mean number density of galaxies in a sample is given as
\begin{align}
\bar{n}_{\rm g}=\int\!\mathrm{d}M~{\color{blue}{\frac{\mathrm{d}n_{\rm h}}{\mathrm{d}M}}}\left[\avrg{N_{\rm c}}\!(M)+\avrg{N_{\rm s}}\!(M)\right],
\end{align}
where $\mathrm{d}n_{\rm h}/\mathrm{d}M$ is the halo mass function which gives the mean number density of halos in the mass range $[M,M+\mathrm{d}M]$.
In the following we use ``blue'' fonts to denote a quantity that can be supplied by \textsc{Dark Emulator}. In the above case 
we use \textsc{Dark Emulator} to compute the halo mass function for $w$CDM cosmology, while we input the HOD model, as described, to compute 
the galaxy number density.

\subsubsection{Galaxy-galaxy weak lensing profile}
\label{sec:model_lensing}
\begin{figure}
    \begin{center}
        \includegraphics[width=1.0\columnwidth]{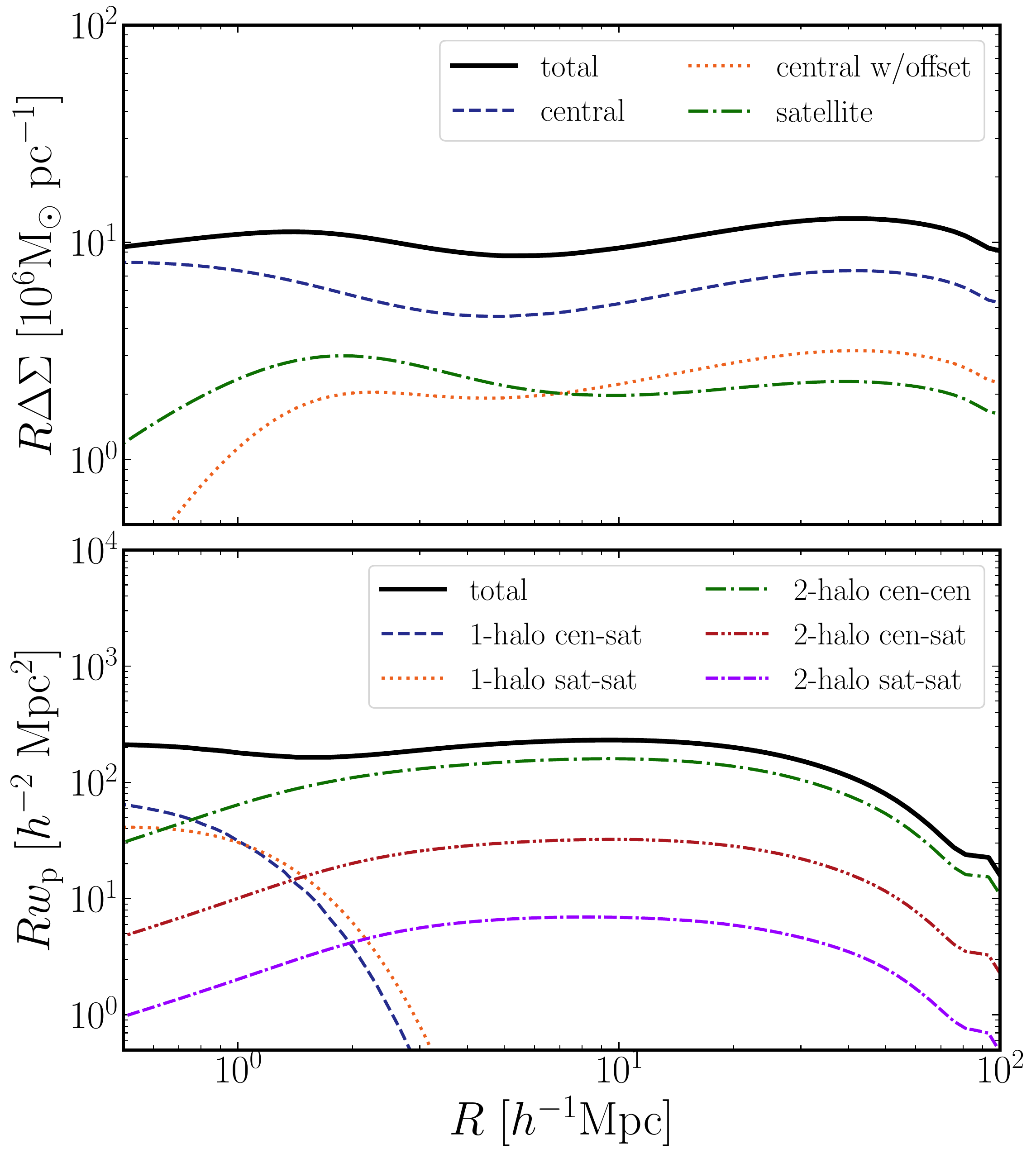}
        \caption{Break down of different contributions to the galaxy weak lensing profile 
        ($\dSigma(R)$; upper panel) and the projected correlation function of galaxies ($\wgg$; lower) for the SDSS LOWZ-like galaxies at $z=0.251$, 
        which are computed using \textsc{Dark Emulator} and our model ingredients of the halo-galaxy connection for 
        the {\it Planck} cosmology.
        For illustrative purpose 
        we multiply each observable by $R$ so that their dynamic range ($y$-axis) becomes narrower.
        {\it Upper panel}: The dashed line shows a contribution arising from 
        the cross-correlation of central galaxies with the surrounding matter distribution, while the dot-dashed line denotes 
        a contribution from the cross-correlation of satellite galaxies with matter. For comparison, we also show how a possible off-centering 
        of central galaxies affects the lensing profile, although we do not include this effect in the theoretical templates.
        Here, as a working example, we consider the off-centering parameters $p_{\rm off}=0.3$ and ${\cal R}_{\rm off}=0.4$ irrespective of
        halo mass, where 
        $p_{\rm off}$ models a fraction of off-centered central galaxies in halos of mass $M$, while 
        ${\cal R}$ is the off-centering radius relative to the scale radius of NFW profile of halos. 
        The upper solid lines is the total power. 
        {\it Lower}: The similar break down of different contributions to the projected correlation function of the galaxies, for the same model 
        in the upper panel. 
        }
        \label{fig:model_signals}
    \end{center}
\end{figure}
As we described above, the galaxy-galaxy weak lensing arises from the cross-correlation of (spectroscopic) lensing galaxies with the surrounding 
matter distribution in large-scale structure, $\xi_{\rm gm}(r;z_{\rm l})$. Hence we need to model $\xi_{\rm gm}(r)$ as a function of the parameters of halo-galaxy connection and cosmological parameters.

Since the Fourier space gives a somewhat simpler form of the expression, we mainly write down the power spectrum that is the Fourier-transformed counterpart of the two-point correlation function, here $\xi_{\rm gm}$. Under the halo model approach, 
the cross-power spectrum of galaxies and matter is given as 
\begin{align}
P_{\rm gm}(k)&=\frac{1}{\bar{n}_{\rm g}}
\int\!\mathrm{d}M{\color{blue}{\frac{\mathrm{d}n_{\rm h}}{\mathrm{d}M}}}
\left[\avrg{N_{\rm c}}\!(M)+\avrg{N_{\rm s}}\!(M)\tilde{u}_{\rm s}(k;M,z)\right]\nonumber\\
&\hspace{5em}\times {\color{blue}{P_{\rm hm}(k)}},
\end{align}
where $P_{\rm hm}(k)$ is the halo-matter cross-power spectrum that can be computed by \textsc{Dark Emulator} as a function of cosmological model 
within $w$CDM cosmologies. The quantity $\tilde{u}_{\rm s}(k;M)$ is the Fourier transform of the averaged radial profile of satellite galaxies 
in host halos of $M$ at redshift $z$, which we need to specify. Note that all the quantities in the above equation are 
evaluated at the lens redshift 
$z_{\rm l}$, but we omit to denote in the argument of each function for notational simplicity.
\textsc{Dark Emulator} was built to calibrate
the halo-matter cross-correlation by measuring the averaged mass profile around halos 
with mass $M$ 
in $N$-body simulation outputs \citep{2018arXiv181109504N}. By construction, the halo-matter cross-correlation satisfies 
the mass conservation around halos. We also note that 
\textsc{Dark Emulator} already includes both the 1- and 2-halo term contributions in 
$P_{\rm hm}$, which correspond to the cross-correlations of halos with matter in the same halo and the surrounding matter outside the halo, respectively. 

More exactly speaking, \textsc{Dark Emulator} outputs $\xi_{\rm hm}(r; M)$ for an input set of parameters (halo mass, separation and 
cosmological parameters) that is the Fourier transform of $P_{\rm hm}(k;M)$. In order to obtain $\xi_{\rm gm}$ or $\dSigma$ for the assumed model
we extensively 
use the publicly-available {\tt FFTLog}\footnote{\url{https://jila.colorado.edu/~ajsh/FFTLog/index.html}} code to perform the Hankel transforms 
when going back and forth between real- and Fourier-space (see Eqs.~\ref{eq:xigm_def} and \ref{eq:dsigma}). 

For the radial profile of satellite galaxies, we assume that satellite galaxies follow a Navarro-Frenk-White (NFW) profile \citep{Navarroetal:97}
that is an approximated model of the averaged dark matter profile in halos. To compute the NFW profile as a function of halo mass, redshift and cosmological model, we need to specify the scaling relation of mass concentration with halo mass in the halo matter profile.  For this
 we employ the fitting formula given in Ref.~\cite{2015ApJ...799..108D}; more exactly, we use the publicly-available python code, ``{\tt Colossus}''\footnote{\url{https://bdiemer.bitbucket.io/colossus/}} \citep{2018ApJS..239...35D}, to compute the concentration parameter for a given set of parameters (halo mass, redshift and cosmological parameters).

As for the default model of the halo-galaxy connection, we consider neither the off-centering effect nor 
the incompleteness effect of galaxies, where 
the latter
describes a possibility that some fraction of even very massive halos might not host a SDSS-like galaxy due to 
an incomplete selection after the specific color and magnitude cuts \citep[][for details of the model]{2015ApJ...806....2M}. Rather than introducing additional nuisance parameters to model these effects, we employ a minimum halo model  as our baseline model. However, we use different types of mock catalogs of SDSS-like galaxies including the off-centering effects and the incompleteness effect, and then will use the mock catalogs to validate and assess the performance of the baseline method. If our baseline model can recover the underlying cosmological parameters, we claim that the method is validated. If the method shows any failure to recover the cosmological parameters, we will start to introduce more parameters. However, if the cosmological constraints turn to be sensitive to such details of a treatment of galaxies inside the halo, this is a sign of the failure or limitation of the halo model based method, because such a small-scale distribution of galaxies is very difficult to accurately model due to complexities of physical processes inherent in formation and evolution of galaxies. Hence, rather than including such model parameters, we employ the 
minimum halo model  to assess its performance. 

The upper panel of Fig.~\ref{fig:model_signals} shows how different terms of $P_{\rm gm}$ contributes to the $\dSigma$ profile for the SDSS-like galaxies 
at $z=0.251$ as we will describe in more detail.
Even though satellite galaxies tend to reside in massive halos, all the curves have a similar shape ($R$-dependence) in large $R$ bins, 
$R\gtrsim 10~\hiMpc$, for a fixed cosmology. All the small-scale physics involved in the halo-galaxy connection affects $\dSigma$ at 
$R\lesssim 10~\hiMpc$. Note that, due to the non-local nature of $\dSigma$, the small-scale physics affects $\dSigma$ up to relatively large scales, compared to a virial 
radius of massive halos. 
Also importantly, the integrated lensing signal up to a few Mpc scales gives an estimate of the average halo mass, which has a close tie to the large-scale amplitudes of $\dSigma$ and $\wgg$ via the scaling relation of halo bias with halo mass (see below). For comparison, the figure also shows the impact of off-centering effects of central galaxies on $\dSigma$, assuming that some fraction of central galaxies might be off-set from the true halo center as a result of the assembly history of galaxies in their host halos \citep{2013MNRAS.435.2345H,2013MNRAS.433.3506M}.

\subsubsection{Projected correlation function of galaxies}
\label{sec:model_wgg}

To model $\wgg$ we need to model the real-space correlation function of galaxies, $\xi_{\rm gg}$, or the Fourier transform $P_{\rm gg}$ for an input 
set of parameters. 
The auto-power spectrum of galaxies in a sample is decomposed into the two contributions, the 1- and 2-halo terms, 
and those are given within 
the halo model framework as
\begin{align}
P_{\rm gg}(k)\equiv P^{\rm 1h}_{\rm gg}(k)+P^{\rm 2h}_{\rm gg}(k)
\end{align}
with
\begin{widetext}
\begin{align}
P^{\rm 1h}_{\rm gg}(k)&= \frac{1}{\bar{n}_{\rm g}^2}
\int\!\mathrm{d}M{\color{blue}{\frac{\mathrm{d}n_{\rm h}}{\mathrm{d}M}}}\avrg{N_{\rm c}}\!(M)
\left[
2\lambda_{\rm s}(M)\tilde{u}_{\rm s}(k;M)+\lambda_{\rm s}(M)^2\tilde{u}_{\rm s}(k;M)^2\right], \nonumber\\
P^{\rm 2h}_{\rm gg}(k)&= \frac{1}{\bar{n}_{\rm g}^2}
\left[\int\!\mathrm{d}M~{\color{blue}{\frac{\mathrm{d}n_{\rm h}}{\mathrm{d}M}}}
\avrg{N_{\rm c}}\!(M)\left\{1+\lambda_{\rm s}(M)\tilde{u}_{\rm s}(k;M)
\right\}
\right]\nonumber\\
&\hspace{10em}\times
\left[\int\!\mathrm{d}M'~{\color{blue}{\frac{\mathrm{d}n_{\rm h}}{\mathrm{d}M'}}}
\avrg{N_{\rm c}}\!(M')\left\{1+\lambda_{\rm s}(M')\tilde{u}_{\rm s}(k;M')
\right\}
\right]{\color{blue}{P_{\rm hh}(k;M,M')}},
\label{eq:Pgg_1h2h}
\end{align}
\end{widetext}
%
where $P_{\rm hh}(k;M,M')$ is the power spectrum between two
halo samples
with masses $M$ and $M'$. 
\textsc{Dark Emulator} outputs the real-space correlation function of halos, 
$\xi_{\rm hh}(r;M,M')$, that is the Fourier transform
of $P_{\rm hh}$.
Similarly to the case of $\dSigma$, 
we use the {\tt FFTLog} code to perform the Fourier transform to obtain the prediction 
of $\xi_{\rm gg}(r)$ (see Eq.~\ref{eq:xigg_def}) and then perform the line-of-sight 
integral to obtain the prediction of $\wgg(R)$ (see Eq.~\ref{eq:wp_def}).
We note that, strictly speaking, our standard halo model implementation of the 1-halo term behaves as a shot noise like term of $k^0(={\rm const.})$ 
at the limit of $k\rightarrow 0$, and this gives a subtle violation of the mass and momentum conservation \citep{1980lssu.book.....P}. 
One could modify the halo model at very small $k$ to enforce the conservation laws, as done in Ref.~\cite{MohammedSeljak:14}, but we found that our treatment, practically, 
gives a sufficiently accurate model prediction over scales of separations that we are interested in, for the SDSS-like galaxies 
\citep{2020arXiv200506122K} \citep[also see][for the similar discussion]{2020PhRvD.101l3520P}.
As given by Eq.~(\ref{eq:wp_def}), we employ $\pi_{\rm max}=100~\hiMpc$, as our default choice to integrate $\xi_{\rm gg}(r=\sqrt{R^2+\pi^{2}})$
to obtain the model prediction for $\wgg(R)$.

The lower panel of Fig.~\ref{fig:model_signals} shows 
different contributions in the projected correlation function 
$\wgg$ for the {\it Planck} cosmology, assuming the same model as in the upper panel. As can be found from Eq.~(\ref{eq:Pgg_1h2h}), the 1-halo term 
arises from the central-satellite and satellite-satellite correlations, while the 2-halo term is from the correlations of central-central, central-satellite and satellite-satellite galaxies in different halos, respectively. The figure clearly shows that all the contributions to the 1-halo term are confined to small scales, at
$R\lesssim {\rm a~few~Mpc}$, a
virial radius of massive halos, reflecting the local nature of $\wgg$ in contrast to $\dSigma$. All the different contributions to the 2-halo term have a similar shape ($R$-dependence); the different terms differ from each other only by a multiplicative factor for a fixed cosmology. This means that the shape of the 2-halo term is not sensitive to details of the halo-galaxy connection. The features around $R\simeq 90~\hiMpc$ are the BAO features, which appear at smaller scales than the BAO scale of $100~\hiMpc$ in the three-dimensional correlation function, due to the line-of-sight projection. However, 
we note that we will not include the BAO information in the parameter estimation in the following; we will use the $\wgg$ information up to 
$R\simeq 30~\hiMpc$ as our default choice.

\subsection{Joint probes cosmology: 
mitigation of galaxy bias uncertainty}
\label{sec:joint_probes}

\begin{figure}
    \begin{center}
        \includegraphics[width=1.0\columnwidth]{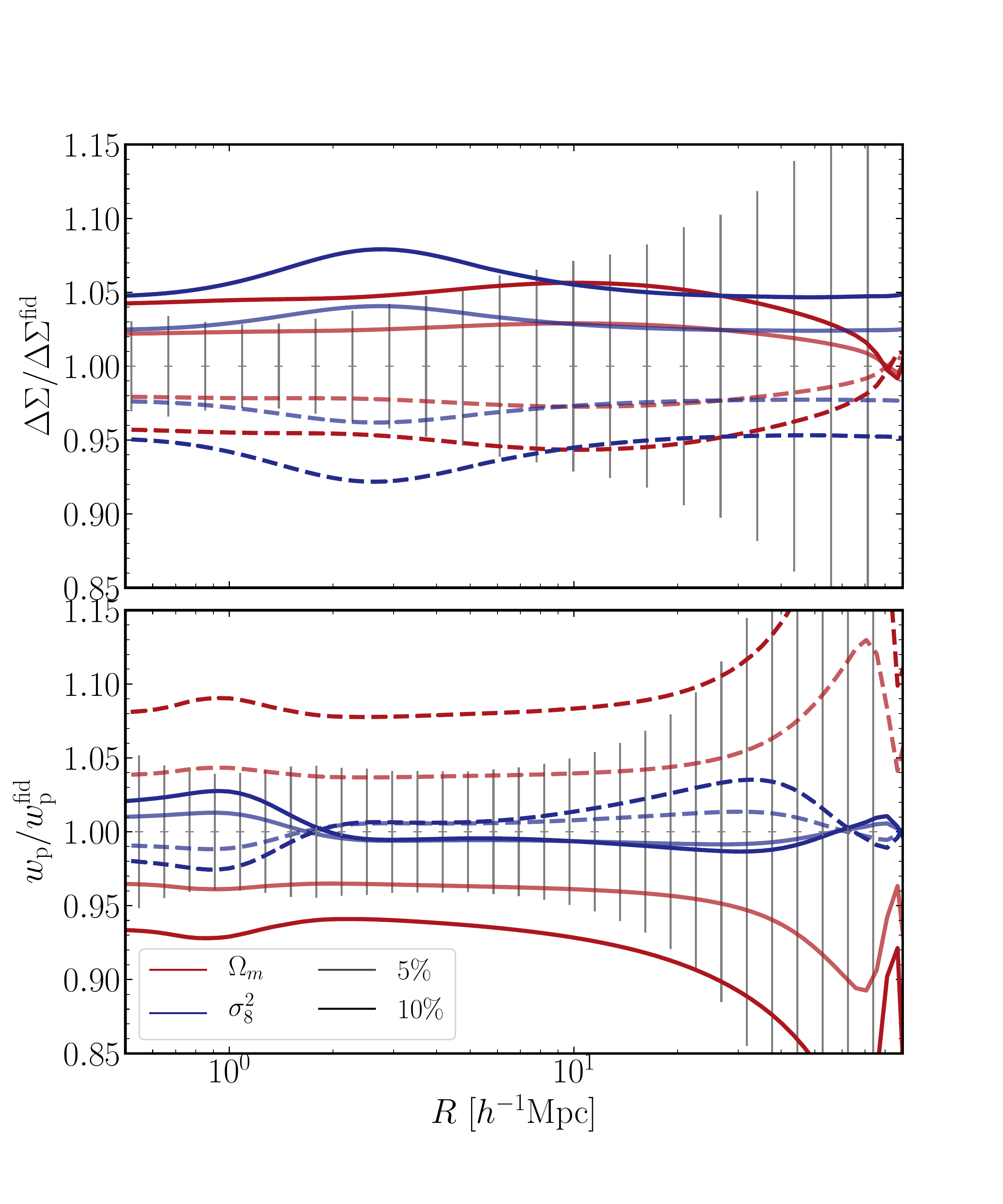}
        \caption{Dependences of $\dSigma$ and $\wgg$ on cosmological parameters, $\sigma_8$ (blue lines) and $\Omega_{\rm m}$ (red), computed using 
        the \textsc{Dark Emulator} based halo model. Here we consider the HOD parameters for the LOWZ sample at $z=0.251$. Here we consider 
        fractional changes of $\sigma_8^2$ (not $\sigma_8$) or $\Omega_{\rm m}$ by $\pm 5 \, \%$ or $\pm 10 \, \%$, respectively, where the other parameters are fixed to their fiducial values. 
        The solid and 
        dashed, respective lines show the fractional changes in $\dSigma$ or $\wgg$ relative to that for the fiducial model when $\sigma_8^2$ or 
        $\Omega_{\rm m}$ is changed to a positive or negative side from its fiducial value.
        The error bars around unity denote $1\sigma$ statistical errors that are computed from the diagonal terms of the covariance matrix expected  
        for the SDSS and HSC-Y1 data, although the neighboring bins are highly correlated with each other. 
        }
        \label{fig:signal_cosmo_dep}
    \end{center}
\end{figure}
\begin{figure*}
    \begin{center}
        \includegraphics[width=1.9\columnwidth]{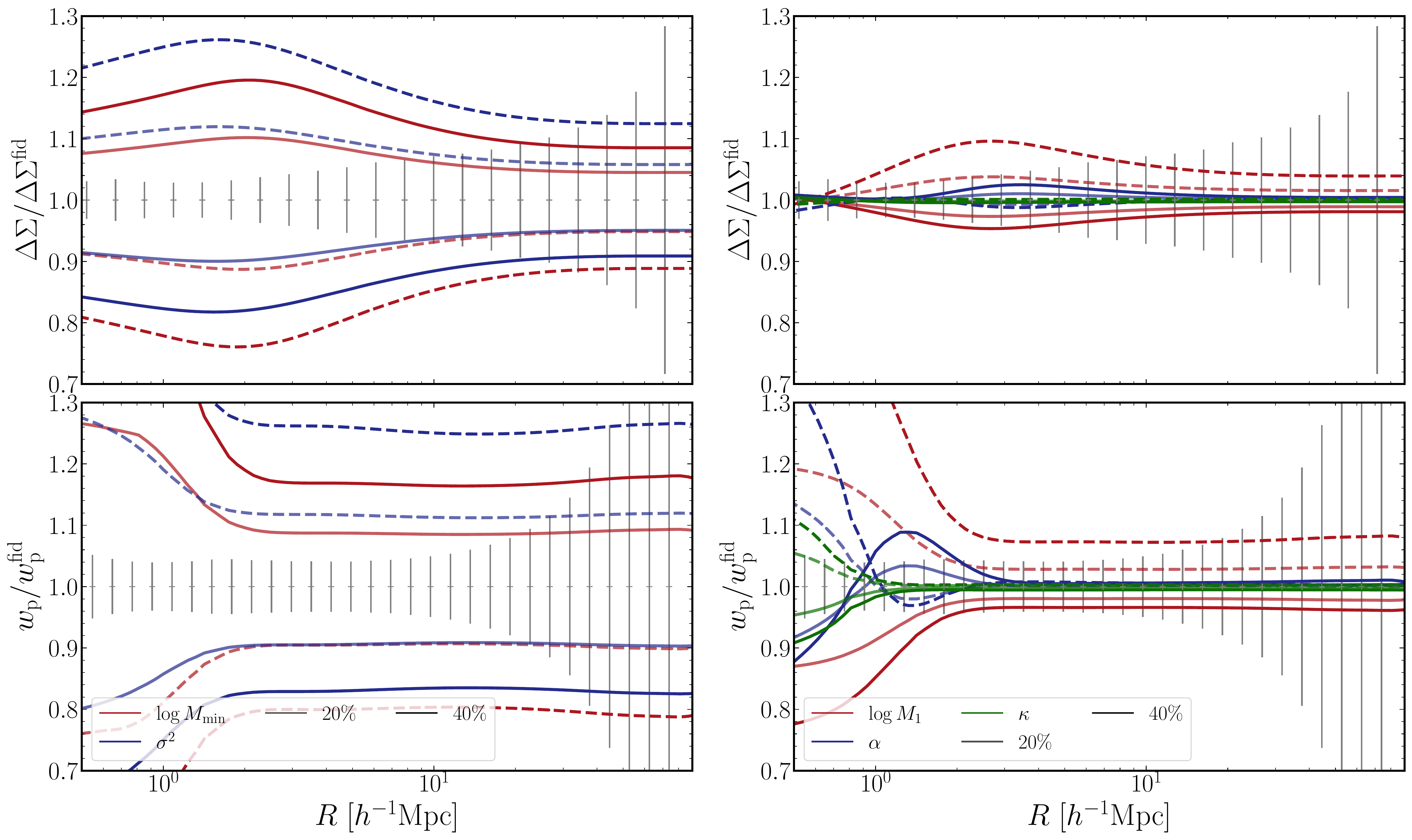}
        \caption{Dependence of $\dSigma$ and $\wgg$
        on changes to HOD parameters can be seen in the top and the bottom panels, respectively.
        Here the fiducial HOD parameters are chosen to resemble 
        the SDSS LOWZ-like galaxy sample at $z=0.251$ for the {\it Planck} cosmology. When one HOD parameter is changed, the other parameters are fixed to their fiducial values.  Solid line shows the result for an increase (positive-side change)
        of each HOD parameter from its fiducial value, while dashed line is the result for the decrease. From light- to dark-color lines 
        show the results 
        for the fractional change of each HOD parameter by $\pm 20 \, \%$ and $\pm 40 \, \%$, respectively.
        The left panel shows the dependences of the central HOD parameters, 
        $M_{\rm min}$ and $\sigma_{\log M}$, while the right panel shows the results for the satellite HOD parameters, $M_1, \kappa$ and 
        $\alpha$. The error bars are the same in the previous figure.
        \label{fig:signal_hod_dep}}
	\end{center}
\end{figure*}

For convenience of our discussion let us define 
the cross-correlation coefficient function
for halo correlation function:
\begin{align}
r_{\rm hm}(r)\equiv \frac{\xi_{\rm hm}(r)}{[\xi_{\rm hh}(r)\xi_{\rm mm}(r)]^{1/2}}. 
\end{align} 
As shown in Fig.~31 of Ref.~\cite{2018arXiv181109504N}, \textsc{Dark Emulator} predicts $r_{\rm hm}\simeq 1$ on 
$r\gtrsim 10~\hiMpc$, the scale greater 
than a typical size of massive halos, and 
$r_{\rm hm}$ is close to unity within 5\% or so even at the intermediate scales 
$\mbox{a few Mpc}<r\lesssim 10~\hiMpc$. As stressed in our companion paper \cite{2020arXiv200806873S}, the real-space 
observables have an advantage that all variants due to the halo-galaxy connection are confined to small scales, $r\lesssim 10~\hiMpc$, and the cross-correlation function on large scales
 satisfies $r_{\rm hm}\simeq 1$ \citep[also see][for the similar discussion]{2020arXiv200804913H}.
 This is not necessarily true in Fourier space, because the variations on small scales become extended in Fourier space due to the nature of Fourier transform. The simplest
 example is the shot noise; the shot noise affects the real-space correlations at zero separation, while it behaves like a white noise and 
 affects the power spectrum over all scales ($k$ bins). 

It is also constructive to discuss an asymptotic behavior of the halo correlation functions, $\xi_{\rm gm}$ and $\xi_{\rm gg}$, on large scales.
At scales much greater than the nonlinear scale,  
$r\gg R_\ast$, the galaxy-matter cross correlation  has an asymptotic behavior given as
\begin{align}
\xi_{\rm gm}(r)&\xrightarrow[r\gg R_\ast]{}\frac{1}{\bar{n}_{\rm g}}\int\!\mathrm{d}M \frac{\mathrm{d}n_{\rm h}}{\mathrm{d}M}\avrg{N_{\rm g}}\!(M) \xi_{\rm hm}(r;M)\nonumber\\
&\simeq b_{\rm eff}\xi_{\rm mm}(r),
\end{align}
where we have used $\tilde{u}_{\rm s}(k;M)\rightarrow 1$ for $k\ll 1/R_\ast$, and we defined the effective bias parameter, at a sufficiently large separation satisfying $r\gg R_\ast$, as
\begin{align}
b_{\rm eff}\equiv \frac{1}{\bar{n}_{\rm g}\xi_{\rm mm}(r)}\int\mathrm{d}M~\frac{\mathrm{d}n_{\rm h}}{\mathrm{d}M}\avrg{N_{\rm g}}\xi_{\rm hm}(r;M).
\label{eq:b_eff}
\end{align}
In addition, as can be found from Eq.~(\ref{eq:Sigma}), $\dSigma$, which has a dimension of $[hM_\odot \, {\rm Mpc}^{-2}]$, has an additional dependence 
on $\bar{\rho}_{\rm m0}$
\begin{align}
\dSigma&\propto \bar{\rho}_{\rm m0}\xi_{\rm gm}\propto \Omega_{\rm m}b_{\rm eff}\xi_{\rm mm}.
\label{eq:dsigma_om_dep}
\end{align}
%
Note that the dependence 
on
$h$
is not relevant because all the quantities are measured in units where the dependence of $h$ is factorized out.

Similarly the auto-correlation function of galaxies is found to have an asymptotic behavior at the limit of large scales, $r\gg R_\ast$, as
\begin{align}
\xi_{\rm gg}(r)&\xrightarrow[r\gg R_\ast]{}\frac{1}{\bar{n}_{\rm g}^2}
\left[\int\!\mathrm{d}M~\frac{\mathrm{d}n_{\rm h}}{\mathrm{d}M}\avrg{N_{\rm g}}\!(M)\right]\nonumber\\
&\hspace{2em}\times
\left[\int\!\mathrm{d}M'~\frac{\mathrm{d}n_{\rm h}}{\mathrm{d}M'}\avrg{N_{\rm g}}\!(M')\right]\xi_{\rm hh}(r;M,M')\nonumber\\
&\simeq b_{\rm eff}^2\xi_{\rm mm}(r).
\end{align}
This is not an exact relation, but we found that the relation $r_{\rm gg}\equiv \xi_{\rm gm}/[\xi_{\rm gg}\xi_{\rm mm}]^{1/2}\simeq 1$ holds for scales of 
$r\gg R_\ast$, as shown in Fig.~32 of Ref.~\cite{2018arXiv181109504N}.
Thus 
the combination of
$\xi_{\rm gm}$ and $\xi_{\rm gg}$, which are inferred from $\dSigma$ and $\wgg$, respectively, can be used to infer the underlying 
matter correlation function $\xi_{\rm mm}$, and in turn we can use it to extract cosmological information. Furthermore, 
as implied from Eq.~(\ref{eq:dSigma_1halo}), an amplitude of $\dSigma$ at scales around a transition scale between the 1- and 2-halo terms 
gives an estimate of the average mass of halos hosting galaxies:
\begin{align}
\left.\dSigma_{\rm gm}(R)\right|_{R\simeq {\rm a~few~Mpc}}\rightarrow \bar{M}_{\rm h}
\end{align}
In turn  this can put a strong constraint on the halo bias, $b_{\rm eff}$, via the scaling relation of halo bias with halo mass (or the dependence 
of $\xi_{\rm hm}$ and $\xi_{\rm hh}$ amplitudes on halo mass). Hence, combining the small- and large-scale information of $\dSigma$ with $\wgg$ helps break degeneracies between cosmological parameters and the galaxy bias, and then gives useful constraints on cosmological parameters. This is 
the basic picture of how the joint probes cosmology using $\dSigma$ and 
$\wgg$ can constrain cosmological parameters. 
However, if the scaling relation of the large-scale galaxy bias amplitude with the average halo mass is broken,
this method does not work. This would be the case for the assembly bias effect, where the galaxy bias depends on secondary parameter(s) 
related to the assembly history of host halos, in addition to halo mass. 

In Fig.~\ref{fig:signal_cosmo_dep}, we study how $\dSigma$ and $\wgg$ depend on the cosmological parameters, $\Omega_{\rm m}$ and
$\sigma_8$, because large-scale structure probes are most sensitive to these parameters. Here we consider the expected signals
of $\dSigma$ and $\wgg$ for the SDSS LOWZ-like galaxies at $z=0.251$ (as we will describe in detail in Section~\ref{sec:mocks_for_signals}) 
which are computed using \textsc{Dark Emulator} based on the halo model. 
We note that, due to the non-local nature of $\dSigma$, the lensing profile at a particular scale, say $R_0$, 
is sensitive to the matter-galaxy cross-correlation at $r\sim R_0/2$. 
Note that we employ $R_{\rm  cut}=2$ and $3~h^{-1}{\rm Mpc}$ for the scale cuts of $\dSigma$ and $\wgg$ as our default choice, respectively, and will use the information of $\dSigma$ and $\wgg$ at scales greater than the scale cuts for the cosmology challenges. 

Let us first consider the dependence of $\sigma_8$. 
There are two competing effects in the change of our target observables originating from that of
$\sigma_8$ for models with a fixed $\Omega_{\rm m}$. First, an increase of $\sigma_8$ boosts the amplitude of $\xi_{\rm mm}$ by definition. 
Second, it leads to a decrease of halo bias ($b_{\rm eff}$) for 
massive halos hosting SDSS-like galaxies ($\sim 10^{13}h^{-1}M_\odot$), because a model with 
higher $\sigma_8$ leads to more evolved large-scale structure at an observed redshift, 
thus the
abundance of massive halos 
gets increased
at the redshift,  
and such halos 
become less biased tracers, 
resulting in
a lowered bias amplitude compared to that for the fiducial model. For $\dSigma$, the first effect is more significant; an increase of $\sigma_8$ leads to an increase of the amplitude of $\dSigma$ over all the scales we consider. For $\wgg$, these competing effects almost cancel out on large scales, $R\gtrsim{\rm a~few~Mpc}$ in the 2-halo term, and the increase of $\sigma_8$ does not largely change the amplitude on the large scales.
Nevertheless we should emphasize that the change of $\sigma_8$ causes a scale-dependent modification in these observables, especially $\wgg$, in
contrast to the linear theory that predicts a constant (scale-independent) shift in the ratio at these large scales (also see Figure~6 in \cite{2020arXiv200806873S} for the similar discussion). 
We checked that the scale-dependent changes are from a
combination of the effects of $\sigma_8$ changes on matter clustering ($\xi_{\rm gg}$) and the halo bias function (see Fig.~\ref{fig:xi2h_cosmo_dependence} in Appendix~\ref{sec:2h_cosmo}).
These effects are automatically built in the \textsc{Dark Emultor} predictions. Also note that the positive- and negative-side changes of $\sigma_8$ cause an asymmetric change in $\wgg$ at large scales. 
On the other hand, 
the 1-halo term amplitude of $\wgg$ is boosted because the increase of $\sigma_8$ leads to 
the increased abundance of massive halos, more satellite galaxies reside in such halos, and their correlations add.
Thus the change of $\sigma_8$ leads to characteristic 
modifications of the amplitude and scale dependence in 
$\dSigma$ and $\wgg$. The notable features around $R\sim 90~\hiMpc$ are due to the effect on BAO features \footnote{Due to the line-of-sight projection, the BAO features, which are around $100~h^{-1}{\rm Mpc}$ 
in the three-dimensional correlation function, appear at shorter projected separation.}, 
but this is not relevant for our results because we use the information of $\dSigma$ and $\wgg$ on scales up to $30~\hiMpc$ for the fiducial choice, 
so do not include the BAO information in parameter estimation.

Next we discuss the results for the change of $\Omega_{\rm m}$, for a fixed $\sigma_8$.
First, an increase of $\Omega_{\rm m}$ leads to a faster evolution of clustering growth for models with a fixed $\sigma_8$ (the fixed 
normalization today), so leads to the smaller amplitude of $\xi_{\rm mm}$ 
at this redshift ($z=0.251$). However, recalling that $\dSigma\propto \Omega_{\rm m}\avrg{\gamma_+}\sim \Omega_{\rm m}\delta_{\rm m}$, 
an increase of $\Omega_{\rm m}$ leads to higher amplitudes in $\dSigma$ due to the prefactor, while it leads to smaller amplitudes in 
$\wgg$. In addition, the change of $\Omega_{\rm m}$ causes a scale-dependent modification in both $\dSigma$ and $\wgg$. 

In Fig.~\ref{fig:signal_hod_dep} we study 
how a change in each of the HOD parameters alters $\dSigma$ and $\wgg$ for the {\it Planck} 
cosmology. Here we employ the same HOD parameters for the LOWZ galaxies as in Fig.~\ref{fig:signal_cosmo_dep},
and vary only one HOD parameter for each result, where other parameters are fixed to their fiducial values. 
The figure shows that, when each of the central HOD parameters is varied, it causes a significant change over all the scales including both the 1- and 
2-halo terms, because the change in the parameter leads to a change in  the mean halo mass. More precisely, an increase in $M_{\rm min}$ or a decrease in $\sigma_{\log M}$ leads to an increase in the mean halo mass, leading to the increased amplitudes in $\dSigma$ and $\wgg$. 
The fractional changes in $\dSigma$ and $\wgg$ at large separations in the 2-halo term regime are roughly given as $\delta \wgg/\wgg \simeq 2 \delta\dSigma/\dSigma$,
reflecting the facts $\xi_{\rm gg}\simeq b_{\rm eff}^2\xi_{\rm mm}$ and $\xi_{\rm gm}\simeq b_{\rm eff}\xi_{\rm mm}$ at the large scale limit as we discussed above. 
In addition, it is clear that the 1-halo term amplitude of $\dSigma$ is sensitive to the central HOD parameters, physically to the mean mass of halos hosting the galaxies in a sample. Hence, combining the 2-halo term amplitudes of $\dSigma$ and $\wgg$ allows one to break degeneracies between the 
bias parameter and other parameters, and then adding the 1-halo term information of $\dSigma$ can constrain the HOD parameters and then 
tighten the determination of the large-scale bias. 
The satellite HOD parameters change 
the relative contribution of massive halos to $\dSigma$ and $\wgg$ as satellite galaxies preferentially reside in massive halos. In particular, the parameter $M_1$, which determines the amplitude of the satellite HOD, causes a decent change in the large-scale amplitudes of $\dSigma$ and $\wgg$. The effects of other parameters are relatively mild. 

Comparing Figs.~\ref{fig:signal_cosmo_dep} and \ref{fig:signal_hod_dep} manifests that the cosmological parameters and the HOD parameters lead to 
different changes in $\dSigma$ and $\wgg$. Hence combining these observables in the 1- and 2-halo term regimes allows for an efficient determination of the cosmological parameters, by breaking the parameter degeneracies. 
These figures nicely illustrate 
the
complementarity of $\dSigma$ and $\wgg$, which is the main focus of this paper.

\subsection{Observational effects: geometrical cosmology dependence, 
photo-$z$, multiplicative shear bias, and RSD}
\label{sec:observational_effects}

In this section we discuss the three observational effects, i.e. the geometrical dependence of the observables on cosmology, 
photometric redshift errors of source galaxies, multiplicative shear bias and redshift-space distortion effect. When comparing the 
measured signals with the model templates, we need to include these effects in the model templates, and here we describe how to do this. 

\subsubsection{Geometrical cosmology dependence}
\label{sec:omegam_dependence}

The observables we consider in this paper are $\dSigma(R)$ and $\wgg(R)$, which are different from other possible choices of the lensing and clustering 
observables such as $\gamma_+(\theta)$ and $\wgg(\theta)$. 
Measurements of $\dSigma$ and $\wgg$ require an observer to assume a reference cosmology 
to perform a correction of the lensing efficiency 
$\Sigma_{\rm cr}(z_{\rm l},z_{\rm s})$ (see Eq.~\ref{eq:dSigma_def}) as well as 
the conversion
of angular scales ($\theta$)
and redshift differences 
to the projected separation ($R$) and radial separation ($\pi$). However, the assumed cosmology generally differs from the true underlying cosmology, and this dependence needs to be taken into account. We use the method in Ref.~\cite{2013ApJ...777L..26M} to include this effect. 

For a flat-geometry $\Lambda$CDM model, the relevant parameter is only $\Omega_{\rm m}$ (or $\Omega_{\rm de}$ that is the density parameter of dark energy), because it affects the angular and 
radial distances (the lensing efficiency also depends on the combination of angular diameter distances and the overall factor of $\Omega_{\rm m}$).
Note  that the dependence 
on $h$ is taken out from the observables because all the quantities are measured in units of $h^{-1}$Mpc or $hM_\odot~{\rm Mpc}^{-2}$ for $\wgg$ and $\dSigma$, respectively. The $\Omega_{\rm m}$ dependences of 
$\dSigma$ and $\wgg$ through this effect turn out to be very small, but
add a slight sensitivity of $\Omega_{\rm m}$ in the theoretical templates. 

\subsubsection{Photometric redshift errors}
\label{sec:photoz}

Photometric redshift errors of source galaxies are one of the most serious systematic errors in the weak lensing measurements. An accuracy of photometric redshifts (hereafter often simply photo-$z$), delivered from a set of broad-band filters ($grizy$ in the HSC data), is limited, and can never be perfect, compared to spectroscopic redshifts, 
although the photo-$z$ accuracies are calibrated using the COSMOS catalog \citep{2018PASJ...70S...9T}. 
For lens redshift we assume spectroscopic redshifts as we will focus on the spectroscopic SDSS-like galaxies.
Here we discuss how we can treat a possible uncertainty of source redshifts in the observable $\dSigma$. 
In this paper we use the method proposed in \citet{OguriTakada:11}. In this method we use a ``single'' population of source galaxies, selected based on the photo-$z$,
and then use the relative strengths of
$\dSigma$ for multiple lens samples at different redshifts
to calibrate the photo-$z$ uncertainty of source galaxies in a statistical sense.

In the presence of a redshift distribution of source galaxies, an estimator of $\dSigma(R)$ from the measured ellipticity component of each source galaxy is given, e.g. by Eq.~(11) in Ref.~\cite{Miyatakeetal:15}, as
\begin{align}
\widehat{\dSigma}(R)=\frac{1}{2{\cal R}}\frac{\sum_{{\rm l,s}}w_{\rm ls}
\left[\avrg{\Sigma^{-1}_{{\rm cr}}}_{{\rm ls}}\right]^{-1}e_{s+}}{\sum_{{\rm l,s}}w_{\rm ls}}
\end{align}
where ${\cal R}$ is the ``responsivity'' that is needed to convert the measured galaxy ellipticity, defined in terms of $(a^2-b^2)/(a^2+b^2)$
($a,b$ is the major, minor axes when the galaxy shape is approximated by an ellipse), to the lensing shear, defined in terms 
of $(a-b)/(a+b)$ \citep{BernsteinJarvis:02,Mandelbaumetal:13,HSCDR1_shear:17}; $e_{s+}$ is the tangential ellipticity component of 
the $s$-th source galaxy with respect to the $l$-th lensing galaxy; 
$w_{\rm ls}$ is the weight (see below).
The summation $\sum_{{\rm l,s}}$ runs over all pairs of source and lens galaxies that  are in a given bin of  projected separation, $R=\chi(z_{\rm l})\Delta \theta$. The measured lensing signal is for the effective redshift of lens galaxies, $\bar{z}_{\rm l}$. 

In a case that we have
only the posterior distribution of photometric redshift for each source galaxy, the ``effective'' lensing efficiency needs to be estimated as
\begin{align}
\avrg{\Sigma_{\rm cr}^{-1}}_{{\rm ls}}\equiv \int_0^{\infty}\!\mathrm{d}z_{\rm s}~ \Sigma^{-1}_{\rm cr}(z_{\rm l},z_{\rm s})p(z_{\rm s}),
\label{eq:Sigma_cr_eff}
\end{align}
where $p(z_{\rm s})$ is the posterior distribution of redshift for the $s$-th source galaxy that satisfies the normalization condition $\int_0^{\infty}\!\mathrm{d}z_{\rm s}~p(z_{\rm s})=1$. 
Note that we set $\Sigma_{\rm cr}^{-1}(z_{\rm l},z_{\rm s})=0$ when $z_{\rm s}<z_{\rm l}$.
We employ the weight $w_{\rm ls}$, motivated by the inverse-variance weight, to have a higher signal-to-noise ratio for the weak lensing measurement 
\cite{2004AJ....127.2544S,Mandelbaumetal:05,2018MNRAS.478.4277S}:
\begin{align}
w_{\rm ls}=\frac{\left[\avrg{\Sigma_{\rm cr}^{-1}}_{({\rm l,s})}\right]^2}{e_{{\rm rms}}^2+\sigma_{e}^2}
\end{align}
where $e_{\rm rms}$ is the rms intrinsic ellipticity per component and $\sigma_e$ is the measurement error of ellipticity 
for the $s$-th source galaxy. 

Thus, in an estimator of $\dSigma$, the quantities $\avrg{\Sigma_{\rm cr}^{-1}}_{{\rm ls}}$ and $w_{\rm ls}$ depend on photometric redshifts of source galaxies via $p(z_{\rm s})$ for individual source galaxies. 
If the estimated 
redshift 
distribution of source galaxies
is systematically offset from the underlying true distribution, the measured $\dSigma$
has a systematic bias (offset) from the true one (even if the assumed cosmological model for the $\dSigma$ estimation is the true cosmology). 
Following the method in Refs.~\cite{Hutereretal:06,OguriTakada:11}, we take into account the systematic offset by introducing additional nuisance parameter, $\Delta z_{\rm ph}$, to shift the posterior distribution of source redshift for all source galaxies as 
\begin{align}
p(z_{\rm s}) \longrightarrow p(z_{\rm s}+\Delta z_{\rm ph}).
\label{eq:photoz_shift}
\end{align}
Then we can repeat the computations of $\avrg{\Sigma_{\rm cr}^{-1}}_{{\rm ls}}$ and $w_{\rm ls}$ for the pairs of source and lens galaxies in actual SDSS and HSC-Y1 datasets. 
We found that the lensing profile after shifting the source redshift distribution is well approximated by the following multiplicative form 
as
\begin{align}
\widehat{\dSigma}(R;\bar{z}_{\rm l},\Delta z_{\rm ph})\simeq f_{\rm ph}\!(R; \bar{z}_{\rm l},\Delta z_{\rm ph})\widehat{\dSigma}(R; \bar{z}_{\rm l},\Delta z_{\rm ph}=0),
\label{eq:fphz_def}
\end{align}
where $f_{\rm ph}(R; \bar{z}_{\rm l}, \Delta z_{\rm ph})$ 
is the multiplicative factor to model the effect of systematic photo-$z$ error, and $\bar{z}_{\rm l}$ is the mean of lens redshifts.  
Here we stress, by notation ``$\widehat{\hspace{1em}}$'', that the above correction is made for the measured $\dSigma$.
By using a single population of source galaxies, we notice that the shift $\Delta z_{\rm ph}$ leads to changes in the amplitudes of $\dSigma$ for each of the multiple lens samples (LOWZ and the two subsamples of CMASS, divided into two redshift bins) depending on the lens redshift ($\bar{z}_{\rm l}$). 
Conversely, we can use the relative variations in the $\dSigma$ amplitudes at different lens redshifts to calibrate out
$\Delta z_{\rm ph}$, simultaneously with cosmological parameter estimation. 
This is a 
 self-calibration method of photo-$z$ errors as proposed in \citet{OguriTakada:11}.
In the above equation, we explicitly include the $R$-dependence in the calibration factor, which could arise because the shift could change relative contributions of different 
lens-source pairs to the lensing efficiency $\avrg{\Sigma_{\rm cr}^{-1}}_{\rm ls}$. However, the $R$-dependence, albeit very weak, appears only for the highest-redshift lens sample (CMASS2 in our sample, as we will define later).

For our method, we  multiply the inverse of the calibration factor in Eq.~(\ref{eq:fphz_def}) with the theoretical template of $\dSigma$, rather than correcting for the measurement, for an assumed $\Delta z_{\rm ph}$:
\begin{align}
\dSigma(R)\rightarrow \dSigma(R)/f_{\rm ph}(R;\Delta z_{\rm ph}). 
\end{align}
We then 
include $\Delta z_{\rm ph}$ as additional nuisance parameter when carrying out the parameter inference. This is better because we will use the same covariance matrix in our parameter inference, which allows for apple-to-apple comparison of
the performance for different setups (because this method does not change the error bars of 
$\dSigma$ in each $R$ bin).

\subsubsection{Multiplicative shear errors}
\label{sec:multiplicative_shear}

An accurate weak lensing measurement requires an exquisite, accurate characterization of individual galaxy shapes. This is not straightforward \cite{HSCDR1_shear:17}, and an imperfect shape measurement leaves a residual systematic error in the weak lensing measurements. 
Systematic errors in shape measurements are usually modeled by ``multiplicative'' and ``additive'' biases in the measured galaxy ellipticities, given as $\gamma \rightarrow (1+m)\gamma+c$, where $m$ and $c$ are the multiplicative and additive bias parameters, respectively \citep{Bridleetal:10}.  
Since the spatial positions of lensing galaxies on the sky are considered random with respect to the positions of source galaxies, the galaxy-galaxy weak lensing is not affected by
the additive bias \cite{HSCDR1_shear:17}. Hence, in the presence of the multiplicative shear bias, we modify the theoretical template as
\begin{align}
\dSigma(R) \longrightarrow (1+m)\dSigma(R; m=0).
\end{align}
By working on the {\it single} sample of source galaxies, we can employ the single $m$ parameter for all the lensing profiles at multiple lens redshifts. This is another advantage of the method of Ref.~\cite{OguriTakada:11}. This is a good approximation as long as source galaxies are well separated from lens redshifts; this is the case as long as we can ignore a contamination of source galaxies into lens redshifts due to photo-$z$ errors. To make this method work, in actual data, we will employ a conservative cut of source galaxy redshifts to ensure that source galaxies are well separated from lens redshifts (Miyatake et al., in prep.). 

\subsubsection{Redshift-space distortion}
\label{sec:rsd}

To model the projected correlation function of galaxies, we ignored the redshift-space distortion (RSD) effect that is caused by peculiar velocities of galaxies. For our default choice of the projected length, $\pi_{\rm max}=100~\hiMpc$, however, the RSD effect is not negligible. We follow the method 
proposed in Ref.~\cite{2013MNRAS.430..725V}.
We employ the linear Kaiser formula \cite{Kaiser:87} to model the RSD effect. We follow Eqs.~(51)-(54) in Ref.~\cite{2013MNRAS.430..725V} to
model the redshift-space two-point correlation function of galaxies using the Kaiser RSD $\beta$ factor, given by $\beta\equiv (1/b_{\rm eff})\mathrm{d}\ln D/\mathrm{d} \ln a$, where $b_{\rm eff}$ is the effective linear bias for a sample of galaxies (see Eq.~\ref{eq:b_eff}) and 
$D$ is the linear growth 
factor. Then we modify the theoretical template of $\wgg$ as
\begin{align}
\wgg(R) \rightarrow f_{\rm RSD}(R,\pi_{\rm max})\wgg(R;\beta=0),
\end{align}
where $f_{\rm RSD}(R, \pi_{\rm max})$ is the correction multiplicative factor to account for the RSD effect. This factor depends on the projected 
separation $R$ and the projection length $\pi_{\rm max}$ for an assumed cosmology. For our default choice of $\pi_{\rm max}=100~\hiMpc$, the linear Kaiser factor is a good approximation to model the RSD effect on $\wgg$ \citep[also see][]{2020arXiv200806873S}.

\section{$N$-body simulations, Dark Emulator, and Mock Catalogs of HSC and SDSS-like galaxies}
\label{sec:simulations}

In this paper we use two types of mock catalogs of SDSS- and HSC-like galaxies. First, we use high-resolution $N$-body simulations, with periodic boundary conditions, and the halo catalogs to 
generate the mock catalogs of SDSS galaxies.
We then measure $\dSigma$ and $\wgg$ from the mock catalogs to define the mock signals that we use in cosmology challenges.   
Second, we also use the mock catalogs of SDSS- and HSC-like galaxies built in the light-cone simulations, including the simulated 
lensing signals on the HSC-like source galaxies, and use those catalogs to estimate  the covariance matrix of the observables. 
These mocks are the same as those used in our companion paper, \citet{2020arXiv200806873S}.
In this section, we describe details of $N$-body 
simulations and the mock catalogs.


\subsection{$N$-body simulations and Dark Emulator}
\label{sec:nbody_DE}

\begin{table}
\begin{center}
\caption{The value of each cosmological parameter for the fiducial 
{\it Planck} cosmology which we use in the following cosmology challenges 
(we will study whether the method can recover the true values within the credible interval). 
The column, labeled as ``supporting range'', gives the supporting range of each 
parameter in \textsc{Dark Emulator} that outputs the halo clustering quantities for a flat-geometry $w$CDM model specified 
by a set of 6 cosmological parameters $\{\Omega_{\rm de},\ln(10^{10}A_{\rm s}),\omega_{\rm b},\omega_{\rm c},n_{\rm s},w\}$
each of which should be in the supporting range. Here 
$\sigma_8$ and $S_8\equiv \sigma_8(\Omega_{\rm m}/0.3)^{0.5}$ are derived 
parameters, and the values in table are those 
for the fiducial {\it Planck} cosmology.
 \label{tab:cosmological_parameters_supportingrange} }
\begin{tabular}{l|l|l} \hline\hline
Parameters & fiducial value & supporting range [min,max]  
\\ \hline
$\Omega_{\rm de}$ & 0.6844 & $[0.54752,0.82128]$\\ 
$\ln (10^{10}A_{\rm s})$ & 3.094 & $[2.4752,3.7128]$\\ \hline
$\omega_{\rm b}$& 0.02225& $[0.0211375,0.0233625]$\\
$\omega_{\rm c}$& 0.1198 & $[0.10782,0.13178]$\\
$n_{\rm s}$ & 0.9645& $[0.916275,1.012725]$\\
$w$ & $-1$ & $[-1.2,-0.8]$\\ \hline
$\sigma_8$ & 0.831 & derived\\
$S_8$ & 0.852 & derived \\
\hline\hline
\end{tabular}
\end{center}
\end{table}

In this paper we extensively use \textsc{Dark Emulator} developed in \citet{2018arXiv181109504N}, which is a software package enabling fast, accurate computations of halo clustering quantities for a given cosmological model. Here we will briefly review \textsc{Dark Emulator}. They constructed an ensemble of cosmological of $N$-body simulations, each of which was performed with $2048^3$ particles for a box with $1$ or $2~{\rm Gpc}/h$ on a side length, for 101 cosmological models within the flat $w$CDM cosmologies. The $w$CDM cosmology is parametrized by 6 cosmological parameters, 
${\bf p}=\{\omega_{\rm b},\omega_{\rm c}, \Omega_{\rm de},\ln(10^{10}A_{\rm s}),n_{\rm s}, w\}$, where 
$\omega_{\rm b}(\equiv \Omega_{\rm b}h^2)$ 
and $\omega_{\rm c}(\equiv \Omega_{\rm c}h^2)$ are the physical density parameters of baryon and CDM, respectively, $h$ is the Hubble parameter, $\Omega_{\rm de}\equiv 1-(\omega_{\rm b}+\omega_{\rm c}+\omega_\nu)/h^2$ is the density parameter of dark energy for a flat-geometry universe, 
$A_{\rm s}$ and $n_{\rm s}$ are the amplitude and tilt parameters of the primordial curvature power spectrum normalized at $k_{\rm pivot}=0.05~{\rm Mpc}^{-1}$, and $w$ is the equation of state parameter for dark energy, respectively. For the $N$-body simulations, they included the neutrino effect fixing the neutrino density parameter $\omega_\nu\equiv \Omega_\nu h^2$ to 0.00064 corresponding to 0.06~eV for the total mass of three neutrino species. They included the effect of massive neutrinos only in the initial linear power spectrum \citep[see][for details]{2018arXiv181109504N}. 
To carry out ``cosmology challenges'' in the following, we employ the fiducial {\it Planck} cosmology that is characterized by the parameter values in Table~\ref{tab:cosmological_parameters_supportingrange}.
We use the $N$-body simulation realizations of $1~h^{-1}{\rm Gpc}$ box size 
for the fiducial {\it Planck} cosmology
 to construct the mock catalogs for SDSS-like galaxies.
The mass of simulation particle for the fiducial {\it Planck} simulations is $m=1.02\times 10^{10}~h^{-1}M_\odot$. 
In the following we use halos with mass greater than $10^{12}~h^{-1}M_\odot$, corresponding to about 100 simulation particles.

For each $N$-body simulation realization (each redshift output) for a given cosmological model, they constructed a catalog of halos using \textsc{Rockstar} \citep{Behroozi:2013} that 
identifies halos and subhalos based on clustering of $N$-body particles in phase space (position and velocity space). The spherical overdensity mass, with respect to the halo center that is defined from the maximum mass density, 
$M\equiv M_{\rm 200 m}=(4\pi/3)R_{\rm 200m}^3\times (200\bar{\rho}_{\rm m0})$, is used for definition of  halo mass, where 
$R_{\rm 200m}$ is the spherical halo boundary radius within which the mean mass density is 200 times $\bar{\rho}_{\rm m0}$.  
By combining the outputs of $N$-body simulations and the halo catalogs at multiple redshifts in the range $z=[0,1.48]$, they built an emulator, 
named as \textsc{Dark Emulator}, which enables fast and accurate computations of the halo mass function, halo-matter cross-correlation, and halo auto-correlation as a function of halo masses, redshift, separations and cosmological models. 
It was shown that \textsc{Dark Emulator} achieves a sufficient accuracy for these statistical quantities for halos of 
$10^{13}M_\odot$, which is a typical mass of host halos of SDSS galaxies, compared to the statistical measurement errors of $\dSigma$ and $\wgg$
expected from the HSC and SDSS data, as shown in Fig.~31 of the paper. 
In summary, \textsc{Dark Emulator} outputs
\begin{itemize}
\item $\frac{\mathrm{d}n_{\rm h}}{\mathrm{d}M}(M; z,{\bf p})$: the halo mass function for halos in the mass range $[M,M+\mathrm{d}M]$ 
at redshift $z$
\item $\xi_{\rm hm}(r; M, z, {\bf p})$: the halo-matter cross-correlation function for a sample of halos in the mass range $[M,M+\mathrm{d}M]$
 at redshift $z$
\item $\xi_{\rm hh}(r; M, M', z,{\bf p})$: the halo-halo auto-correlation function for two samples of halos with masses $[M,M+\mathrm{d}M]$ 
and $[M',M'+\mathrm{d}M']$ at redshift $z$
\end{itemize}
for an input set of parameters, halo mass $M$ (and $M'$ for the correlation function of two halo samples), redshift $z$, and cosmological parameters 
${\bf p}$. 
In addition, the \textsc{Dark Emulator} package outputs auxiliary quantities, based on emulation, such as 
the linear halo bias (the large-scale limit of the halo bias), the linear mass power spectrum, the linear rms mass fluctuations of halo mass scale 
$M$ ($\sigma^L_{\rm m}(M)$), and 
$\sigma_8$. 
The supporting range of each of cosmological parameters for \textsc{Dark Emulator} is given in Table~\ref{tab:cosmological_parameters_supportingrange}.
These ranges are sufficiently broad, e.g. to cover the range of cosmological constraints from the current state-of-the-art large-scale structure probe 
such as the Subaru HSC cosmic shear results \cite{2019PASJ...71...43H,2020PASJ...72...16H}.
In this paper we will use \textsc{Dark Emulator} to perform MCMC analyses of cosmological parameters by comparing the model templates of $\dSigma$
and $\wgg$ with the mock signals expected from the SDSS and HSC data.

\subsection{Mock catalogs for the SDSS-like galaxies for the mock signals of $\dSigma$ and $\wgg$}
\label{sec:mocks_for_signals}
\begin{table}
	\centering
    \caption{
		Specifications of the mock galaxy catalogs that resemble the LOWZ and CMASS galaxies for the spectroscopic SDSS DR11 data. 
		For the CMASS sample we consider two subsamples divided into two redshift ranges. We give the redshift range and the comoving volume for each sample, assuming $8300$~deg$^2$ for the area coverage. The column denoted as ``representative redshift'' is a representative redshift of each sample which we assume to represent the clustering observables for each sample (i.e. ignore redshift dependences of the observables within the redshift bin). 
    The lower columns, 
    below double lines, denote the HOD parameters used to build the mock catalogs of each sample from $N$-body simulations for the {\it Planck} cosmology. Note that $M_{\rm min}$ 
    and $M_1$ 
    are in units of $h^{-1}M_\odot$.
	}
    \label{tab:sample_HODparameters}
    \begin{tabular}{l|ccc} \hline \hline
         & LOWZ & CMASS1 & CMASS2 \\ 
		\hline
        redshift range  & $[0.15,0.35]$ & $[0.47,0.55]$ & $[0.55,0.70]$ \\ 
        representative redshift  & $0.251$ & $0.484$ & $0.617$ \\ 
		volume [$(h^{-1}{\rm Gpc})^3$] & 0.67 & 0.81 & 2.00 \\
	    \hline \hline
		HOD parameters & \multicolumn{3}{c}{fiducial values}\\ \hline
		$\log M_{\rm min}$ & $13.62$   & $13.94$   & $14.19$  \\
		$\sigma_{\log{M}}$ & $0.6915$ & $0.7919$ & $0.8860$  \\
		$\kappa$ & $0.51$   & $0.60$   & $0.066$ \\
		$\log M_1$         & $14.42$   & $14.46$   & $14.85$  \\
		$\alpha$           & $0.9168$  & $1.192$   & $0.9826$ \\ \hline 
    \end{tabular}
\end{table}

\subsubsection{Mock catalogs of SDSS LOWZ- and CMASS-like galaxies}

We use the $N$-body simulation realizations and the halo catalogs for the {\it Planck} cosmology to build mock catalogs of galaxies 
that resemble spectroscopic galaxies in the SDSS-III BOSS DR11
sample \footnote{\url{https://www.sdss.org/dr11/}}.
We consider three galaxy samples in three redshift bins: ``LOWZ'' galaxies in the redshift range $z=[0.15,0.30]$ and 
two subsample of ``CMASS'' galaxies that are obtained from subdivision of CMASS galaxies into two redshift bins, $z=[0.47,0.55]$
and $z=[0.55,0.70]$. Here we consider luminosity-limited samples rather than flux-limited samples for these galaxies (see Miyatake et al. in preparation for details)
\citep[also see][for the similar discussion on the stellar-mass limited sample]{Miyatakeetal:15}.
The luminosity-limited sample is considered as a nearly volume-limited sample in each redshift bin, and we expect that 
properties of galaxies do not strongly evolve within the redshift bin, which is desired because we ignore the redshift evolution of clustering observables within the redshift bin for each sample. Table~\ref{tab:sample_HODparameters} summarizes characteristics of each galaxy sample. 

To build the mock catalogs for each galaxy sample, we first
perform a fitting of the HOD model predictions to the projected correlation function 
$\wgg$ that is actually measured from the SDSS data for each sample, assuming the {\it Planck} cosmology model, and then estimate the HOD parameters for the best-fit model. 
Note, however, that we here 
use the analytical halo model method used in \citet{2015ApJ...806....2M}, and 
did not use the weak lensing profile to further constrain the HOD parameters.
The fiducial HOD model is the same as that described in Section~\ref{sec:halo_model}. Table~\ref{tab:sample_HODparameters} gives the best-fit HOD parameters for each sample. 

We use the outputs of $N$-body simulations at $z=0.251$, 0.484 and 0.617 as representative redshifts of the galaxy samples to build the mock catalogs. 
We then populate galaxies into halos of each realization for the {\it Planck} cosmology, using the best-fit HOD parameters in Table~\ref{tab:sample_HODparameters} \citep[also see][for details of the method]{2019arXiv190708515K,2020arXiv200506122K}. 
For our default mocks, we assume the NFW profile for the radial distribution of satellite galaxies. 
The default mock is the same as the default HOD model used in the theoretical template (see Section~\ref{sec:halo_model}), so we will use the default mock to perform a sanity check of whether we can recover the true cosmological parameters,
i.e. the parameter values of {\it Planck} cosmology, in the parameter estimation by comparing model templates with the mock signals measured 
from the default mocks. In the default mocks, we neither include the off-centering effect of central galaxies, 
an 
incompleteness effect of central galaxies, nor the redshift-space distortion effect. For other mocks, we include these effects one by one, and then 
study the impact of each effect on parameter estimation. 

\subsubsection{Mock signals of $\dSigma$ and $\wgg$}
\label{sec:mock_signal_measuerements}

\begin{figure*}
    \begin{center}
        \includegraphics[width=2.0\columnwidth]{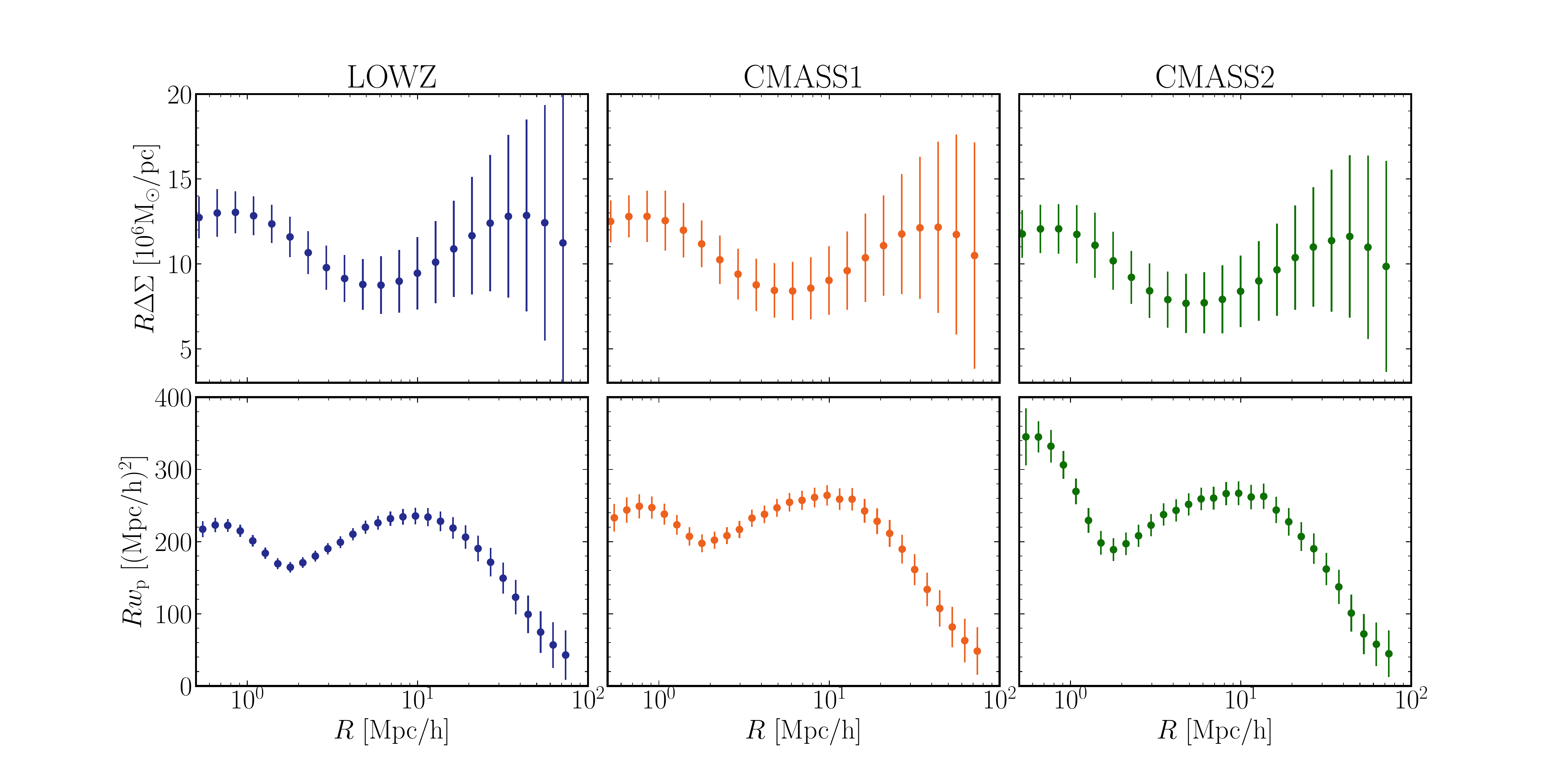}
        \caption{Mock signals of $\dSigma$ and $\wgg$ for the SDSS LOWZ, CMASS1 and CMASS2 galaxies at the representative 
        redshifts, $z=0.251$, 0.484 and 0.617, respectively. The signals are computed from the mock catalogs of the galaxies that are constructed by applying 
        the HOD model (Table~\ref{tab:sample_HODparameters}) to halo catalogs in high-resolution $N$-body simulations (see text for details). 
        The error bars in each bin denote the statistical errors expected from the SDSS DR11 data
        (8,300~sq.~deg.) and the HSC-Y1 data (140~sq.~deg.), which are the square root of the diagonal components of each 
        covariance matrix that is estimated from the mock catalogs of HSC and SDSS galaxies in the light-cone simulations.
        Here we multiply $R$ by $\dSigma(R)$ and $\wgg(R)$ for illustrative purpose. 
        }
        \label{fig:mock_signals}
    \end{center}
\end{figure*}
We generate the mock signals of $\dSigma$ and $\wgg$ for the SDSS-like galaxies, by measuring those clustering observables from each 
of the mock catalog realizations for the {\it Planck} cosmology as we described in the preceding section. 
In doing this, we take an advantage of the periodic boundary conditions in 
each mock,  which allows for a fast computation of the clustering quantities using the FFT algorithm. 
In addition, the measured clustering quantities are not affected by the window function thanks to the periodic boundary conditions. 
We here describe details of the measurement method of $\dSigma$ and $\wgg$ from 
each mock catalog \citep{2019arXiv190708515K,2020arXiv200506122K}. 

For $\dSigma$, 
we first project the matter ($N$-body) particles and galaxies along one axis of $N$-body cubic box assuming that the axis is along the line-of-sight 
direction,  
and 
assign the matter particles and galaxies to the $46,332^2$ two-dimensional grids using the Nearest Grid Point (NGP) interpolation kernel. 
We then Fourier transform the matter and galaxy density fields and take a product of the two, $\Re \left[ \tilde{\delta}^{\rm 2D}_{\rm g}(\bk_\perp) \tilde{\delta}^{\rm 2D}_{\rm m}(-\bk_\perp) \right]$, where $\bk_\perp$ is the two-dimensional wavevector. 
We perform 
the inverse 2D Fourier transform to obtain
\begin{align}
    {\xi}_\mathrm{gm}^\mathrm{2D}(\mathbf{R}) = \int \frac{\mathrm{d}^2 {\bf k}_\perp}{(2\pi)^2} e^{i\bk_\perp \cdot \mathbf{R}} 
    \Re \left[ \tilde{\delta}^{\rm 2D}_{\rm g}(\bk_\perp) \tilde{\delta}^{\rm 2D}_{\rm m}(-\bk_\perp) \right],
\end{align}
where $\mathbf{R}$ is the two-dimensional separation vector perpendicular to the projection direction. 
We perform the azimuthal angle average in an annulus of each radial bin and then obtain 
the two-dimensional galaxy-matter cross correlation function, $\xi_\mathrm{gm}^\mathrm{2D}(R)$.

We compute the projected surface mass density profile around galaxies from $\xi_\mathrm{gm}^{\mathrm{2D}}(R)$, according to 
Eq.~(\ref{eq:Sigma}), as
\begin{align}
    \Sigma_{\rm gm}(R) \simeq \bar{\rho}_{\rm m0}\xi^{\rm 2D}_\mathrm{gm}(R).
\end{align}
The above surface mass density differs from Eq.~(\ref{eq:Sigma}) by a constant additive term (spatially-homogeneous term), as this is okay 
because such a constant term is irrelevant to the weak lensing shear or the excess surface mass density profile.
To achieve higher spatial resolution, we measure $\Sigma_{\rm gm}(R)$ using the folding method of the FFT box \cite{Jenkins1998}.
Following the method described in Refs.~\cite{2011A&A...527A..87V,Takahashi_2012}, we perform multiple measurements in which we fold the box different times (we denote the folding times as $n_{\rm fold}$).
We have six measurements of $n_{\rm fold} = \{0,1,2,3,4,5\}$ for each catalog, while the grid number $46,332^2$ is kept unchanged at each FFT step.
It means that we have measurements which have finer resolutions by up to a factor of $2^5 = 32$.
We then combine the $\Sigma_{\rm gm}(R)$ signals obtained by stitching the six measurements between
the five boundary scales $\{0.125,0.25,0.5,0.1,0.2\}\,h^{-1}\,\mathrm{Mpc}$. Then we compute the excess surface mass density profile, $\dSigma(R)$ from $\Sigma(R)$, 
according to Eq.~(\ref{eq:dSigma_def}).
In each measurement, to achieve better statistics we perform three measurements using the 
projection along $x$-, $y$- or $z$-axis directions, and use the average of the three results as the measured signal of $\dSigma$ for the realization.

For $\wgg$,
we first assign the mock galaxies to the $1024^3$ three-dimensional FFT grids using the NGP kernel.
We then Fourier transform the number density field, and take its square amplitudes $|\tilde{\delta}_{\rm g}(\bk)|^2$ for each wavevector $\bk$.
By performing the inverse Fourier transform, we obtain the estimate of the correlation function at each spatial separation ${\bf r} = (R, \pi)$,
\begin{align}
\label{eq:estimate_corr}
    {\xi}_{\rm gg}({\bf r}) = \int \frac{\mathrm{d}^3{\bf k}}{(2\pi)^3} \, e^{i\bk \cdot {\bf r}}\,\left|\tilde{\delta}_{\rm g}(\bk)\right|^2.
\end{align}
Then we estimate the projected correlation function, $\wgg(R)$, from the azimuthal angle average in the circular annulus of each radial bin $R$ and 
the line-of-sight projection over $\pi=[0,\pi_{\rm max}]$, according to Eq.~(\ref{eq:wp_def}).

We use 19 and 22 independent realizations that are built using the different seeds of the initial conditions for $\dSigma$ and $\wgg$, respectively
\citep{2018arXiv181109504N} \footnote{We use the slightly different number of realizations for $\dSigma$ and $\wgg$, because the $N$-body simulation data for the 4 realizations in the difference were lost in the middle of this work. The difference is small compared to the statistical measurement errors for the SDSS and HSC data, and the main results of this paper are not changed.}. 
We then measure the average
mock signals of $\dSigma$ and $\wgg$ for each of the galaxy samples (Table~\ref{tab:sample_HODparameters}).
These correspond to 19 and 22~$(h^{-1}{\rm Gpc})^3$ volumes, respectively, that are larger than the volume of any of the three redshift slices 
in Table~\ref{tab:sample_HODparameters} by at least a factor of 11. Recalling that the overlapping area of HSC-Y1 data and SDSS is only about 
140~sq. deg., compared to the SDSS area coverage of 8,300 sq. deg., the effective volume for the $\dSigma$ measurement of each sample is smaller than 
that of $\wgg$ by a factor of 60. Hence the simulation volume used for the mock signal of $\dSigma$ is larger than that of the $\dSigma$ measurement by at least a factor of 650. Thus by using the mock signals for the larger volumes in cosmology challenges, we can minimize any unwanted bias in estimated 
parameters due to sample variance.
In this way we can evaluate the performance of each method, i.e. the ability to recover the true cosmological 
parameters, without being affected by the sample variance. 

The data points in each panel of Fig.~\ref{fig:mock_signals} show the mock signals of $\dSigma$ and $\wgg$ for each of the LOWZ, CMASS1 and 
CMASS2 samples (see Table~\ref{tab:sample_HODparameters}) for the {\it Planck} cosmology. The error bars in each bin are the statistical 
errors expected for the SDSS and HSC-Y1 surveys as we will explain below. The figure shows that a sufficient number of  the realizations of the mock catalogs 
lead to well-converged, smooth signals in each bin, and the statistical scatters appear to be negligible. This allows us to robustly evaluate the performance of the method for cosmological parameter estimation. 

\subsection{Light-cone mock catalogs of HSC- and SDSS-like galaxies for estimating the covariance}
\label{sec:light_cone_mock}

\begin{table}
	\centering
\caption{
		The cumulative signal-to-noise ($S/N$) ratios of $\dSigma$, $\wgg$ and the joint measurements for the LOWZ, CMASS1 and CMASS2 samples, which are estimated using the mock signals and the covariance matrices. 
		Here we define the ``cumulative'' $S/N$ over the ranges of 
		$R/[h^{-1} \, {\rm Mpc}]=[3,30]$ and $[2,30]$ for $\dSigma$ and $\wgg$, respectively, which are our baseline choices of the radial range (see text for details). For the ``total'' $S/N$ of $\dSigma$ we take into account the cross-covariances between $\dSigma$'s of different galaxy samples. 
    We assume that
    the $\wgg$-signals for the three samples are independent from each other, and ignore the cross-covariances between $\dSigma$ and $\wgg$.
	}
	\begin{tabular}{c|ccc|c} \hline\hline
		& LOWZ & CMASS1 & CMASS2 & total \\
		\hline
		$\dSigma$ & 8.64 & 8.86 & 8.48 & 14.7 \\
        $\wgg$ & 32.9 & 32.1 & 30.4 & 54.7 \\
        joint ($\dSigma+\wgg$) & 34.0 & 33.3 & 31.6 & 56.6 \\ \hline
	\end{tabular}
	\label{tab:signal-to-noise-ratio}
\end{table}
As we described in Section~\ref{sec:observables}, 
$\dSigma$ is independent of source redshift, and depends only on the galaxy-matter 
cross correlation at lens redshift. Based on this fact, we construct the best-available, accurate mock signal for $\dSigma$ from the mock catalogs of SDSS-like galaxies as we described above. However, the lensing effects on the same population of source galaxies by foreground structures at different redshifts, from the redshifts of SDSS galaxies, -- cosmic shear causes statistical scatters in the observed galaxy ellipticitiies. We need to properly take into account these effects. In this subsection, we describe the mock catalogs of HSC- and SDSS-like galaxies that are built in the light-cone simulations, and then use the mock catalogs to model the covariance matrices of $\dSigma$ and $\wgg$.

To construct the mock catalogs in a light-cone volume, we use the full-sky, light-cone simulations generated in \citet{2017ApJ...850...24T}.
The light-cone simulation consists of multiple spherical shells with an observer being at the center of the sphere, and each spherical shell contains 
the lensing fields and the halo distribution, where the lensing fields at the representative redshift of the shell
can be used to simulate the lensing distortion effect on a galaxy at the position by foreground structure if the galaxy is located within the shell. 
The halo distribution in each shell reflects a realization of halos in large-scale structure at the redshift corresponding to the radius of the shell (the distance from an observer to the shell). In this paper we use 108 realizations of the full-sky simulations. 

As described in Appendix~\ref{sec:covariance} in detail, we populate HSC- and SDSS-like galaxies in the full-sky, light-cone simulation. For the HSC galaxies, we use the actual  HSC shape catalog \cite{HSCDR1_shear:17}, used for the HSC-Y1 weak lensing measurements \cite{2019PASJ...71...43H}, and populate each 
galaxy into the corresponding shell in the light-cone simulation according to its angular position (RA and dec) and photometric redshift (best-fit 
photo-$z$). Then we simulate the lensing signal on each galaxy using the lensing information of light-cone simulation. Thus the mock HSC catalog includes properties of actual data (the angular positions and the distributions of ellipticities and photo-$z$'s) as well as the geometry and masks of the HSC footprints. Since the HSC-Y1 data still has a small area coverage (140~sq.deg.), we 
identify 21 footprints of the HSC-Y1 data in each of the all-sky, light-cone simulation realizations \cite{2017MNRAS.470.3476S,2019MNRAS.486...52S}.
We thus generate 2268 mock catalogs of the HSC data in total. 

For the SDSS galaxies, we populate galaxies into halos in the light-cone simulation based on the HOD method. We built mock catalogs for each of the three SDSS-like galaxies (LOWZ, CMASS1, and CMASS2) in their corresponding redshift ranges in the assigned survey regions of the SDSS DR11 survey footprints. 
Given the large survey area of SDSS data  (about 8,000~sq.~deg.), we identify only one SDSS region in each realization of the light-cone simulations, and thus build 108 mock catalogs for each of the SDSS galaxies in total. 

Each of our mock catalog realizations in the light-cone volume not only contains the angular and redshift (radial) distributions of HSC and SDSS galaxies, but also contains
the lensing effects on each source galaxy by the SDSS galaxies and other foreground structures at different redshifts from the SDSS redshift. 
We then perform measurements of $\dSigma$ and $\wgg$ from each realization using the same analysis pipelines that are used in the actual measurements of the real HSC and SDSS data. Finally we estimate the covariance for the galaxy-galaxy weak lensing $\dSigma$ for each of the SDSS LOWZ, CMASS1 and CMASS2 galaxies, from the scatters among the 2268 measurements. The covariance matrices of $\dSigma$ for the different galaxy samples 
at different lens redshifts have cross-covariance components because the measurements use the same source galaxies and are affected by weak lensing due to the same foreground structure (cosmic shear) in each light-cone simulation realization. Our covariance properly takes into account the cross-covariance. 
On the other hand, for the covariance of $\wgg$, we apply the jackknife method to 
estimate the covariance of $\wgg$ for each realization, and then use the average of the 108 covariances as the estimate of 
the $\wgg$ covariance used in this paper. 

The error bars in each bin in Fig.~\ref{fig:mock_signals} are estimated from the covariance matrices we described above. The figure shows that 
the HSC-Y1 data allows for a significant detection of the lensing signals at each bin. 
Table~\ref{tab:signal-to-noise-ratio} gives the cumulative signal-to-noise ($S/N$) ratios expected for measurements of $\dSigma$
and $\wgg$ from the HSC-Y1 and SDSS data and the joint measurements. Here, as for our default choice of the range of separation scales, 
we adopt $3\le R/[h^{-1}{\rm Mpc}]\le 30$ for $\dSigma$ and $2\le R/[h^{-1}{\rm Mpc}]\le 30$ for $\wgg$, respectively. The table gives the 
cumulative $S/N$ integrated over the separation ranges properly taking into account the covariance and the cross-covariance matrices. 
It is clear that $\wgg$ has a grater $S/N$ value than $\dSigma$ does by a factor of 4, because of the much wider area coverage of SDSS data compared to the HSC-Y1 data by a factor of 60. 
Nevertheless, we will show later that combining $\dSigma$ and $\wgg$ is crucial to 
lift parameter degeneracies and obtain useful cosmological constraints. 

In the following, we do not include the cosmology dependence of the covariance matrix, motivated by the discussion in \citep{2019OJAp....2E...3K}. 
Hence the differences in the performance of different setups/methods that we will show below 
are purely from the differences in the model or setups.

\section{Methodology for cosmology challenges}
\label{sec:methodology}

The purpose of this paper is to study whether the halo model based method can recover the true cosmological parameters from a hypothetical parameter inference, i.e.  comparing 
the theoretical templates with 
the mock signals, taking into account the error covariance in the likelihood analysis. 
Here we describe our methodology to perform cosmology challenges.  We employ different setups of the analysis method to quantify the impact of various effects on cosmological parameter estimation, and describe each setup and its purpose.

\begin{table}
\begin{center}
\caption{Prior range of each model parameter. In cosmology challenges of parameter estimation, we consider
 two cosmological parameters, $\Omega_{\rm de}$, and $\ln(10^{10}A_{\rm s})$, and consider 
 5 HOD parameters for each of the LOWZ, CMASS1 and CMASS2 samples. Hence we have 17 model parameters in total for the baseline method. 
 For an extended method, we also include the nuisance parameters to model the effects of photo-$z$ errors ($\Delta z_{\rm ph})$ 
 and the multiplicative shear bias ($m_\gamma$) for which we employ the Gaussian prior with width given in the number in the table. 
 \label{tab:priors} }
\begin{tabular}{l|l|l} \hline\hline
Parameters & prior range [min,max]  
\\ \hline
$\Omega_{\rm de}$ & [0.54752, 0.82128] \\ 
$\ln (10^{10}A_{\rm s})$ & [2.4752, 3.7128] \\ \hline
$\log M_{\rm min}$& [12.0,14.5]\\ 
$\sigma_{\log M}^2$& [0.01,1.0]\\
$\log M_1$& [12.0,16.0] \\
$\kappa $ & [0.01,3.0] \\
$\alpha$& [0.5,3.0] \\ \hline
$\Delta z_{\rm ph}$ & Gauss: 0.04 or 0.1\\
$m_\gamma$ & Gauss: 0.01\\
\hline\hline
\end{tabular}
\end{center}
\end{table}

\subsection{Parameter estimation method}
\label{sec:MCMC}

We assume that the likelihood of the mock signals given the model parameters 
follows a multivariate Gaussian distribution: 
\begin{align}
\ln {\cal L({\bf d}|\boldsymbol{\theta})}&=
-\frac{1}{2}\sum_{i,j}\left[d_i-t_i(\boldsymbol{\theta})\right]({\bf C}^{-1})_{ij} \left[
d_j-t_j(\boldsymbol{\theta})
\right]\nonumber\\
&\hspace{3em}+{\rm const.}, 
\label{eq:likelihood}
\end{align}
where ${\bf d}$ is the mock data vector, ${\bf t}$ is the model vector computed from the theoretical template as a function of the 
model parameters ($\boldsymbol{\theta}$), ${\bf C}^{-1}$ is the inverse of the covariance matrix, and the summation runs over the indices, $i,j$, corresponding to the dimension of the data vector. 
In our baseline analysis we use, as the data vector, $\dSigma(R)$ given in 
9 radial bins, logarithmically-evenly spaced 
over $3\le R/[\hiMpc]\le 30$, and 
$\wgg(R)$ in 16 radial bins over $2\le R/[\hiMpc]\le 30$, respectively, for each galaxy sample
\footnote{Precisely speaking, the binning of $R$ does not have an exact cut at $2$, 3  or 30$~\hiMpc$ for the lower and upper cuts. We employ the 30 bins  
logarithmically spaced over 
$0.05\le R/[\hiMpc]\le 80$, and use the bins residing the fitting range of $\dSigma$ and $\wgg$.}. 
Thus we use 75 data points in total ($3\times (9+16)=75$). 
Note that, for our default analysis, we do not include the abundance information of galaxies ($\bar{n}_{\rm g}$) in parameter inference.
Even if we employ a weak prior on the abundance (e.g., 50\% of the number density), the following results remain almost unchanged, but 
will come back to a question of whether a mild use of the abundance information can improve the parameter estimation.

For the model parameters ($\boldsymbol{\theta}$) in Eq.~(\ref{eq:likelihood}), we consider the two cosmological parameters, $\Omega_{\rm de}$ and $\ln(10^{10}A_{\rm s})$, 
and 5 HOD parameters for each of the LOWZ, CMASS1, and CMASS2 samples (Table~\ref{tab:sample_HODparameters}).
Hence we have 17 parameters ($2+3\times 5 = 17$) in total. Since the clustering observables are primarily sensitive to the amplitude parameters 
$\Omega_{\rm de}$ and $A_{\rm s}$ for the flat $\Lambda$CDM cosmology, we consider only the two parameters, and fix other cosmological parameters to their values of the {\it Planck} cosmology in parameter inference. Hence the degrees of freedom for the fiducial analysis is $N_{\rm dof}=N_{\rm d}-N_{\rm p}=75-17=58$. If we include further nuisance parameters to model the photo-$z$ errors and/or the shear multiplicative bias, 
we include up to 19 parameters (i.e. 1 or 2 additional parameters).
For a given set of the model parameters, we can compute the model vector ${\bf t}$ at each radial bin. 

We then perform parameter estimation based on the Bayesian inference: 
\begin{align}
{\cal P}(\boldsymbol{\theta}|{\bf d})\propto {\cal L}({\bf d}|\boldsymbol{\theta})\Pi(\boldsymbol{\theta}),
\end{align}
where ${\cal P}(\boldsymbol{\theta}|{\bf d})$ is the posterior distribution of $\boldsymbol{\theta}$ and $\Pi(\boldsymbol{\theta})$ is the 
prior distribution. Throughout this paper we employ a flat prior on each model parameter as given in Table~\ref{tab:priors}.
We checked that the priors of the cosmological parameters are wide enough, and the following results for the posterior distribution are not affected by the prior range.

We draw samples from the posterior distribution of parameters, given the mock signals, with the help of nested sampling as implemented 
in the publicly available package \textsc{Multinest} (Multi-Modal Nested Sampler) \cite{2009MNRAS.398.1601F} together with the package 
\textsc{Monte Python} \cite{2013JCAP...02..001A}
to sample the posterior distribution of the parameters. 
In the following we mainly focus on  the posterior distributions of $\Omega_{\rm m}$, $\sigma_8$ and $S_8\equiv \sigma_8(\Omega_{\rm m}/0.3)^{0.5}$. 
These are derived parameters from the cosmological parameters we use ($\Omega_{\rm de}$, $\ln(10^{10}A_{\rm s})$), and 
we employ the definition of $S_8$ following 
\citet{2019PASJ...71...43H} so that the forecast of $S_8$ estimation from our method is compared to the previous results. 

In this paper we adopt the mode of the marginalized 1D or 2D posterior distribution to infer the central value(s) of the parameter(s), and the 
highest density interval of the marginalized posterior to infer the credible interval of the parameter(s). We often report the best-fit parameters 
that correspond to a model at the maximum likelihood in a multi-dimensional parameter space.
As stressed in 
\cite{2020arXiv200806873S} \citep[also see][]{2020arXiv200701844J}, 
a point estimate of parameter is not useful, because 
it
is sensitive to the degree of degeneracies between parameters. 
For example, even if we consider an ideal case that the input signals are from the model predictions, 
the central value of a parameter, estimated from the mode of the marginalized posterior distribution, does not necessarily recover the true 
value as a result of marginalization of the parameters, if the target parameter is highly degenerate with other parameters. Rather a more useful quantity is the credible interval. Hence in the following we will mainly focus on the credible interval, and evaluate each method/setup to study whether the true value of the cosmological parameters are recovered to within the 68\% credible interval. We are not interested in an accuracy of recovery of the HOD parameters, and we will not pay much attention to the HOD parameters.

\begin{table*}
\caption{A summary of the analysis setups. \label{tab:analysis_setups}}
\begin{center}
\begin{tabular}{l|ccl}\hline\hline
setup & scale cut & sample parameters & Note \\ 
& $[h^{-1}{\rm Mpc}]$ &  
\\ \hline 
baseline & $(2,3)$ &
$(\Omega_{\rm m},\sigma_8)$+HOD (17 paras.) & fiducial mock ($R_{\rm max}=30~\hiMpc$) \\ \hline
+RSD & (2,3) & -- & include RSD effect in the fiducial mock 
\\ \hline 
scale cuts & $(0.5,0.75)$ & -- & -- \\
& $(1,1.5)$ & -- & -- \\
& $(8,12)$ & -- & -- \\ \hline
$R_{\rm max}=70~\hiMpc$ & (2,3) & -- & use up to $R_{\rm max}=70~\hiMpc$ \\
\hline
$\dSigma$ alone & (2,3) & -- & -- \\ 
$w_{\rm p}$ alone & (2,3) & -- & -- \\ \hline
$\Omega_{\rm m}$-goem. & (2,3) & -- & via ($\Sigma_{\rm cr},R$) \\
photo-$z$ error ($\Delta z_{\rm ph}$) & (2,3) & +$\Delta z_{\rm ph}$ & only in model\\ 
shear-$m$ ($\Delta m_\gamma$) & (2,3) & + $\Delta m_\gamma$ & only in model \\
$\Omega_{\rm m,geom}$+$\Delta z_{\rm ph}$+$\Delta m_\gamma$ & (2,3) & + ($\Omega_{\rm m, geom},\Delta z_{\rm ph}, \Delta m_\gamma) $ & only in model\\ \hline
full HSC & (2,3) & -- & use $0.1\times {\rm Cov}_{\Delta\Sigma}$ mimicking the full HSC data\\ 
\hline\hline
\end{tabular}
\end{center}
\end{table*}

\subsection{Validation strategy against analysis setups: 
scale cuts, parameter degeneracies, observational effects and RSD}
\label{sec:analysis_setups}

An advantage of the emulator based halo model is that \textsc{Dark Emulator}
gives an accurate prediction of the halo correlation functions 
($\xi_{\rm hm}$ and $\xi_{\rm hh}$) including all the nonlinear effects down to small scales (nonlinear clustering, nonlinear bias, and the halo exclusion effect) and their dependences on cosmological models within $w$CDM framework. 
To evaluate the performance of the emulator based method,
we study various setups as summarized in Table~\ref{tab:analysis_setups}. 

One question we want to address in this paper is; which scale cuts for $\dSigma$ and $\wgg$ are adequate in parameter estimation? Since $\dSigma$ and $\wgg$ have higher signal-to-noise ratios at smaller scales, we want to include the information of $\dSigma$ and $\wgg$
down to smaller scales in the nonlinear 1-halo term regime.  
However, such smaller scales are more affected by nonlinear physics, especially galaxy physics, so it would be difficult to accurately model the clustering signals on very small scales. 
Including such small-scale information might cause a bias in the estimated cosmological parameters. 
We should avoid such a failure situation as much as possible. 
To estimate appropriate scale cuts, we will study the performance of the method adopting different scale cuts of $(0.5,0.75)$, (1,1.5), $(2,3)$ or 
(8,12) (in units of $h^{-1}{\rm Mpc}$) for $\wgg$ and $\dSigma$, respectively, where $(2,3)$ is our fiducial choice unless explicitly stated. For all the cases 
we adopt $R_{\rm max}=30~h^{-1}{\rm Mpc}$ for the maximum scale up to which we include the information of $\dSigma$ and $\wgg$ for parameter estimation. 
Thus we do not include the BAO information for all the analyses.
To study the impact of the maximum scale, we also study the setup for $R_{\rm max}=70~h^{-1}{\rm Mpc}$, fixing other parameters to those in the fiducial setups. 

For the setups labeled as ``$\dSigma$ alone'' and ``$\wgg$ alone'', we study the parameter constraints if using either of $\dSigma$ or $\wgg$ alone. Comparing this result with the baseline method manifests complementarity of $\dSigma$ and $\wgg$ in the cosmological parameter
estimation. 

To
study the impact of the observational effects on parameter estimation, 
we include  the geometrical 
dependence of $\Omega_{\rm m}$ and 
introduce additional parameter to model 
the photo-$z$ error and/or multiplicative shear bias,
as discussed in Sections~\ref{sec:omegam_dependence}--\ref{sec:multiplicative_shear}.

For the setup labeled as ``RSD'', we study the impact of RSD effect. In the theoretical template we model the RSD effect using the linear RSD 
model.
Then we compare the theoretical template with the mock catalog including the full RSD effects, and then assess whether the theoretical model 
is still applicable to a realistic setup, without any significant bias or degradation in the estimated parameter. 

Finally, we show a forecast of how the anticipated full HSC dataset covering 1400~sq.~deg., about factor of 10 larger area than the HSC-Y1 data, can improve the cosmological constraints.
To do this forecast, we simply scales the covariance of $\dSigma$ by the area factor between the HSC-Y1 and full datasets.

\subsection{Validation strategy against uncertainties in halo-galaxy connection}
\label{sec:validation_halo_galaxy_connection}

\begin{table*}
\caption{A summary of the mock catalogs we use in this paper to assess ``robustness'' of the baseline method 
against variations in halo-galaxy connection. All the mock catalogs, except for ``{\tt cent-imcomp.}'' and 
``{\tt FoF-halo}'' catalogs, have the same HOD in average sense, but leads to modifications in $\dSigma$
and $\wgg$ in their amplitudes and scale-dependences. The column ``satellite gals.'' denotes a model of the spatial 
distribution of satellite galaxies in the host halo. 
In the columns of $\dSigma$ and $\wgg$,
``$\checkmark$'' or ``--'' denote whether they are modified from the fiducial mock or not, respectively.
\label{tab:mocks}}
\begin{center}
\begin{tabular}{l|lllll}\hline\hline
Model & HOD & satellite gals. & $\Delta\!\Sigma$& $\wgg$ & description \\ \hline
{\tt fiducial} & fid. & NFW & -- & -- \hspace{1em} &fiducial model \\
{\tt RSD} & fid.& NFW & -- & $\checkmark$ & include the RSD effect in $w$\\ \hline
{{\tt sat-mod}} & fid. & NFW & $\checkmark$
 & $\checkmark$ & populate satellites irrespectively of centrals \\
{\tt sat-DM} & fid. & DM part. & $\checkmark$ & $\checkmark$ & populate satellites according to $N$-body particles  \\ 
{\tt sat-sub} & fid. & subhalos & $\checkmark$ & $\checkmark$ & populate satellites according to subhalos \\ 
{\tt off-cent1} & fid. & NFW & $\checkmark$ & $\checkmark$ & all centrals off-centered, with Gaussian profile\\
{\tt off-cent2} & fid. & NFW & $\checkmark$ & $\checkmark$ & a fraction ($0.34$) of ``off-centered'' centrals, assuming 
Gaussian profile\\
{\tt off-cent3} & fid. & NFW & $\checkmark$ & $\checkmark$ & similar to ``off-cent1'', but with NFW profile\\
{\tt off-cent4} & fid. & NFW & $\checkmark$ & $\checkmark$ & similar to ``off-cent2'', but with NFW profile\\
{\tt cent-incomp.} & $\avrg{N_{\rm c}}$ mod. & NFW & $\checkmark$ & $\checkmark$& include an ``incomplete'' selection of centrals \\
{\tt FoF-halo} & mod. & FoF halos & $\checkmark$ & $\checkmark$ & use FoF halos to populate galaxies \\ 
{\tt assembly-$b$-ext} & fid. & NFW & $\checkmark$ & $\checkmark$ & populate galaxies according to 
concentrations of host halos\\ 
{\tt assembly-$b$} & fid. & NFW&  $\checkmark$ & $\checkmark$ & similar to ``assembly-$b$-ext'',
but introduce scatters\\ 
{\tt baryon} & fid. & NFW & $\checkmark$ & -- & mimic the baryonic effect of {\tt Illustris} on the 
halo mass profile\\
\hline\hline
\end{tabular}
\end{center}
\end{table*}
\begin{figure*}
\centering
\includegraphics[width=\textwidth]{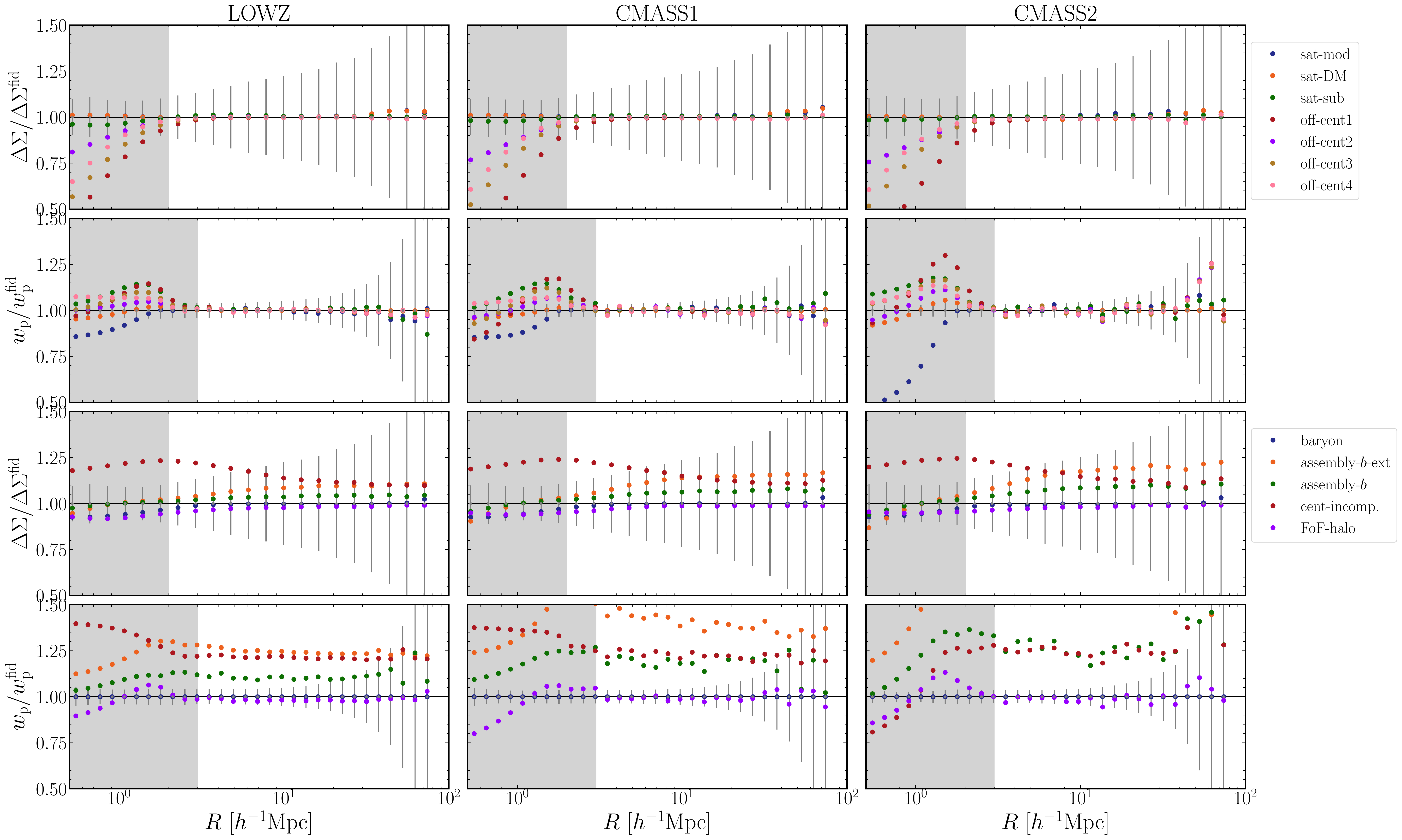}
\caption{The mock signals of $\dSigma$ and $\wgg$ relative to those for the fiducial mock, for each of variants of the mock catalogs 
in Table~\ref{tab:mocks}. The error bars are the same as Fig.~\ref{fig:signal_cosmo_dep}.
\label{fig:mock_signals_variants}}
\end{figure*}
The HOD model is an empirical prescription of the halo-galaxy connection. Our expectation is that we could recover the underlying cosmological parameters as long as a sufficient number of 
halo-galaxy connection parameters are introduced and then the effects are marginalized over when estimating cosmological parameters. 
To assess the ``robustness'' of our emulator based halo model against uncertainties in halo-galaxy connection, we use a wide variety of mock galaxy catalogs to study whether the baseline method can recover the true cosmological parameters for the different catalogs, as summarized in 
Table~\ref{tab:mocks}. 

Fig.~\ref{fig:mock_signals_variants} compares the mock signals of $\dSigma$ and $\wgg$ for 
different
mocks relative to those of the fiducial mock. 
Here all the mocks, except for the ``{\tt cent-incomp.}'' and ``{\tt FoF-halo}'' mocks, are built using the same HOD as that of the fiducial mocks, but using different ways of populating galaxies into individual halos, as described below. 
The variety in the target observables for a fixed HOD is possible, because the HOD 
only specifies
the average number of galaxies per halo 
in each mass bin, but there are many different ways to populate galaxies into halos leading to different spatial distributions (clustering properties) of galaxies, 
even for the same HOD. 
The different mock catalogs lead to modifications in the amplitudes and 
scale-dependences of $\dSigma$ and $\wgg$ in a complex way, 
compared to the fiducial mock.
One of the most important
systematic effects is the assembly bias effect, and we also use the mock catalogs, labeled as 
``{\tt assembly-$b$-ext}'' and ``{\tt assembly-$b$}'', to test the performance of our method against the assembly bias effect. 

In the following we describe details of each mock. Readers, who are interested in the results, can skip this section and directly go to Section~\ref{sec:results}.

\subsubsection{Satellite galaxies}
\label{sec:satellites}

Even if the HOD model is fixed, there are several ways of populating satellites in halos in each simulation realization. To study the impact of variations 
in the distribution of satellite galaxies, we construct several mocks, for the same HOD as that of the {\tt fiducial} mock. 

The  ``{\tt sat-mod}'' mock is a slight modification from the {\tt fiducial} mock. In this mock we populate satellite galaxies in halos irrespective 
of whether each halo already hosts a central galaxy. In this mock there are halos which host only satellite galaxy(ies) inside, without a central galaxy. 
Here we assume that the radial distribution of satellite galaxies follows the NFW profile as 
in the {\tt fiducial} mock. 

For the ``{\tt sat-DM}'' mock, we populate satellite galaxy(ies) in each host halo by randomly assigning each satellite to dark matter 
particles in the halo, in contrast to the NFW profile. 

For the  ``{\tt sat-subhalo}'' mock, we populate satellite galaxy(ies) in each host halo by randomly assigning each satellite to subhalo(s) in the 
host halo, which are taken from the \textsc{Rockstar} output.

\subsubsection{Off-centering effects of ``central'' galaxies}
\label{sec:off-centering}

For the {\tt fiducial} mocks, we assume that ``central'' galaxies are located at the center (the highest mass density) of each host halo. However, a central galaxy in a host halo can be ``off-centered'' as a result of 
merger or accretion in a hierarchical structure formation \citep{2013MNRAS.435.2345H,2013MNRAS.433.3506M}. To mimic this possible effect, we generate mock catalogs including the off-centering effects of central galaxy in each halo. 
Note here that we mean, by ``central'' galaxies, galaxies that are populated into halos according to the central HOD, and the central galaxies can be off-centered from the true halo center. More physically speaking, 
a galaxy which resides in the most massive subhalo can be considered as a central galaxy, but the galaxy can be off-centered due to the physical effects we discussed above. 

We generate four kinds of mock catalogs including the off-centering effects, following 
the method in \citet{2020arXiv200506122K} \cite[also see][]{OguriTakada:11}. The {\tt off-cent1} mock is designed to include 
the maximum possible
amount of 
off-centering effect, 
where we assume that all central galaxies are off-centered from the true halo center of each host halo. We assume that 
the average radial profile of off-centered galaxies with respect to the halo center follows a Gaussian profile with width 
$R_{\rm off}=2.2$, i.e. given by $\tilde{p}_{\rm off}(k)\propto \exp[-k^2(R_{\rm off}r_s)^2/2]$ for the radial profile in Fourier space, 
where $R_{\rm off}$ is a parameter to characterize 
the
typical off-centering radius relative to the scale radius ($r_s$) of the NFW 
profile.
When we populate a central galaxy in each halo, we randomly draw an off-centering radius from the Gaussian profile, and then place the galaxy into the spherical shell of the off-centering 
radius 
in the host halo (with randomly choosing the angular position for the azimuthal angles). 
Then we populate satellite galaxies in the same way as that for the {\tt fiducial} mock.
 
For the  ``{\tt off-cent2}'' mock, we assume that a fraction of central galaxies 
are off-centered. 
Following the implication found in \citet{2015ApJ...806....2M} \cite{2013MNRAS.435.2345H} for the SDSS galaxies, we assume $q_{\rm off}=0.34$ of the central galaxies are off-centered, while the remaining galaxies ($1-q_{\rm off}=0.66$) are at the halo center. 
We then populate satellite galaxies in the same way as that for the {\tt fiducial} mock. 

The  ``{\tt off-cent3}'' mock is very similar to the {\tt off-cent1} mock, but we populate the off-centered ``central'' galaxies into halos assuming 
that the off-centered galaxies follow the NFW profile of the host halo, similarly to satellite galaxies. 

The ``{\tt off-cent4}'' mock is very similar to the {\tt off-cent2} mock, but we populate the off-centered galaxies assuming the NFW profile 
as in {\tt off-cent3} mock. 

\subsubsection{FoF halos}
\label{sec:fofhalo}
\begin{figure}
\begin{center}
    \includegraphics[clip, width=0.90\columnwidth]{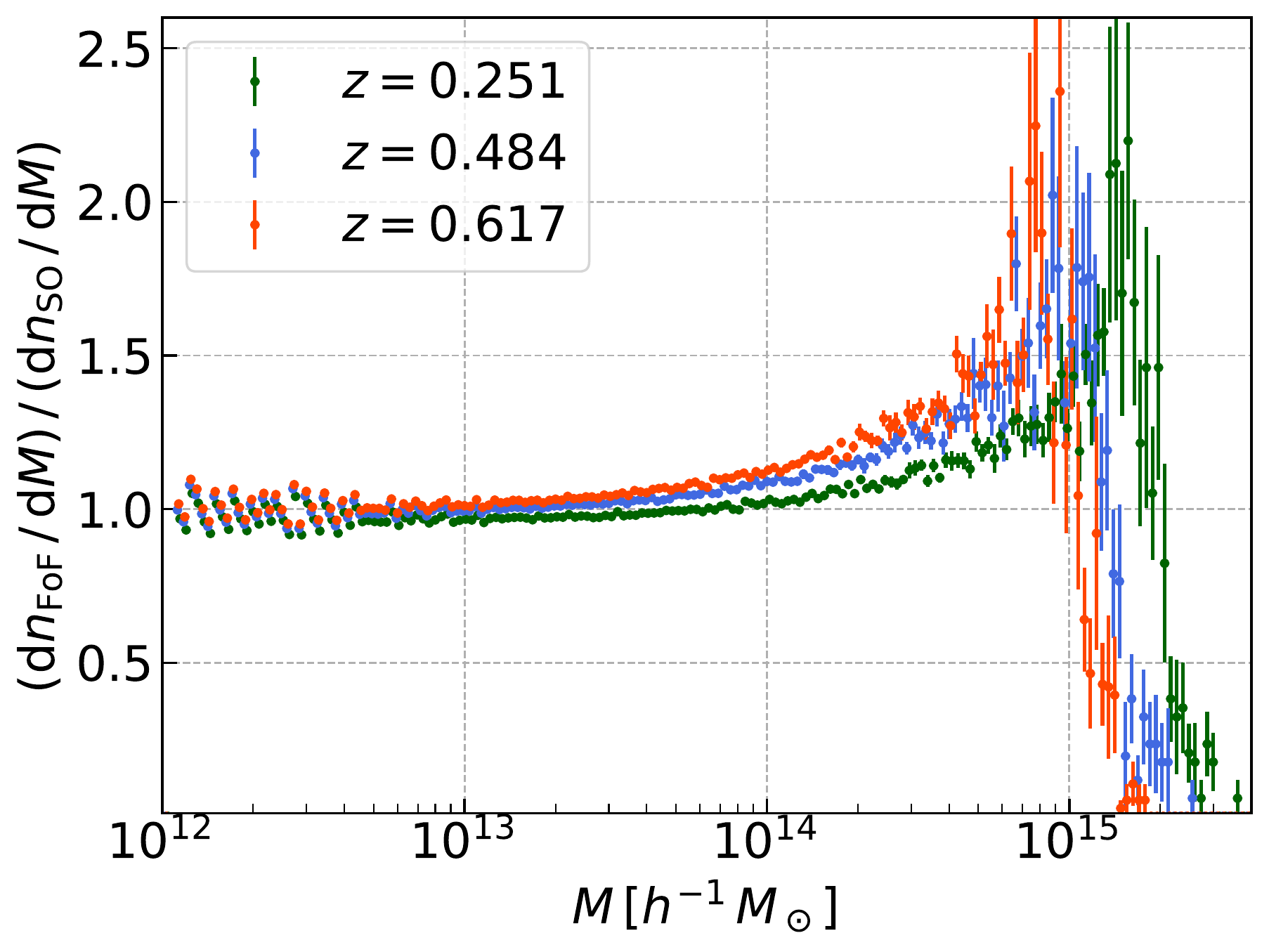}
    \caption{The ratio of the halo mass function of FoF halos to that of 
    SO halos in each mass bin. The data points show the results at
    three redshifts $z=0.251,0.484\text{ and }0.617$, corresponding to the representative redshifts of 
    LOWZ, CMASS1 and CMASS2, respectively. The data point in each bin is the 
    the mean of the ratios among 19 realizations of $1~(h^{-1}{\rm Gpc})$-size box 
    simulations, and 
    the error bar is the error on the mean among 19 realizations, which is computed 
    from the scatters among 19 realizations, divided by $\sqrt{19}$. 
    }
    \label{fig:massfunc_FoF_SO}
\end{center}
\end{figure}
Dark matter halos are neither uniquely-defined objects nor have a clear boundary with the surrounding structure. In this paper 
we use 
those identified by the \textsc{Rockstar} algorithm with the spherical-oversensity (SO) masses 
in each simulation realizations as our default choice (see Section~\ref{sec:nbody_DE}). 
For the  ``{\tt FoF halo}'' mock, 
we use halos that are identified by the friends-of-friends (FoF) method with linking length $b_{\rm FoF}=0.2$, in simulation realizations. 
The FoF halos do not necessarily have 
a one-to-one correspondence to the fiducial 
halos. The mass of each FoF halo is different from the SO 
mass
even if the corresponding halos are identified, because their boundaries are different. Fig.~\ref{fig:massfunc_FoF_SO} compares 
the mass function of halos measured using the SO halo mass definition or the FoF definition from the same $N$-body simulation realization, 
at three redshifts corresponding to those for LOWZ, CMASS1, and CMASS2, respectively. The figure shows a sizable difference 
in the halo mass functions over the range of mass scales we consider. 

To generate the  ``{\tt FoF-halo}'' mock we treat the FoF halo mass of individual halos, as the mass argument for the mean
HOD functions, and then populate galaxies in FoF halos using the same HOD method as that for the {\tt fiducial} mock.

\subsubsection{Assembly bias effect}
\label{sec:assembly_bias}

The ``assembly bias'' effect refers to the fact that the clustering amplitudes of galaxies or halos at large scales depends on a secondary parameter other than the halo mass, especially depending on the assembly history of galaxies/halos 
\citep{2005MNRAS.363L..66G,2006ApJ...652...71W,2008ApJ...687...12D}. The assembly bias is one of the most important
physical effects 
causing a violation of 
the simple halo model picture which assumes that clustering properties of halos are determined solely by halo mass. 
To study whether or not cosmology inference based on our method is robust against the assembly bias effect, 
we use the 
mocks generated following the method in 
\citet{2019arXiv190708515K}. This is one of the most important tests we address in this paper. 

The ``{\tt assembly-$b$-ext}'' mock is intended to study the worst case scenario for the assembly bias effect. To make this mock, we first, for each halo, 
calculate the fraction of mass enclosed within a sphere of 50\% of the halo radius $R_{200}$ to the whole halo mass $M_{200}$. 
We denote this inner mass fraction as $f_{\rm in}$. We use 
this quantity 
as a proxy of the halo concentration, i.e., the higher $f_{\rm in}$ means the higher concentration.
Then we make a ranked list of halos in which we sort the halo by ascending order of $f_{\rm in}$, in each narrow bin of halo masses.
We populate central galaxies, according to the central HOD, into halos from the top of the list (from the lowest-concentration halo) in each mass bin. We then populate satellite galaxies in halos that already host central galaxies using the satellite HOD. For the mock generated in this way, 
we can have a maximum effect of the assembly bias in the large-scale clustering amplitudes for all the three galaxy samples. As can be found from Fig.~\ref{fig:mock_signals_variants}, $\wgg(R)$ has 
larger amplitudes at large separations, by up to a factor of 1.6 than that of $\wgg$ in the {\tt fiducial} mock, which is quite substantial. 

For the mock {\tt assembly-$b$}, we introduce a scatter to $f_{\rm in}$ of each halo: 
\begin{align}
    \log_{10} f_{\rm in}^{\rm scatter} = \log_{10} f_{\rm in} + \epsilon. 
\end{align}
We assign a random scatter, $\epsilon$, to each halo drawn from 
a zero-mean Gaussian distribution with $\sigma$, where $\sigma$ is a parameter to control the amount of the scatter.  
We adopt $\sigma=0.1$ to generate the {\tt assembly-$b$} mock for all the three galaxy samples, which still leads to a significant boost in the 
clustering amplitudes in $\wgg$ by up to a factor of 1.3 (therefore about halved strength compared to the {\tt assembly-$b$-ext} mock) than
that of $\wgg$ in the {\tt fiducial} mock. This might be more realistic for host halos of the SDSS galaxies ($\sim 10^{13}~h^{-1}M_\odot$), although the 
assembly bias effect has not been detected at a high significance from the real data. 
Note that the mean HOD is not modified from the fiducial mocks with these procedures. 

\subsubsection{Baryonic effect}
\label{sec:baryonic_effect}

\begin{figure}
\begin{center}
       \includegraphics[clip, width=0.90\columnwidth]{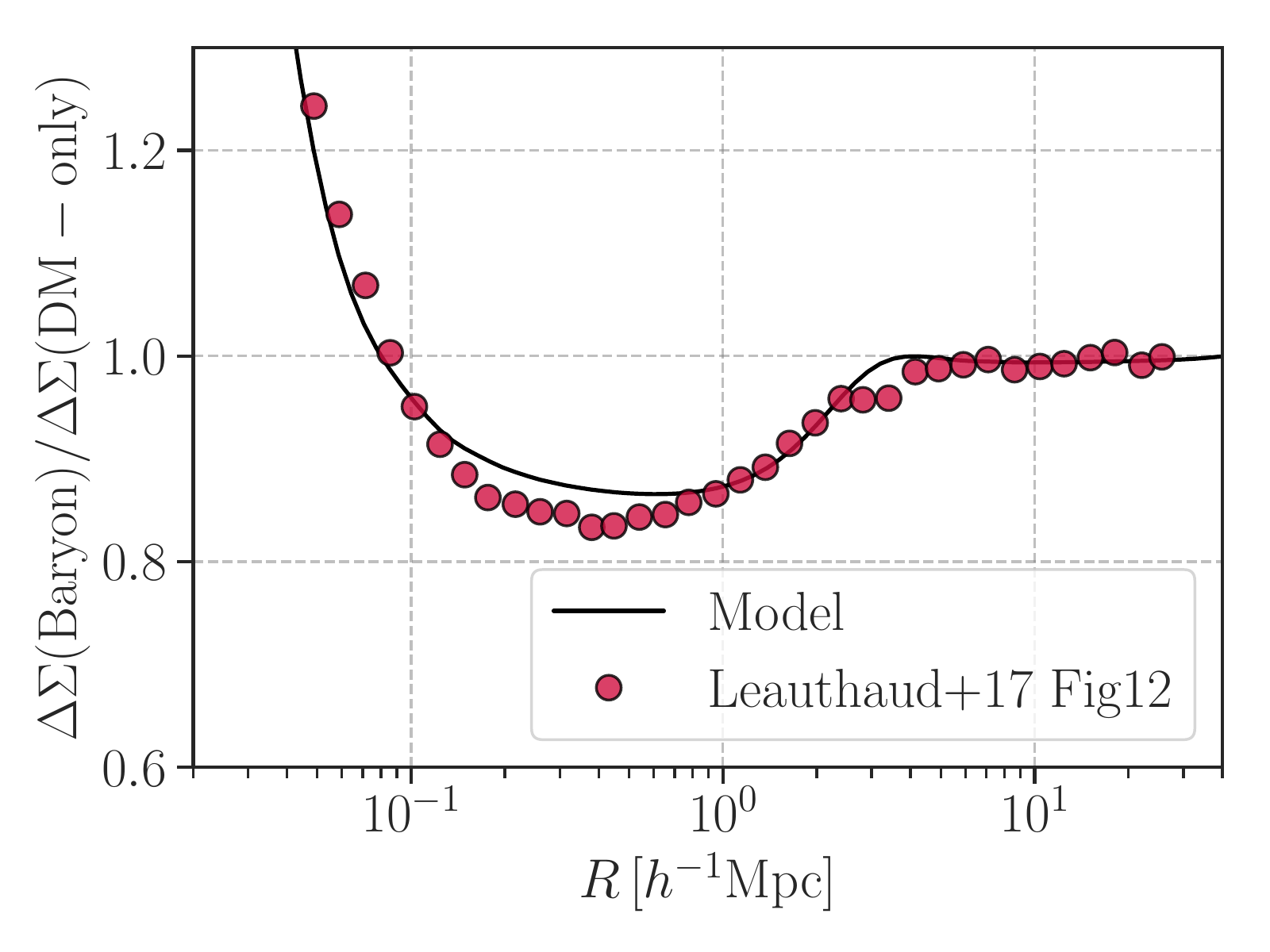}
     \caption{We use the method in \citet{2015JCAP...12..049S} to model 
 the baryonic effect on the galaxy-galaxy weak lensing profile, $\dSigma$, where the mass conservation around halos hosting galaxies is explicitly imposed. We 
tune
 the model parameters to reproduce the baryonic effect that is seen in 
 the \textsc{Illustris} hydrosimulation \cite{2014Natur.509..177V}. Shown is the ratio of $\dSigma$ for the matched host halos of SDSS-like galaxies in the simulations with and without the baryonic effect, which is taken from 
 Fig.~12 in \citet{2017MNRAS.467.3024L}. Our method nicely capture the baryonic effect, but we note that this would be a worst case of the baryonic effect, because the \textsc{Illustris} (not \textsc{IllustrisTNG}) employed the too large baryonic 
 effect (especially AGN feedback). 
     \label{fig:dsigma_baryon}} 
    \end{center}
\end{figure}

The baryonic effects inherent in galaxy formation/evolution, which we are missing in our $N$-body simulations, 
are
another important physical systematic effect, and we need to quantify 
their
impact 
on the
parameter inference in our cosmology challenges. Although the baryonic effects 
are still difficult to accurately model from the first principles, one should keep in mind some conservation properties in the distribution of galaxies. First, massive galaxies like the SDSS LOWZ/CMASS galaxies are likely to form at the same peaks of primordial density 
fluctuations even in the presence of baryonic physics. Hence the distribution of massive galaxies relative to the total matter distribution, i.e. 
the bias function, is not largely changed by baryonic physics \citep{2013MNRAS.434..148S,2018MNRAS.475..676S}. On the other hand, the baryonic physics 
causes a redistribution of matter around each galaxy, e.g. due to various effects such as dissipative contraction and supernova/AGN feedbacks. For this reason, the radial profile of matter distribution around a galaxy would likely be modified. 

We follow the method 
in \citet{2015JCAP...12..049S} \citep[also see][]{2019JCAP...03..020S} to include baryonic effects to the mock signals. The notable feature of this model is that
the model explicitly imposes the mass conservation around halos, and models the baryonic effects as a redistribution of the surrounding matter around each halo (more exactly the halo profile). We tuned the model parameters so that the model prediction reproduces the baryonic effect on the lensing profile, $\dSigma$, in the original \textsc{Illustris} hydrosimulation \cite{2014Natur.509..177V}
that implemented too large baryonic effect than implied by observations, as explicitly demonstrated in Fig.~\ref{fig:dsigma_baryon}. Hence the {\tt baryon} mocks are considered as a worst case of the baryonic effect on the weak lensing profile. For $\wgg$, we use the mock signal for the {\tt fiducial} mocks; that is, we do not include a possible effect of the baryonic physics on the distribution of galaxies in the host halo.

\begin{figure*}
\centering
\includegraphics[width=\textwidth]{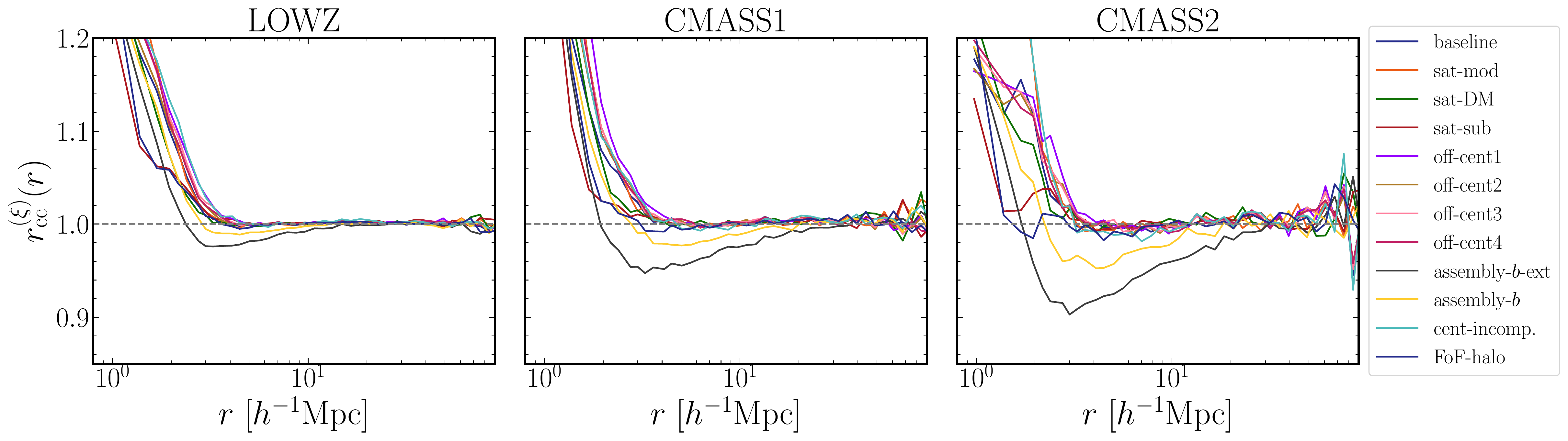}
\caption{Cross correlation coefficients, defined as $r^{(\xi)}_{\rm cc}(r)\equiv \xi_{\rm gm}(r)/\sqrt{\xi_{\rm gg}(r)\xi_{\rm mm}(r)}$, for 
different mock catalogs of SDSS-like galaxies we use in this paper. 
\label{fig:rcc}}
\end{figure*}
\subsection{Cross-correlation coefficients in the mocks}
\label{sec:cross_correlation}

All the mock catalogs we use in this paper have a property that the cross-correlation coefficient, $r_{\rm cc}(r)\equiv \xi_{\rm gm}/[\xi_{\rm gg}\xi_{\rm mm}]^{1/2}\simeq 1$ at the limit of large scales, greater than 
the size of massive halos, say at $\gtrsim 10~h^{-1}{\rm Mpc}$, as shown 
in Fig.~\ref{fig:rcc}.
This property is expected if galaxy physics is confined to local, small scales, and because the clustering amplitudes at larger scales than the scale of galaxy physics are governed by gravity alone (and properties of primordial fluctuations) and then behave 
as a linear biasing relation to the 
underlying matter distribution at the large scales. Encouragingly this is confirmed recently by using the hydrodynamical simulation IllustrisTNG in 
Ref.~\cite{2020arXiv200804913H}, where various galaxy samples, which are selected based on host halos and the various environment parameters, display 
$r_{\rm cc}\simeq 1$ at $r\gtrsim 10~h^{-1}{\rm Mpc}$. 
The \textsc{Dark Emulator} outputs also predict $r\equiv \xi_{\rm hm}/[\xi_{\rm hh}\xi_{\rm mm}]^{1/2}\simeq 1$ for halo correlation functions
at scales greater than 
the
size of halos. 
However, even for the scales where $r_{\rm cc}\simeq 1$, this does not mean that the linear theory serves as an accurate theoretical template
\citep{2020arXiv200806873S}. It is important to include the nonlinear clustering, 
the nonlinear halo bias and the halo exclusion effect in the theoretical template. 

We also note that $r_{\rm cc}$ can be greater than unity; $r_{\rm cc}\ge 1$, which occurs in the 1-halo term regime where the sub-Poisson nature 
$\avrg{N_{\rm g}(N_{\rm g}-1)}\ne \avrg{N_{\rm g}}^2$ becomes important due to a finite number statistics of galaxies in the same host halo 
\citep{Seljak:00,Scoccimarroetal:01,2015ApJ...806....2M}.

\section{Results}
\label{sec:results}
\begin{figure*}
\centering
\includegraphics[width=1.9\columnwidth]{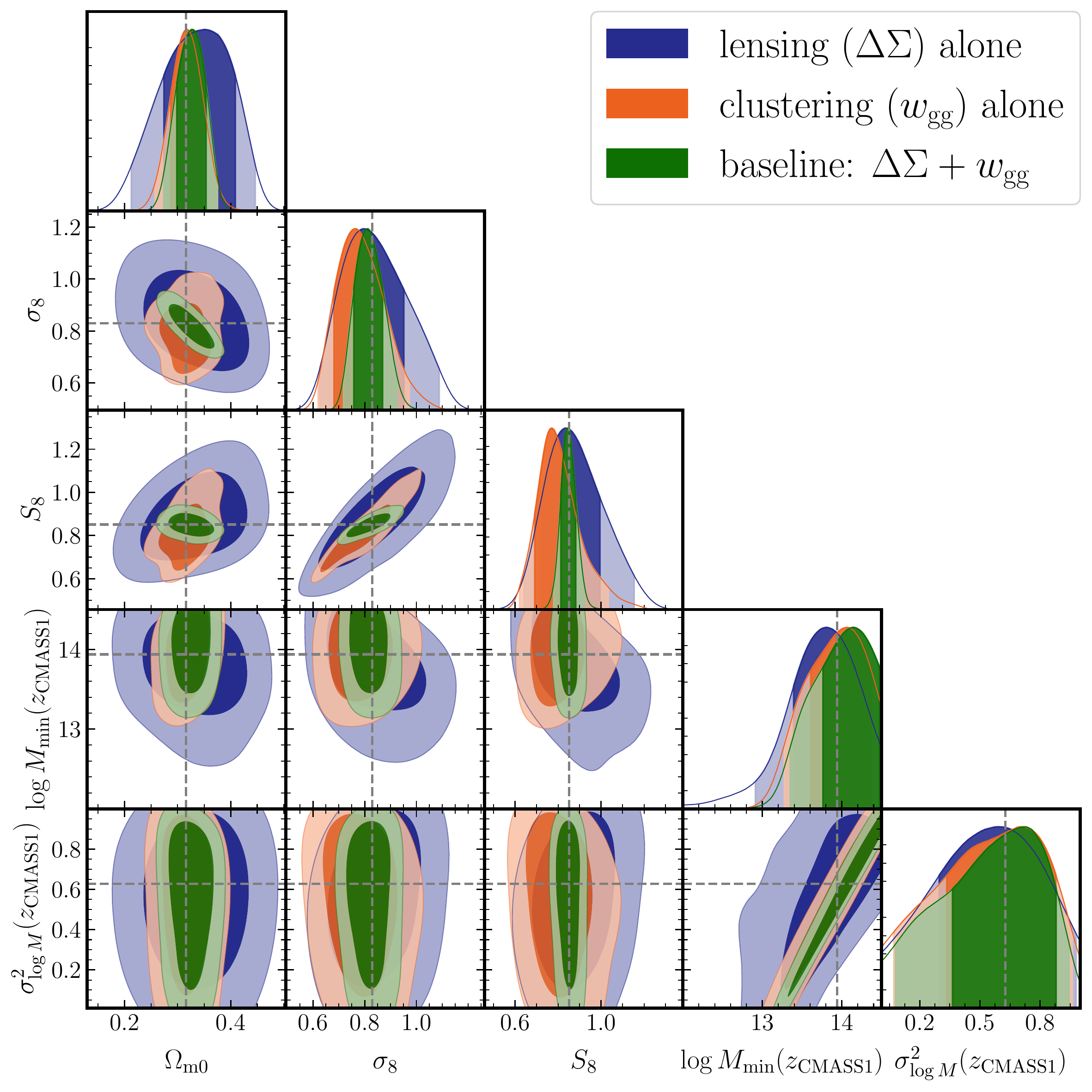}
\caption{Marginalized posterior distribution in each 2D sub-space of the parameters, obtained from the projected correlation function information alone ($\wgg$; orange-color contours), the lensing information alone ($\dSigma$; blue) and the joint constraints (green), respectively.
The inner and outer contours show the $68 \, \%$ and $95 \, \%$ credible regions, respectively.
We 
adopt
the baseline setup for which we employ $R_{\rm cut}=2$ and $3~\hiMpc$ for the scales cuts of $\wgg$ and $\dSigma$, respectively. 
For the mock signals, we 
use
the signals measured from the {\tt fiducial} mocks which are generated using the same HOD model as those in the 
theoretical templates. 
Here we include 17 model parameters in the parameter inference: 2 cosmological parameters ($\Omega_{\rm m}$ and 
$\sigma_8$) plus $5$ HOD parameters for each of the 3 galaxy samples (LOWZ, CMASS1 and CMASS2). Here we show, as an example, the results for the central HOD parameters for the CMASS1 sample at $z=0.484$, i.e. $M_{\rm min}$ and $\sigma_{\log M}$. 
The vertical and horizontal dashed lines denote the input parameter value used in the mock catalog.
\label{fig:2Dpost_baseline_cosmo_hod}}
\end{figure*}
\begin{figure}
\centering
\includegraphics[width=1.\columnwidth]{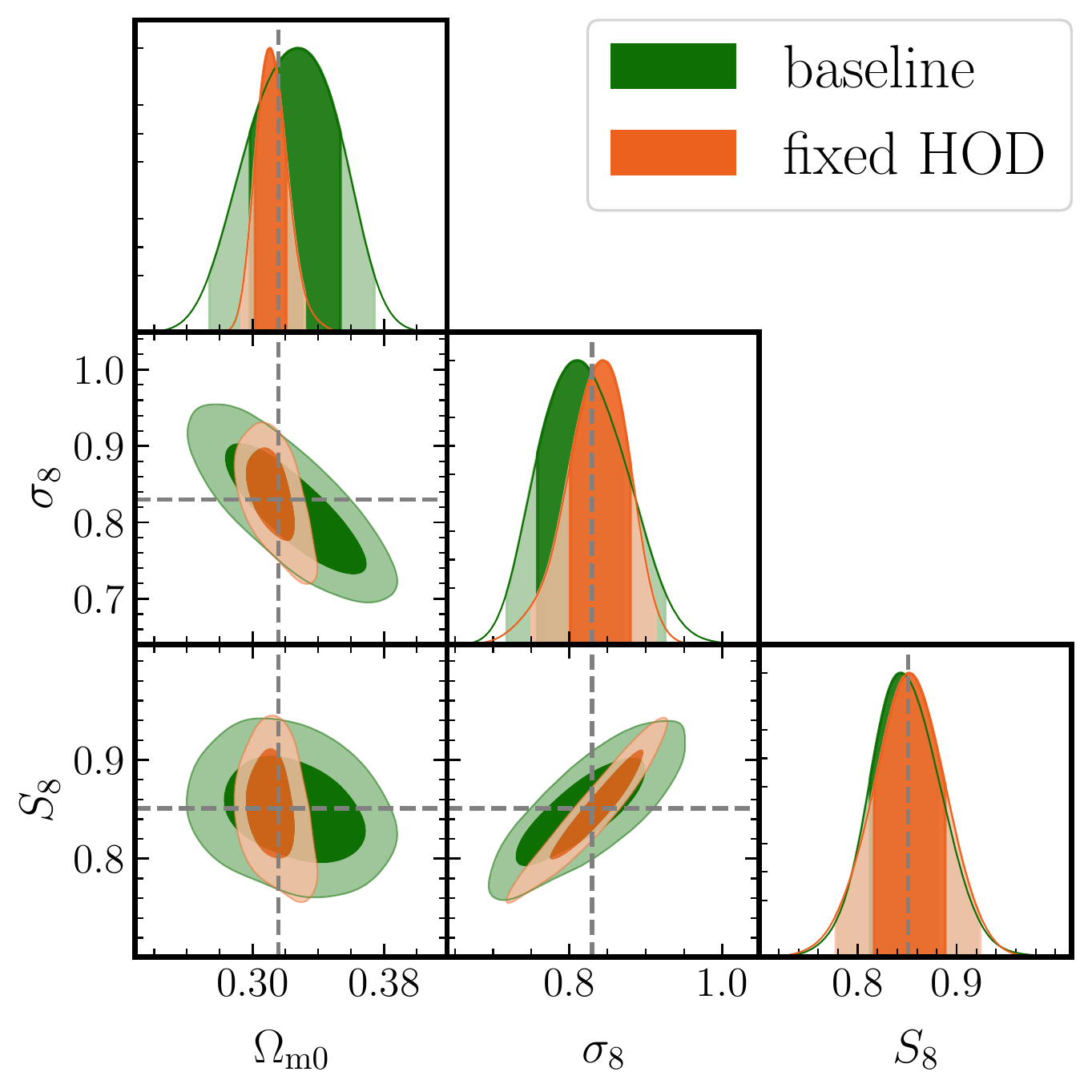}
\caption{Similar to the previous figure, but shown 
are
the results when fixing the HOD parameters in the model templates to their fiducial (input) values. The green contours (labeled as ``baseline'') are the same as those in Fig.~\ref{fig:2Dpost_baseline_cosmo_hod}.
\label{fig:2Dpost_HODfixed}}
\end{figure}
\begin{figure}
\centering
\includegraphics[width=\columnwidth]{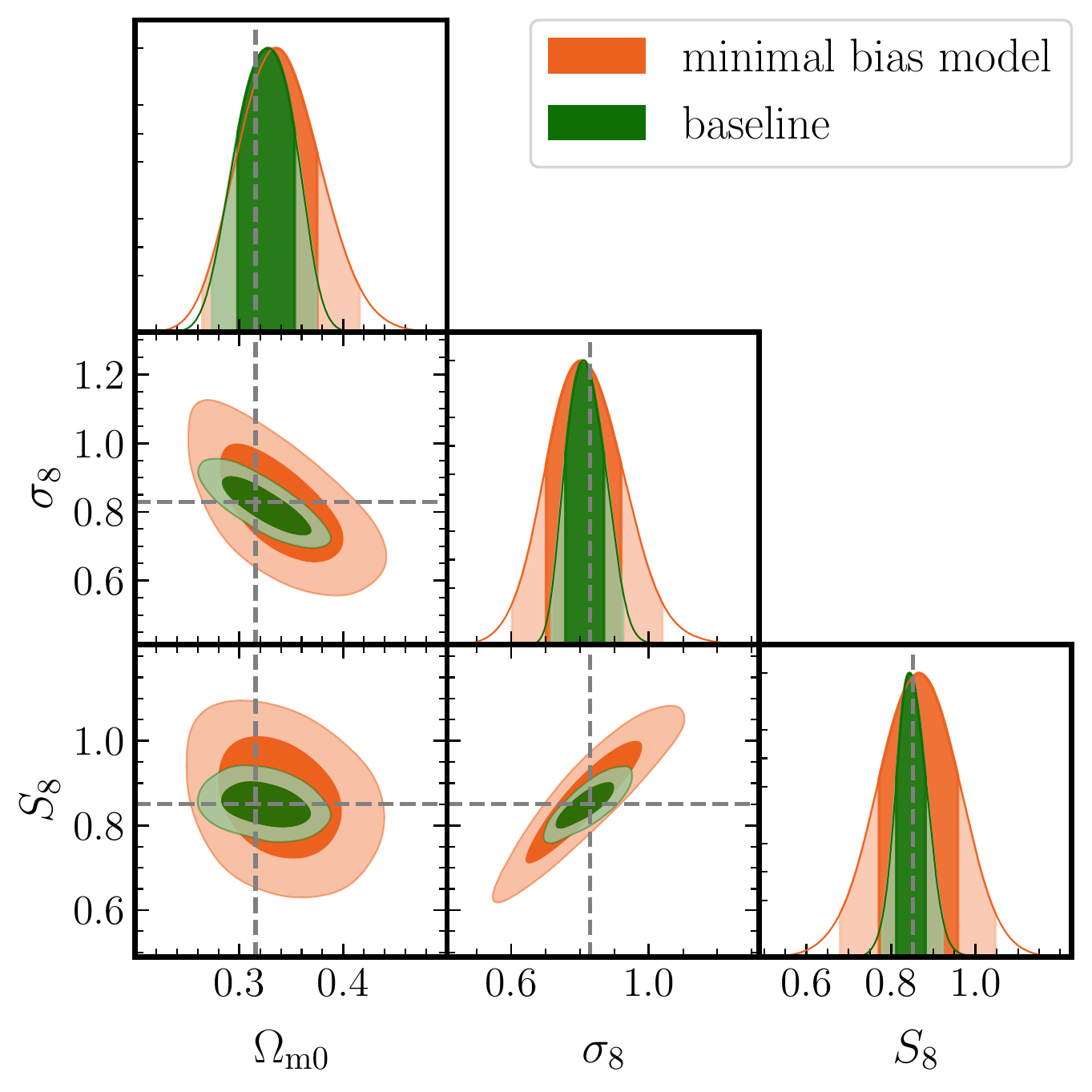}
\caption{Similar to Fig.~\ref{fig:2Dpost_baseline_cosmo_hod}, but the figure compares the marginalized posterior distributions 
obtained from the HOD based method, studied in this paper, and the perturbation theory (PT) inspired method in \citet{2020arXiv200806873S}. 
Here we apply both the methods to the same {\tt fiducial} mock signals. For the PT method, we model the 
matter-galaxy cross-correlation ($\xi_{\rm gm}$)
and the galaxy auto-correlation ($\xi_{\rm gg}$) by the nonlinear matter auto-correlation multiplied by a linear bias parameter: 
$\xi_{\rm gm}=b_1\xi^{\rm NL}_{\rm mm}$ and $\xi_{\rm gg}=b_1^2\xi_{\rm mm}^{\rm NL}$. We then treat $b_1$ as a free parameter in the parameter inference, and 
employ
the scale cuts of $R_{\rm cut}=12$ and $8~\hiMpc$ for $\dSigma$ and $\wgg$, respectively, compared to $R_{\rm cut}=3$ and
$2~\hiMpc$ for the halo model based method. 
The green contours are the same as those in Fig.~\ref{fig:2Dpost_baseline_cosmo_hod}. 
\label{fig:2Dpost_HOD_vs_minimalbias}}
\end{figure}
\begin{figure*}
    \begin{center}
        \includegraphics[width=1.\textwidth]{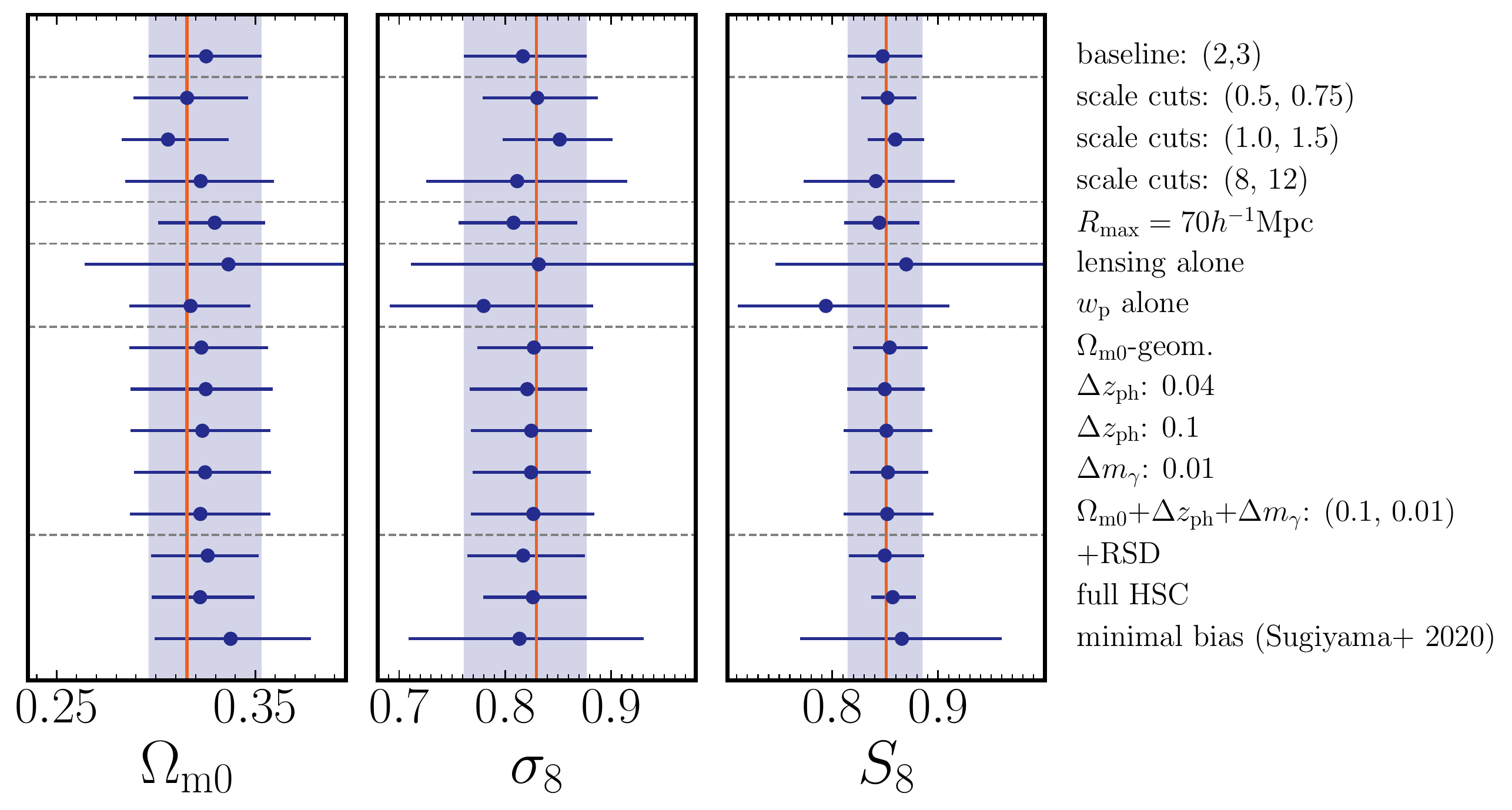}
        \caption{Summary of the estimation of each cosmological parameter, 
        $\Omega_{\rm m}$, $\sigma_8$ or $S_8(\equiv \sigma_8\Omega_{\rm m}^{0.5})$,  for the different setups 
        in Table~\ref{tab:analysis_setups}. The blue dot in each row denotes the mode of the marginalized posterior distribution of 
        each parameter, and the error bar denotes the 68\% credible interval, which is computed from the highest density interval of the 
        marginalized 1D posterior distribution. 
        The vertical red line denotes the true value used in 
        the mock catalog, and the shaded region denotes the 68\% credible interval for the baseline setup for comparison. 
        \label{fig:scores_fiducial}}
    \end{center}
\end{figure*}

In this section we show the main results of this paper, which are assessment and validation of 
the emulator based halo model method for 
cosmology inference. Here we mean by ``validation'' whether the method can recover the true cosmological parameters, $\Omega_{\rm m0},
\sigma_8$ and $S_8$, to within 68\% credible 
intervals, after marginalization over the HOD parameters (galaxy-halo connection parameters). 
Note that we will not focus on the accuracy of recovering the HOD parameters.

\subsection{Validation of the baseline method, complementarity of $\dSigma$ and $\wgg$, and tests with 
different scale cuts }
\label{sec:results_complementarity}

First we perform a sanity check; 
we study whether our baseline method (see Table~\ref{tab:analysis_setups})
can recover the true cosmological parameters when comparing the theoretical templates to the 
mock signals measured from the {\tt fiducial} mock that is based on the same HOD model used in the theoretical template. 
Different model parameters affect the observables in a complex way, so it is not 
obvious whether the baseline method can recover the true cosmological parameters, after projecting the posterior distribution in a multi-dimensional parameter space onto a sub-space including the cosmological parameters. In fact, as discussed in Refs.~\cite{2020arXiv200806873S,2020arXiv200701844J}, 
the modes
of the marginalized posterior distribution of cosmological parameters could be biased from the true values, if the parameters suffer from severe 
degeneracies. 

Fig.~\ref{fig:2Dpost_baseline_cosmo_hod} shows the results. 
Here we employ 
$R_{\rm cut}=2$ and $3~\hiMpc$ for the scale cuts 
for $\wgg$ and $\dSigma$ above which we include the clustering and lensing information in parameter inference (see Table~\ref{tab:analysis_setups}). 
The figure nicely shows that the lensing and clustering information are 
complementary to each other, and combining the two lifts the parameter degeneracies. 
Thus Fig.~\ref{fig:2Dpost_baseline_cosmo_hod} gives a validation of the baseline method; the baseline method can recover the cosmological 
parameters if properties of the galaxy clustering for SDSS-like 
galaxies are close to the fiducial HOD model we employed.
The baseline method achieves a precision of $\sigma(S_8)\simeq 0.035$ for the SDSS and HSC-Y1 data. 
In the following we show mainly the results for the joint probes combining the information of $\dSigma$ and $\wgg$.
For 
completeness
of our discussion, in Fig.~\ref{fig:2Dpost_baseline_all} we show the posterior distribution of the parameters including all the HOD parameters 
for the CMASS1-like galaxy sample. 

What is the impact of HOD parameters on cosmological parameter estimation? To answer this question, in Fig.~\ref{fig:2Dpost_HODfixed} we show the results when fixing the HOD parameters to their fiducial values. It is clear that the HOD parameters cause significant
degradations in the cosmological 
parameters; for example, the marginalized error of $\Omega_{\rm m}$ is enlarged by a factor of 3 when including the HOD parameters. 
However, this is a price one has to pay to obtain robust constraints on cosmological parameters when including the small-scale information. 
If one uses a more aggressive method, e.g. by using a less flexible model of halo-galaxy connection (such as fixing some HOD parameters), one could suffer from severe biases in cosmological parameters.

Our results might also be compared to a more conservative approach, e.g. a method using only the clustering observables at large scales, i.e. not including the small-scale information. In Fig.~\ref{fig:2Dpost_HOD_vs_minimalbias} we compare the results from the halo model method, studied in this paper, with those from the perturbation theory (PT) based method studied in \citet{2020arXiv200806873S}. 
They employed the ``minimal''-bias model using the fully nonlinear matter power spectrum and the linear bias parameter ($b_1$); 
$\xi_{\rm gm}=b_1\xi_{\rm mm}^{\rm NL}$ and
$\xi_{\rm gg}=b_1^2\xi_{\rm mm}^{\rm NL}$, where the halofit model is used to model the nonlinear $\xi_{\rm mm}^{\rm NL}$. 
It was shown that, as long as the conservative scale cuts $R_{\rm cut}=12$ and $8~\hiMpc$ for $\dSigma$ and $\wgg$ are employed and the bias parameter 
is treated as a free parameter, the minimal bias method passes all the validation tests against a variety of the mock catalogs including the assembly bias mock. Hence, the results for the minimal bias method can be considered as conservative, yet robust parameter constraints that can be extracted from the SDSS and HSC-Y1 data. The figure clearly shows that including the small-scale information and the halo bias information can improve cosmological constraints compared to the conservative method. In particular, the halo model based method gives a factor of 2 
or
3 improvement in 
the marginalized error of $\sigma_8$
or
$S_8$, respectively, 
compared to that for the minimal bias method. Thus the halo model method has a potential to obtain the improved cosmological constraints, if the model is flexible enough to describe variations in galaxy clustering at small scales down to a few Mpc, which we will test later. 

In Fig.~\ref{fig:scores_fiducial} we give a summary of the performance for different setups. The second to the fourth row, 
lying between 
the 
horizontal
dashed lines, show the results when using the different scale cuts, $(R^{\dSigma}_{\rm cut},R_{\rm cut}^{\wgg})=(0.5,0.75), 
(1.0,1.5)$ or $(8,12)~\hiMpc$, instead of $(2,3)$ as our fiducial choice. For the {\tt fiducial} mocks, these smaller scale cuts apparently recover the cosmological parameters; the true value of each parameter is within the 68\% credible interval. However, a closer look also reveals that the size of 
the credible interval is not significantly improved by including the smaller information. We also checked that, when applying the smaller scale cuts to other mock catalogs rather than the {\tt fiducial} mocks, those lead to a larger bias in the cosmological parameters compared to the baseline method. 
Hence, with these results, we conclude that the fiducial scale cuts of $(2,3)~\hiMpc$ are reasonable. 
On the other hand, the row labeled as ``$R_{\rm max}=70~\hiMpc$'' shows the results when including the information of $\dSigma$ and $\wgg$ up to 
$R_{\rm max}\simeq 70~\hiMpc$ instead of our default choice $R_{\rm max}=30~\hiMpc$. Note that this maximum scale is still below the BAO scales, and 
we do not include the BAO information. It is clear that the larger-scale information does not improve the cosmological parameter estimation, and 
$30~\hiMpc$ seems sufficient for the signal-to-noise level expected for the HSC-SDSS analysis.

\subsection{Comparison with analytic halo model}
\label{sec:compare_More}
\begin{figure}
\centering
\includegraphics[width=1.\columnwidth]{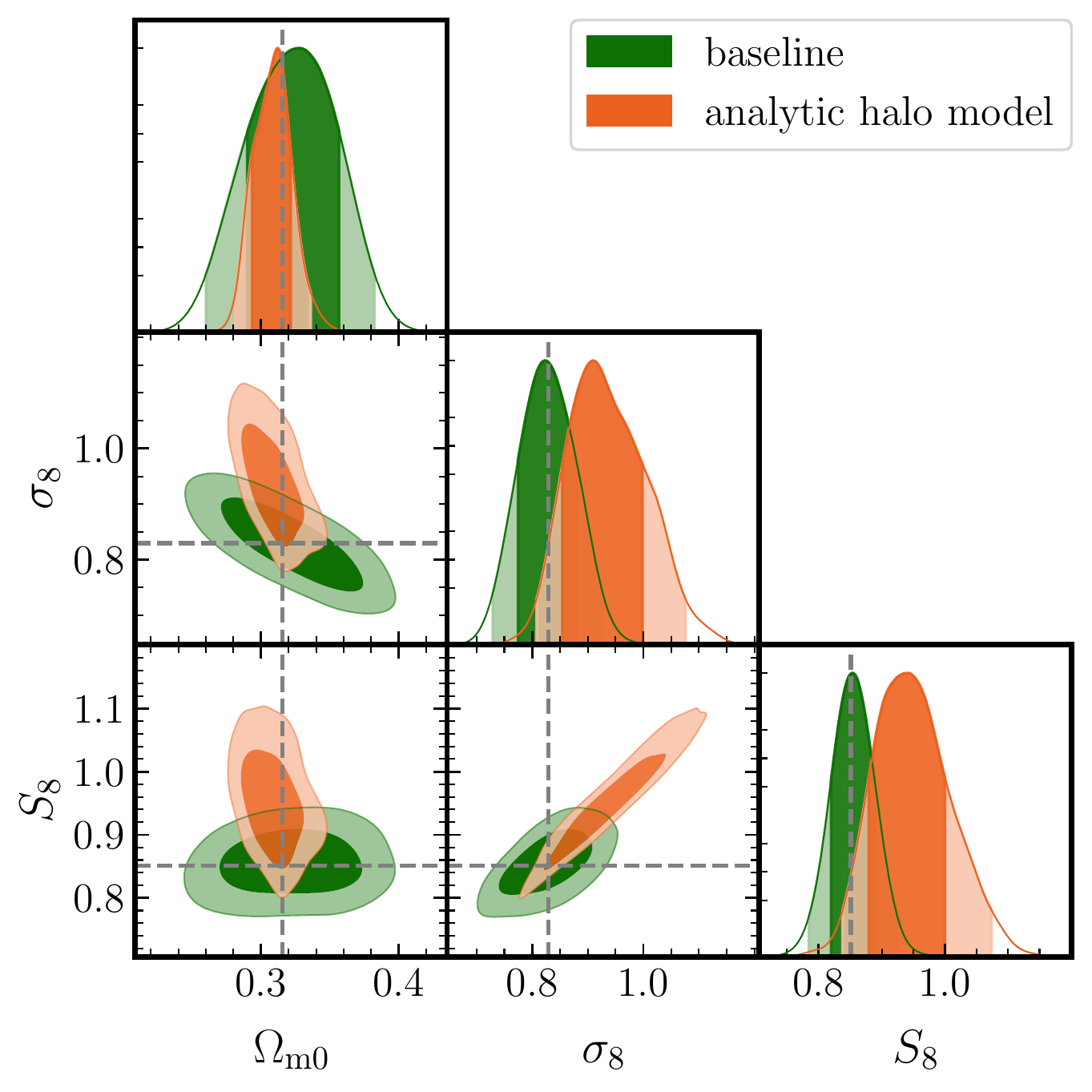}
\caption{Similar to Fig.~\ref{fig:2Dpost_baseline_cosmo_hod}, but the plot shows 
the results for the baseline method with those obtained by using the analytic halo model 
in \citet{2015ApJ...806....2M}. Here we used the same scale cuts, $R_{\rm cut}=2$
and 3~$\hiMpc$, for $\wgg$ and $\dSigma$, respectively, for both methods.}
\label{fig:compare_withMore}
\end{figure}

In Fig.~\ref{fig:compare_withMore}, we compare the results for our baseline method with those obtained by comparing the model predictions of analytical halo model \footnote{The 
code of analytical halo model is available from \url{https://github.com/surhudm/aum}.}
in 
\citet{2015ApJ...806....2M} with the mock signals from the {\tt fiducial} mocks. The analytical 
halo model was constructed by calibrating the standard halo model \citep{Seljak:00} with $N$-body 
simulation results available at that time, e.g. to include the halo exclusion effect and the nonlinear halo bias effect \citep[see Ref.][for details]{2013MNRAS.430..725V}. For example, 
the analytical halo model uses the fitting formula of halo bias developed in Ref.~\cite{2005ApJ...631...41T,2010ApJ...724..878T}, but several recent studies point out an inaccuracy in the fitting formula,
 up to 5\% in the bias amplitude \citep[see the bottom panel of Fig.~22 in Ref.][]{2018arXiv181109504N} \citep[also see Refs.][]{Lietal:16,Baldaufetal:16,Lazeyras:2016}.
For the analytical halo model, we did not include the off-centering effect or incompleteness effect as in our baseline method. 

The figure shows sizable differences in all the cosmological parameters between 
our emulator based method and the analytical halo model. The 1D posterior distribution 
shows that
the analytical halo model cannot recover the true values of $\sigma_8$ and $S_8$ 
to within the 68\% credible interval. 
In addition, the degeneracy 
directions in the projected two parameter subspace are quite different. 
Figuring out the cause of the difference is beyond the scope of this paper, but 
here at least we would like to stress that the emulator-based halo model displays a better performance.

\subsection{The impacts of observational effects: the geometrical correction, photo-$z$ errors, shear multiplicative bias and RSD}
\label{sec:results_observational_effect}

As discussed in Section~\ref{sec:observational_effects}, various observational effects could affect the cosmological inference from 
the measured $\dSigma$ and $\wgg$. For the results discussed up to the preceding subsections, we ignored these effects, and in this section we 
study the impact of these effects. 

The rows with labels starting from ``$\Omega_{\rm m}$-geom.'' to ``$+$RSD'' in Fig.~\ref{fig:scores_fiducial}
show the results when including each or all of these observational effects. First let us discuss the results for ``$+$RSD''. For this test 
we 
use the mock catalogs where we use the fiducial HOD to populate galaxies into the simulation realizations and include the RSD effects due to the peculiar velocities of individual galaxies \citep{2019arXiv190708515K,2020arXiv200506122K}.
Then we perform the cosmology inference by comparing the theoretical templates, including the linear Kaiser effect 
(Section~\ref{sec:rsd}), with the mock signals. The figure shows that the linear Kaiser 
model
can properly take into account the RSD effect for our fiducial 
projection length ($\pi_{\rm max}=100~\hiMpc$), and there is no degradation in the parameter estimation. This is similar to the result in
\citet{2020arXiv200806873S}. Although the RSD effect itself carries the dependence on $\Omega_{\rm m}$ (more exactly via the growth rate), 
it leads to only slight improvement in $\Omega_{\rm m}$.

Now we discuss the results for ``$\Omega_{\rm m}$-geom.'', which refers to the fact that a reference cosmology needs to be assumed to measure 
$\dSigma(R)$ and $\wgg(R)$ from direct observables, and the assumed cosmology generally differs from the underlying true cosmology. 
Exactly speaking, for a flat-geometry $\Lambda$CDM model, we have to ``re-measure'' $\dSigma(R)$ and $\wgg(R)$ every time $\Omega_{\rm m}$ is varied in parameter estimation. This might be time-consuming, and indeed \citet{2013ApJ...777L..26M} showed that this conversion can be safely done 
by a multiplicative factor taking into account the geometrical dependence, which we employ in this paper. Fig.~\ref{fig:scores_fiducial} shows that 
including the $\Omega_{\rm m}$ geometrical dependence does not cause a bias in the cosmological parameters. However, a closer look shows that including this dependence slightly enlarges the credible interval. This happens as follows. If we assume a slightly larger value of $\Omega_{\rm m}$, 
it leads to the smaller amplitude in the model prediction of $\wgg$ (recall that the current constraint is mainly from the $\wgg$ information). 
However, when assuming such a model with increased $\Omega_{\rm m}$ in the measurements, it leads to the smaller amplitudes in the measured $\wgg$ as shown in Fig.~\ref{fig:signal_cosmo_dep}.
Thus the dependences of $\Omega_{\rm m}$ are compensated to some extent in the model and the measurement. This is the reason that the credible interval of $\Omega_{\rm m}$ is slightly degraded. 

Next we discuss the impacts of photo-$z$ errors and the multiplicative shear bias, which are among the most important systematic errors in the weak lensing measurements. As discussed in Sections~\ref{sec:photoz} and \ref{sec:multiplicative_shear}, we introduce nuisance parameters, denoted as 
$\Delta z_{\rm ph}$ and $\Delta m_\gamma$,  to model these effects. To do this we employ the parameters to model the systematic effects that could be present for the actual HSC data. 
For a sample of source galaxies, we assume that source galaxies are selected based on their photo-$z$'s satisfying 
a conservative cut that the photo-$z$ posterior distribution of individual galaxies satisfies $\int_{0.75}^\infty\mathrm{d}z~ p(z)\ge 0.99$ \cite{2018PASJ...70S..20O,2019ApJ...875...63M}, where 
the lower cutoff of the integration, $z=0.75$, is well above the redshifts of CMASS galaxies (the maximum redshift cut of CMASS galaxies 
is 0.7 as shown in Table~\ref{tab:sample_HODparameters}). After using the re-weighting method in \cite{2019PASJ...71...43H} to infer the intrinsic redshift distribution of source galaxies, we compute the average of the lensing critical surface density over the source distribution, 
$\avrg{\Sigma_{\rm cr}^{-1}}$, that is used in the $\dSigma$ measurement for each sample of lensing galaxies (LOWZ, CMASS1 and CMASS2). Then, by shifting the posterior distribution of all the source galaxies by the same amount, $\Delta z_{\rm ph}$, we repeat the same calculation of $\avrg{\Sigma_{\rm cr}^{-1}}$ to estimate a shift in the $\dSigma$ signal.
After computing these quantities for multiple values of $\Delta z_{\rm ph}$,  we include $\Delta z_{\rm ph}$ to compute the shift in $\dSigma$
from an interpolation of the pre-computed values of $\avrg{\Sigma_{\rm cr}^{-1}}$. In this paper we study the impact of photo-$z$ errors on parameter estimation by adopting the Gaussian prior with width $\sigma(\Delta z_{\rm ph})=0.04$ or 0.1. Here 
$\sigma(z_{\rm ph})=0.04$ corresponds to the quoted errors of photo-$z$ estimation for the source galaxies, estimated using the same 
method in \cite{2019PASJ...71...43H}. 
We also adopt a very conservative prior of $\sigma(\Delta z_{\rm ph})=0.1$, which is  
about a factor of 2.5 larger than the quoted errors. 
Fig.~\ref{fig:scores_fiducial} shows that the cosmological parameters remain almost unchanged by the photo-$z$ errors. 
Thus, as claimed in 
\citet{OguriTakada:11}, a self-calibration of the photo-$z$ errors 
seems to work very well, by using the single sample of source galaxies and 
utilizing the dependence of lensing efficiency on lens redshifts.

For a multiplicative shear bias, $\Delta m_\gamma$, we employ the Gaussian prior $\sigma_{\Delta m_\gamma}=0.01$ as 
recommended based on dedicated image simulations of HSC galaxies 
in \citet{HSCDR1_shear:17,2018MNRAS.481.3170M}. Fig.~\ref{fig:scores_fiducial} shows that the multiplicative shear bias does not 
either
affect 
the cosmological parameters. Thus, the method of \citet{OguriTakada:11} allows us to self-calibrate both the photo-$z$ errors and the multiplicative 
bias errors. 
The row denoted as ``$\Omega_{\rm m}+\Delta z_{\rm ph}+\Delta m_{\gamma}$'' shows the results including all these three effects. The results 
are not so different from the baseline method ignoring these effects. Hence we conclude that these observational effects are not a severe
source of systematic errors causing a sizable bias in the cosmological parameters. 

When including these observational effects, the baseline method can achieve a precision of $\sigma(S_8)\simeq 0.042$. 

\subsection{Tests and validations against uncertainties in halo-galaxy connection}
\label{sec:validation_halo-galaxy_connection}
\begin{figure*}
    \begin{center}
        \includegraphics[width=0.95\textwidth]{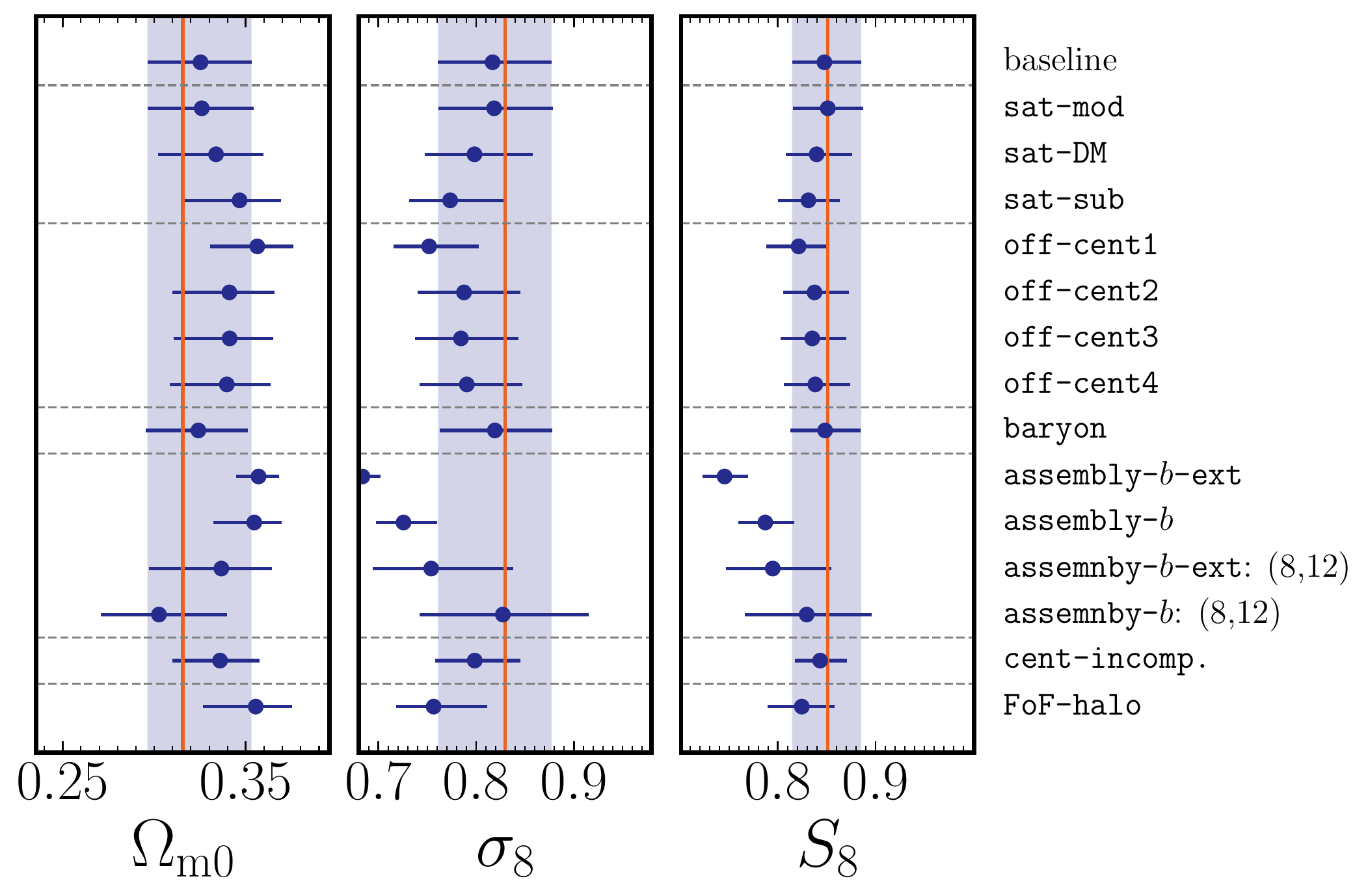}
        \caption{A summary of the performance of the baseline method against different mock catalogs. The circle shows the mode 
        of the marginalized posterior distribution of each cosmological parameter, and the the error bar denotes the 68\% credible interval. The vertical solid line denotes the true value of each parameter. The shaded region denotes the 68\% credible interval for the fiducial mock, for comparison. The columns labeled as ``{\tt assembly-$b$}: (8,12)'' and ``{\tt assembly-$b$-ext}: (8,12)'' show the 
        results when the larger scale cuts of ($8,12$)~$\hiMpc$ are employed for $\wgg$ and $\dSigma$.
        \label{fig:scores_mocks}}
    \end{center}
\end{figure*}
\begin{figure}
\centering
\includegraphics[width=1.\columnwidth]{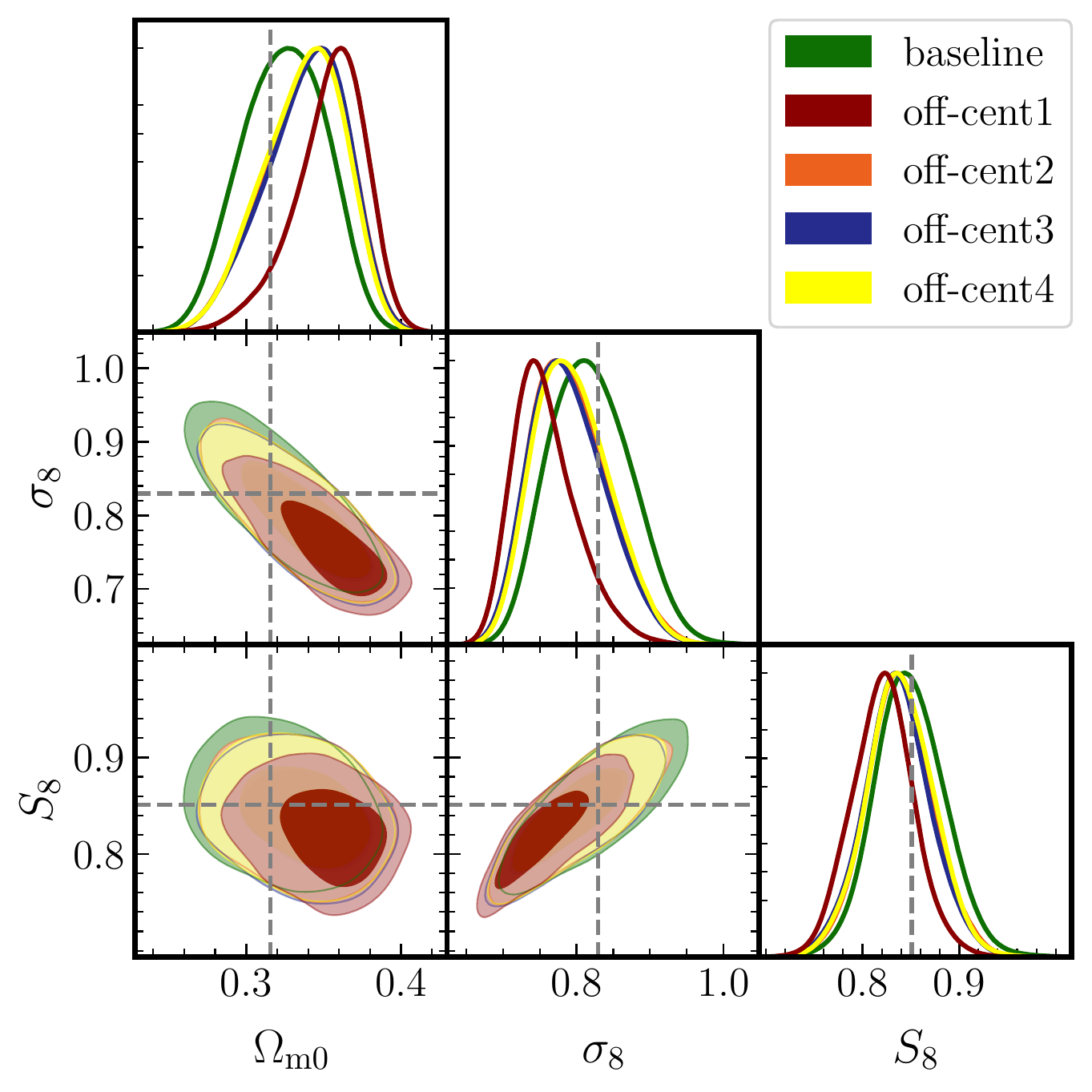}
\caption{Marginalized posterior distributions of the cosmological parameters when applying the halo model method to the mock catalog where the off-centering effect of central galaxies is included (Table~\ref{tab:mocks}). For comparison, we show the result for the {\tt fiducial} mock, 
the same as the green contours in Fig.~\ref{fig:2Dpost_baseline_cosmo_hod}.
\label{fig:2Dpost_offcent}}
\end{figure}
\begin{figure}
\centering
\includegraphics[width=1.\columnwidth]{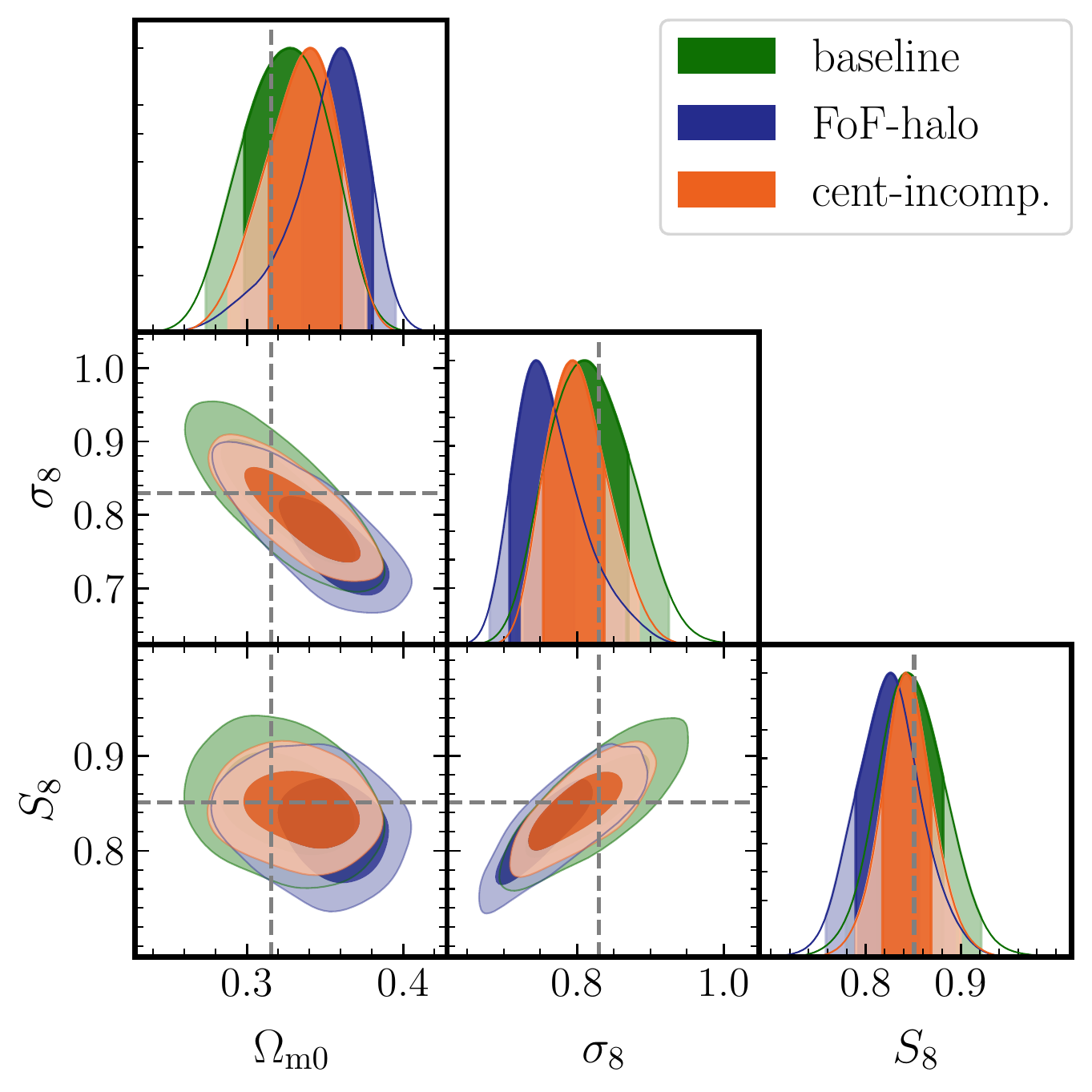}
\caption{Similar to the previous figure, but shown is the result for the {\tt FoF} and {\tt cent-incomp.} mocks. 
\label{fig:2Dpost_fof}}
\end{figure}
\begin{figure}
\centering
\includegraphics[width=1.\columnwidth]{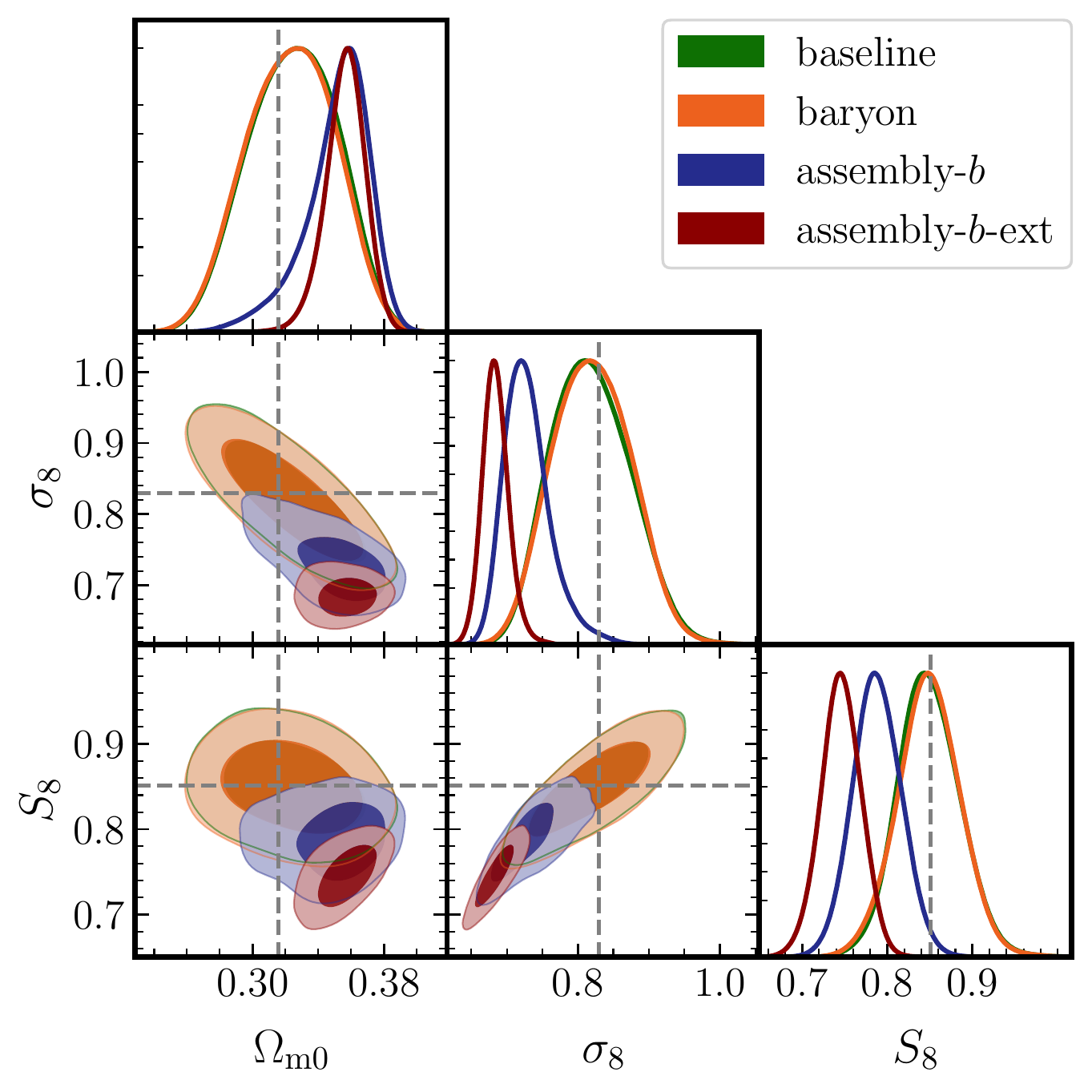}
\caption{Similar to the previous figure, but shown is the marginalized posterior distributions for 
the {\tt baryon} mocks and the assembly bias mocks ({\tt assembly-$b$} and {\tt assembly-$b$-ext}). 
Here the {\tt assembly-$b$-ext} is the worst (extreme) effect of assembly bias, where the large-scale amplitude of $\wgg$ is modified by 
a factor of 1.6 (see Fig.~\ref{fig:mock_signals}). In the {\tt assembly-$b$} mock, the boost in the amplitude  is halved, which is still larger than what is expected for an actual data, even if exists. 
\label{fig:2Dpost_assembly-b}}
\end{figure}
\begin{figure}
\centering
\includegraphics[width=1.\columnwidth]{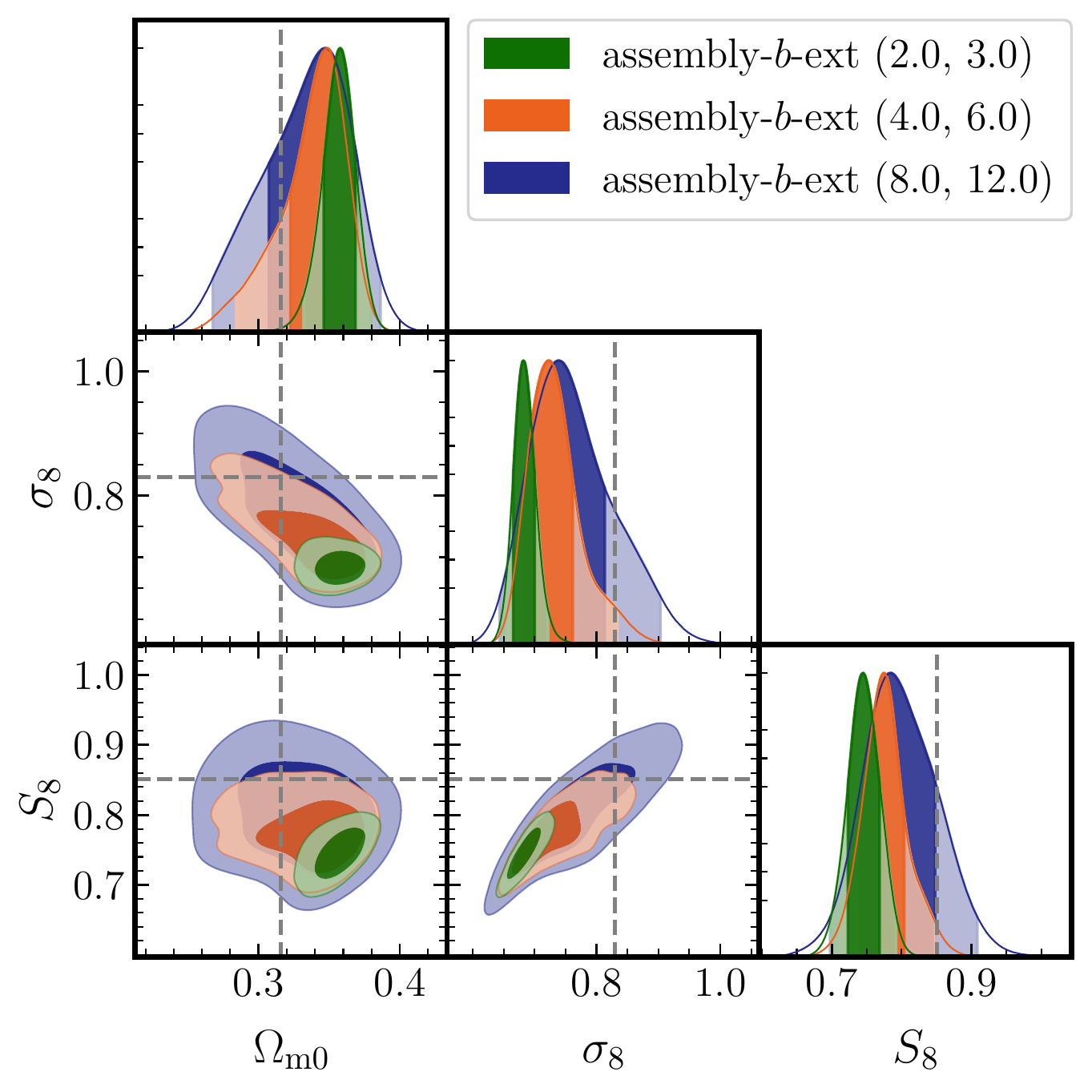}
\caption{Similar to the previous figure, but we here study how biases in the cosmological parameters can be mitigated by using the different 
scale cuts, $(4,6)$ or $(8,12)~h^{-1}{\rm Mpc}$ for $\wgg$ and $\dSigma$, respectively, instead of the baseline setup $(2,3)$. The true cosmological parameters are recovered if using the large-scale cuts such as $(8,12)$; that is, if we do not include the 1-halo term contribution.
\label{fig:2Dpost_assembly-b_scalecuts}}
\end{figure}
In this section, to assess the performance of the halo model method against uncertainties in the halo-galaxy connection, 
we perform the parameter inference against 
various mock catalogs  
(Table~\ref{tab:mocks}). 
Note that we here employ the fiducial HOD model (5 parameters model), which does not include the off-centering effect, the incompleteness effect, the baryonic effect 
nor
the 
assembly bias effect. Here we address whether the fiducial halo model can recover the true cosmological parameters to within the credible intervals
after marginalizing over the HOD parameters. 
Even if we use a weak prior on the number density of galaxies (the abundance), e.g. 50\% or 5\% of the number density, the following results remain
almost unchanged.

Fig.~\ref{fig:scores_mocks} gives a summary of the results.
The figure shows that, except for the assembly bias mocks (and the extreme off-centering mock, {\tt off-cent1}, for $\Omega_{\rm m0}$ and $\sigma_8$), 
the halo model method with the baseline setup recovers the input cosmological parameters to within the 68\% credible interval. 
We should also note that $S_8$ is better recovered compared to $\Omega_{\rm m0}$ or $\sigma_8$. This is an encouraging result, 
because $S_8$ is close to 
the
primary parameter 
combination
that determines the amplitudes of $\wgg$ and $\dSigma$. 
We below discuss the results for some variants of the mocks in more detail.

Figs.~\ref{fig:2Dpost_offcent} and \ref{fig:2Dpost_fof} show the marginalized distributions of the cosmological parameters for the {\tt off-cent}, 
{\tt FoF-halo} and {\tt cent-incomp.} mocks (see Table~\ref{tab:mocks}). For most cases except for the {\tt off-cent1} mock, where all the central galaxies are off-centered, the halo model method recovers the true cosmological parameters within 68\% credible interval. 
As shown in Fig.~\ref{fig:mock_signals}, the {\tt off-cent1} mock gives smaller amplitudes in $\dSigma$ at scales around the scale cut ($R_{\rm cut}=3~h^{-1}{\rm Mpc}$), but does not change the $\wgg$ amplitude at scales greater than the scale cut ($R=2~h^{-1}{\rm Mpc}$). The smaller lensing 
amplitude leads to an underestimation in the average mass of host halos. Since smaller-mass halos 
give
a smaller bias amplitude for 
a fixed cosmology, this would lead to a lower $\sigma_8$ to reproduce the amplitude of $\wgg$ in the mock signal (see Fig.~\ref{fig:signal_cosmo_dep}).
This explains a negative bias in $\sigma_8$ for the {\tt off-cent1} mock. We confirmed that, if we employ the larger scale cut for $\dSigma$, the bias in $\sigma_8$ is 
mitigated.
For the {\tt FoF-halo} mock, our method nicely recovers $S_8$, although 
we find sizable biases in $\Omega_{\rm m0}$ and $\sigma_8$.

In Fig.~\ref{fig:2Dpost_assembly-b}, we show the results for the {\tt baryon}, {\tt assembly-$b$} and {\tt assembly-$b$-ext} mocks. The figure shows that the assembly bias mocks lead to biases in the cosmological parameters, greater than the credible interval. Thus the assembly bias is indeed the most dangerous systematic effect, which violates the scaling relation of halo bias with halo mass. However, we note that, in this paper, the result is considered as the worst-case scenario, because we include 
the maximum possible
effect of the assembly bias in the {\tt assembly-$b$-ext} mock, where 
we populate galaxies into halos assuming 
a fully deterministic assignment of centrals in ascending order
of halo concentration: we populate galaxies from the lowest concentration halos 
in each  mass bin. The 
impact of the assembly bias for more realistic galaxies, even if exists for SDSS galaxies, would be much smaller than what is shown in this paper. Nevertheless, 
this possible impact of the assembly bias needs to be kept in mind. 

To tackle 
the possibility that the actual galaxy sample is suffering from the assembly bias effect,
we here propose a practical solution to 
assess its impact
or mitigate 
it in parameter estimation.
In Fig.~\ref{fig:2Dpost_assembly-b_scalecuts}, we show the results when using different scale cuts, $(4,6)$ or $(8,12)~h^{-1}{\rm Mpc}$, for 
$\wgg$ and
$\dSigma$, instead of our fiducial choice $(2,3)$. Note that $(8,12)$ is the choice of the minimal bias method in \citet{2020arXiv200806873S}
(also see Figs.~\ref{fig:2Dpost_HOD_vs_minimalbias} and \ref{fig:scores_mocks}). The figure clearly shows that the posterior distribution systematically moves with varying the scale cuts, and then the choice of $(8,12)$ recovers the true cosmological parameters. In this case, the method does not use the 1-halo term signal 
of $\dSigma$, and tries to fit the mock signals by the model templates, where $r_{\rm cc}\simeq 1$ is satisfied. Thus, if we observe a similar systematic shift in the parameter constraints with varying the scale cuts, we could identify a signature of the assembly bias effect in actual data.

\section{Discussion and Conclusion}
\label{sec:conclusion}

In this paper we have in detail studied 
the
performance of the halo model based method for cosmological parameter estimation. We used \textsc{Dark Emulator}
to model the halo clustering quantities (halo mass function, halo auto-correlation function and halo-matter cross correlation), where the emulator  includes, by design, all non-trivial effects such as nonlinear clustering, nonlinear halo bias and halo exclusion effect that are otherwise difficult to analytically model. 
We combined \textsc{Dark~Emulator} with the HOD method to model clustering observables of galaxies for which we consider the projected correlation 
function, $\wgg(R)$, and the galaxy-galaxy weak lensing $\dSigma(R)$ in this paper. Then we validate the emulator-based halo model  method by studying whether 
to recover the cosmological parameters from MCMC analyses by comparing the model predictions with the mock signals for 
the spectroscopic SDSS galaxies and the HSC-Y1 galaxies. 

The main results of this paper are summarized as follows. 
\begin{itemize}
	\item Our method using \textsc{Dark~Emulator} allows for computations of $\wgg$ and $\dSigma$ at a few CPU seconds for each model, which is equivalent to a factor of 
	million
	reduction in computation time compared to the standard method (run $N$-body simulations, populate galaxies in halos, and then measure the galaxy clustering observables from the mocks). With this emulator, we can perform the MCMC analysis in practice. 
	\item Changes in the cosmological parameters cause characteristic changes in the amplitudes and scale-dependences of $\wgg$ and $\dSigma$ via a combination of various effects: nonlinear clustering, nonlinear halo bias and halo exclusion effect (Fig.~\ref{fig:signal_cosmo_dep}). 
	\item We showed that the halo model based method can recover the underlying cosmological parameters, especially $S_8=\sigma_8 \Omega_{\rm m0}^{0.5}$,  
	after marginalization over the halo-galaxy connection (HOD) parameters, for a variety of 
	mock catalogs except for the mocks including 
	an
	extreme effect of assembly bias, for the nominal choices of scale cuts, $(R^{\wgg}_{\rm cut},R^{\dSigma}_{\rm cut})=(2,3)~h^{-1}{\rm Mpc}$ (Figs.~\ref{fig:scores_fiducial} and \ref{fig:scores_mocks}). 
	The baseline method can achieve a precision of $\sigma(S_8)\simeq 0.035$--$0.042$ for the SDSS and HSC-Y1 datasets.
	This method allows for tight constraints on the cosmological parameter, because the small-scale information of $\dSigma(R)$ at scales, $3\le R/[h^{-1}{\rm Mpc}]\lesssim 10$, can be used to infer the average mass of host halos of the SDSS galaxies, yielding useful information on the galaxy bias that is sensitive to the large-scale amplitudes of $\wgg$ and $\dSigma$.
	\item The halo model method suffers from sizable biases in the cosmological parameters if the SDSS galaxies have 
	a significant
	assembly bias (although we consider the extreme mocks including  the maximum amount of the assembly bias effect in this paper) (Figs.~\ref{fig:scores_mocks} and \ref{fig:2Dpost_assembly-b}). 
	Even if this is the case, we showed that the cosmological parameters can be recovered
	if employing sufficiently large scales cuts of $R_{\rm cut}\gtrsim 10~h^{-1}{\rm Mpc}$, where the cross-correlation coefficient 
	$r_{\rm cc}(r)\equiv \xi_{\rm gm}(r)/[\xi_{\rm gg}(r)\xi_{\rm mm}(r)]^{1/2}\simeq 1$ 
	(Fig.~\ref{fig:2Dpost_assembly-b_scalecuts}). 
	This is equivalent to 
	the
	case that we do not include the small-scale lensing information in the cosmological parameter estimation. However the price to pay is that the statistical accuracy of cosmological parameter determination
	is degraded, leading to $\sigma(S_8)\simeq 0.07$ (see Fig.~\ref{fig:scores_mocks}). Thus in practice we should monitor whether the cosmological parameters have a systematic shift with different scale cuts, as a test of the assembly bias effect inherent in actual data (Fig.~\ref{fig:2Dpost_assembly-b}).
	\item We used the method using a single population of source galaxies to measure the galaxy-galaxy weak lensing for multiple lens galaxies at different redshifts (LOWZ, CMASS1 and CMASS2) over the range of redshifts, $z=[0.15,0.7]$. This method 
	allows for {\it self-calibration} of  the photometric redshift errors and the multiplicative shear bias (Fig.~\ref{fig:scores_fiducial}).  
\end{itemize}

Hence we conclude that we can safely apply the halo model based method to actual SDSS and HSC-Y1 data. The cosmological constraints from actual observational data are presented in a companion paper (Miyatake et al. in prep.).

\acknowledgments
We would like to thank Ryoma~Murata for his contribution during the early phase of this work. 
This work was supported in part by World Premier International Research Center Initiative (WPI Initiative), MEXT, Japan, by JSPS KAKENHI Grant Numbers JP15H03654,
JP15H05887, JP15H05893, JP15H05896, JP15K21733, JP17H01131, JP17K14273, JP18H04350, JP18H04358, JP19H00677, JP19K14767, JP20H01932, JP20H04723, JP20H05855, and JP20A203, by Japan Science and Technology Agency (JST) AIP Acceleration Research Grant Number JP20317829, Japan.
SS is supported by International Graduate Program for Excellence in Earth-Space Science (IGPEES), World-leading Innovative Graduate Study (WINGS) Program, the University of Tokyo.
KO is supported by JSPS Overseas Research Fellowships.

\bibliography{./refs.bib}

\appendix

\section{Cosmological dependence of the 2-halo term of galaxy correlation functions}
\label{sec:2h_cosmo}

\begin{figure}
    \begin{center}
        \includegraphics[width=1.0\columnwidth]{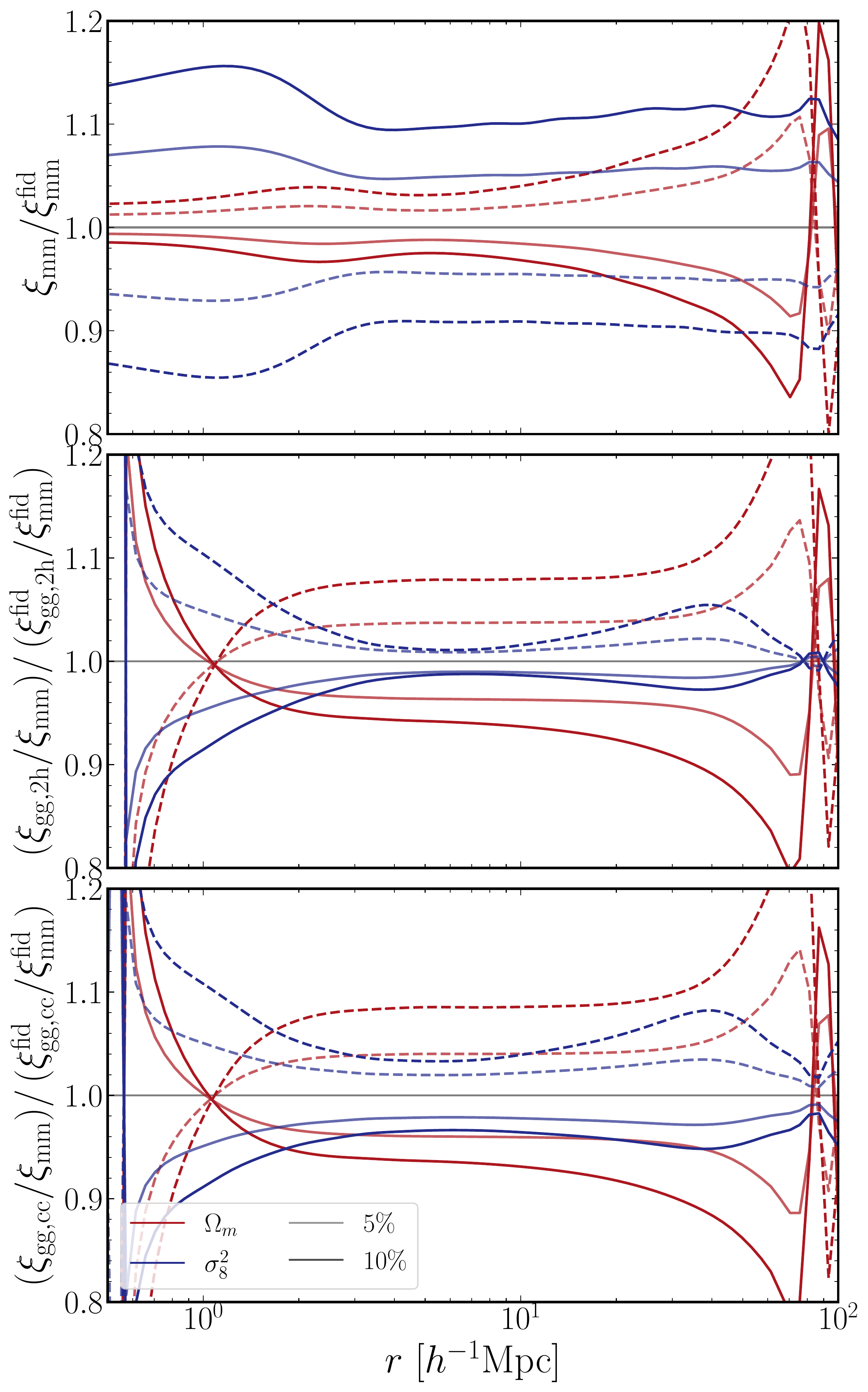}
        \caption{Fractional changes in the 2-halo term of the three-dimensional galaxy-galaxy correlation function relative to 
        that for the fiducial model, when varying either of $\sigma_8$ or $\Omega_{\rm m0}$. Note that 
        other model parameters keep fixed to their fiducial values. The middle panel shows the fractional change for the ratio
        $\xi_{\rm gg, 2h}/\xi_{\rm mm}$, because the ratio becomes scale-independent if the linear theory holds, which predicts $\xi_{\rm gg}=b^2 
        \xi_{\rm mm}$ with a scale-independent coefficient $b$. 
        The lower panel shows the result for the two-point correlation function of central galaxies that includes only the two-halo term by definition.
        The different lines show the results when $\pm5\%$ and 
        $\pm 10\%$ changes in $\sigma_8^2$ and $\Omega_{\rm m0}$, and the solid or dashed lines correspond to the results when increasing or decreasing 
        the parameter by the amount from the fiducial value. Here we consider the HOD for LOWZ-like galaxies at $z=0.25$.
        \label{fig:xi2h_cosmo_dependence}
        }
    \end{center}
\end{figure}

The cosmological information in the joint probes cosmology using $\dSigma$ and $\wgg$ lies in the 2-halo terms of $\dSigma$ and $\wgg$. 
In particular, an advantage in the use of 
\textsc{Dark Emulator} is that it includes all the nonlinear effects such as the nonlinear matter clustering, the nonlinear halo bias and the halo exclusion effect that are very difficult to accurately calibrate unless $N$-body simulations are employed as done in \textsc{Dark Emulator}. In this appendix, we study how these nonlinear effects cause scale-dependent variations in the 2-halo term by changes in the cosmological parameters. 

Fig.~\ref{fig:xi2h_cosmo_dependence} shows how a change in $\sigma_8$ or $\Omega_{\rm m0}$ causes a scale-dependent change in the 2-halo term of 
the three-dimensional correlation function, $\xi_{\rm gg,2h}$, while other parameters including the HOD parameters are kept to their fiducial values. 
Note that we use \textsc{Dark Emulator} to compute these results, and we here focus on the galaxy auto-correlation function because most of the cosmological information for the SDSS and HSC datasets are from $\wgg$ that is obtained from the projection of $\xi_{\rm gg}$. 
First, for comparison, the upper panel shows the change in the matter correlation function, $\xi_{\rm mm}$, relative to that for the fiducial model 
({\it Planck} cosmology). The figure shows that a change in $\sigma_8$ causes a fairly sale-independent change in $\xi_{\rm mm}$ at scales 
$r\lesssim 70~h^{-1}{\rm Mpc}$ which we consider throughout this paper. On the other hand, a change in $\Omega_{\rm m0}$ causes a scale-dependent change in $\xi_{\rm mm}$. 

The middle panel shows the fractional change in $\xi_{\rm gg,2h}/\xi_{\rm mm}$ relative to the ratio for the fiducial model. 
Note that we here consider the fiducial model of HOD for LOWZ-like galaxies at $z=0.25$.
The reason we consider this ratio is that, if $\xi_{\rm gg}$ follows the linear theory prediction as given by $\xi_{\rm gg}=b^2\xi_{\rm mm}$ with a scale-independent coefficient $b$, the change becomes scale-independent 
for the change in the cosmological parameters. Here, for the sake of clarity, we consider only the 2-halo term of $\xi_{\rm gg}$ using Eq.~(\ref{eq:Pgg_1h2h}). The figure shows that a change in $\sigma_8$ causes a scale-dependent change in the ratio, which should arises from the scale-dependent change in the halo bias and the halo exclusion effect by the change in $\sigma_8$. The change in $\Omega_{\rm m0}$ is also found to cause a scale-dependent change in the ratio, again via the dependence of halo bias on $\Omega_{\rm m0}$. Thus \textsc{Dark Emulator} includes these complex 
dependences of $\xi_{\rm gg}$ on the cosmological parameters. 
For further comparison, the lower panel shows the results for $\xi_{\rm gg,cc}$, i.e the two-point correlation function of central galaxies, which includes only the 2-halo term by definition. The results appear to be similar to those in the middle panel.

Thus \textsc{Dark Emulator} includes these complex dependences of the galaxy-galaxy correlation function on the cosmological parameters ($\sigma_8$
and $\Omega_{\rm m0}$). These dependences are difficult to accurately calibrate analytically e.g. by the perturbation theory, and the accurate calibration requires the use of $N$-body simulations as done in the development of \textsc{Dark Emulator}. These complex cosmological dependences 
of $\xi_{\rm gg}$ are an advantage of our method.

\section{Covariance estimation based on mock catalogs}
\label{sec:covariance}

In this appendix, we describe the estimation of statistical uncertainties in 
the galaxy-galaxy weak lensing profile ($\dSigma$) by using a set of synthetic observational datasets, i.e. the covariance matrix.

A robust estimation of the covariance matrix is one of the most important subjects 
in modern cosmology in practice 
\citep[e.g., see Refs.~][]{2007A&A...464..399H, 2013MNRAS.432.1928T, 2013PhRvD..88f3537D, 2015MNRAS.454.4326P, 2018MNRAS.473.4150F}. 
In this paper, we use a set of 
numerical simulations in Ref.~\cite{2017ApJ...850...24T} to 
construct
realistic mock catalogs of galaxy shapes as well as the tracers of large-scale structures. 
We then adopt the same analysis pipeline to measure $\dSigma$ from 
the mock catalogs as we do for actual measurements. 
Our mock measurements of the lensing signals have 2268 realizations in total, allowing 
us to evaluate the covariance matrix of $\dSigma$ for the SDSS-like galaxies
at 
multiple
redshifts in a rigorous way.
In the following, we 
describe
how to produce mock catalogs from the simulations incorporated with observational data and show the validation of our mock lensing analyses.

\subsection{Massive production of mock catalogs}

\subsubsection*{Full-sky simulations of lensing and halos}

We first briefly introduce the full-sky ray-tracing simulations and the halo catalogs in the line-cone simulation realization, developed
in Ref.~\cite{2017ApJ...850...24T}
(The full-sky simulation data are freely available from
\footnote{\url{http://cosmo.phys.hirosaki-u.ac.jp/takahasi/allsky_raytracing/}.}).
The full-sky simulations are based on a set of $N$-body simulations 
with $2048^3$ particles in cosmological volumes. 
Ref.~\cite{2017ApJ...850...24T} adopted the standard $\Lambda$CDM cosmology 
with the following cosmological parameters: 
the CDM density parameter 
$\Omega_{\rm cdm}=0.233$, 
the baryon density 
$\Omega_{\rm b}=0.046$, 
the matter density 
$\Omega_{\rm m}=\Omega_{\rm cdm}+\Omega_{\rm b}
= 0.279$, 
the cosmological constant 
$\Omega_{\Lambda}=0.721$, 
the Hubble parameter
$h= 0.7$, 
the amplitude of density fluctuations
$\sigma_8= 0.82$,
and the spectral index
$n_s= 0.97$.
Note that the cosmological model in the simulation 
is consistent with the WMAP $9$ year cosmology \citep{Hinshaw2013}.

Full-sky weak gravitational lensing simulations
have been performed with the standard multiple lens-plane algorithm \cite{2001MNRAS.327..169H, 2013MNRAS.435..115B, 2015MNRAS.453.3043S}. 
In this simulation, one can take into account the light-ray deflection on the celestial sphere
by using the projected matter density
field given in the format of spherical shell (see the similar approach for Ref.~\cite{2008MNRAS.391..435F}).
The simulations used the projected matter fields in 38 shells in total, 
each of which was computed by projecting 
$N$-body simulation realization over a radial width of $150\, h^{-1}{\rm Mpc}$, 
in order to make the light-cone simulation covering a cosmological volume up to $z=5.3$.
As a result, the lensing simulations
consist of the shear field at each of 38 different source redshifts with angular resolution of 0.43 arcmin.
Each simulation data is given in the \textsc{HEALPix} format \cite{2005ApJ...622..759G}.
The radial depth between nearest source redshifts is set to be $150\, h^{-1}{\rm Mpc}$ in comoving distance, corresponding to the redshift interval of $0.05-0.1$ for $z\simlt1$.

In each output of the $N$-body simulation, 
Ref.~\cite{2017ApJ...850...24T} identified dark matter halos using the \textsc{Rockstar} algorithm
\cite{Behroozi:2013}. 
In the following, we define the halo mass by using the spherical overdensity criterion:
$M_{\rm 200m} = 200\bar{\rho}_{\rm m0}(4\pi/3)R^{3}_{\rm 200m}$. 
Individual halos in $N$-body boxes are assigned to the pixels in the celestial sphere
with the \textsc{HEALPix} software. It should be noted that the $N$-body simulations 
allow us to resolve dark matter halos with masses greater than 
a few times $10^{12}\, h^{-1}M_{\odot}$ with more than 50 
$N$-body particles at redshifts $z<0.7$, which is a typical redshift range of massive galaxies 
in the SDSS BOSS survey.

\subsubsection*{Shapes of background galaxies}

For shapes of background galaxies, we use the mock catalogs produced in Ref.~\cite{2019MNRAS.486...52S}.
The mocks are specific to cosmological analyses with the Subaru HSC-Y1 shape catalog
\cite{2018PASJ...70S..25M} by taking into account various observational effects as
the survey footprints, inhomogeneous angular distribution of source galaxies, 
statistical uncertainties in photometric redshift estimate, 
variations in the lensing weight, and the statistical noise in galaxy shape measurements including both intrinsic shapes and the measurement errors.
We produced 2268 mock catalogs from 108 full-sky ray-tracing simulations of gravitational lensing in Ref.~\cite{2017ApJ...850...24T} by using a similar approach developed in Refs.~\cite{2014ApJ...786...43S, 2017MNRAS.470.3476S}. We properly incorporated with the simulated lensing shear and 
the observed galaxy shape on an object-by-object basis, enabling the mock to share exactly 
the same information of angular positions, redshifts, the lensing weights and 
the shear responsivity with the real catalog.
The further details are found in Ref.~\cite{2019MNRAS.486...52S}.
The mock shape catalogs are publicly available at
\footnote{\url{http://gfarm.ipmu.jp/~surhud/}}.

\subsubsection*{Mock catalogs of lensing galaxies, SDSS-like galaxies}

For the foreground galaxies to be used in the galaxy-galaxy lensing analysis, 
we produce the mock galaxy catalogs assuming the HOD 
method
as summarized in Section~\ref{sec:mocks_for_signals}.
As in the mock galaxy shape catalogs, we first extract 2268 realizations 
of the HSC-Y1 survey windows from 108 full-sky halo catalogs in Ref.~\cite{2017ApJ...850...24T}.
We then populate galaxies into halos using the fiducial HOD method whose parameters are chosen to mimic LOWZ and CMASS galaxies 
in the redshift range of $0.15<z<0.35$, $0.43<z<0.55$, and $0.55<z<0.70$, 
but spanning the entire SDSS BOSS footprint. 
The HOD method is used to populate mock LOWZ and CMASS galaxies in halos of each of the light-cone simulation realization.

Besides, we include the redshift-space distortion effects in mock LOWZ and CMASS galaxies.
Along the line of sight of each host halo, we set the radial velocity of the central galaxy to be the same as one of its host halo, while we assign the random velocities to the satellites by following a Gaussian distribution with width given by 
the virial velocity dispersion.

\subsection{Validation of our covariance estimation}

\begin{figure}
\begin{center}
       \includegraphics[clip, width=0.90\columnwidth]{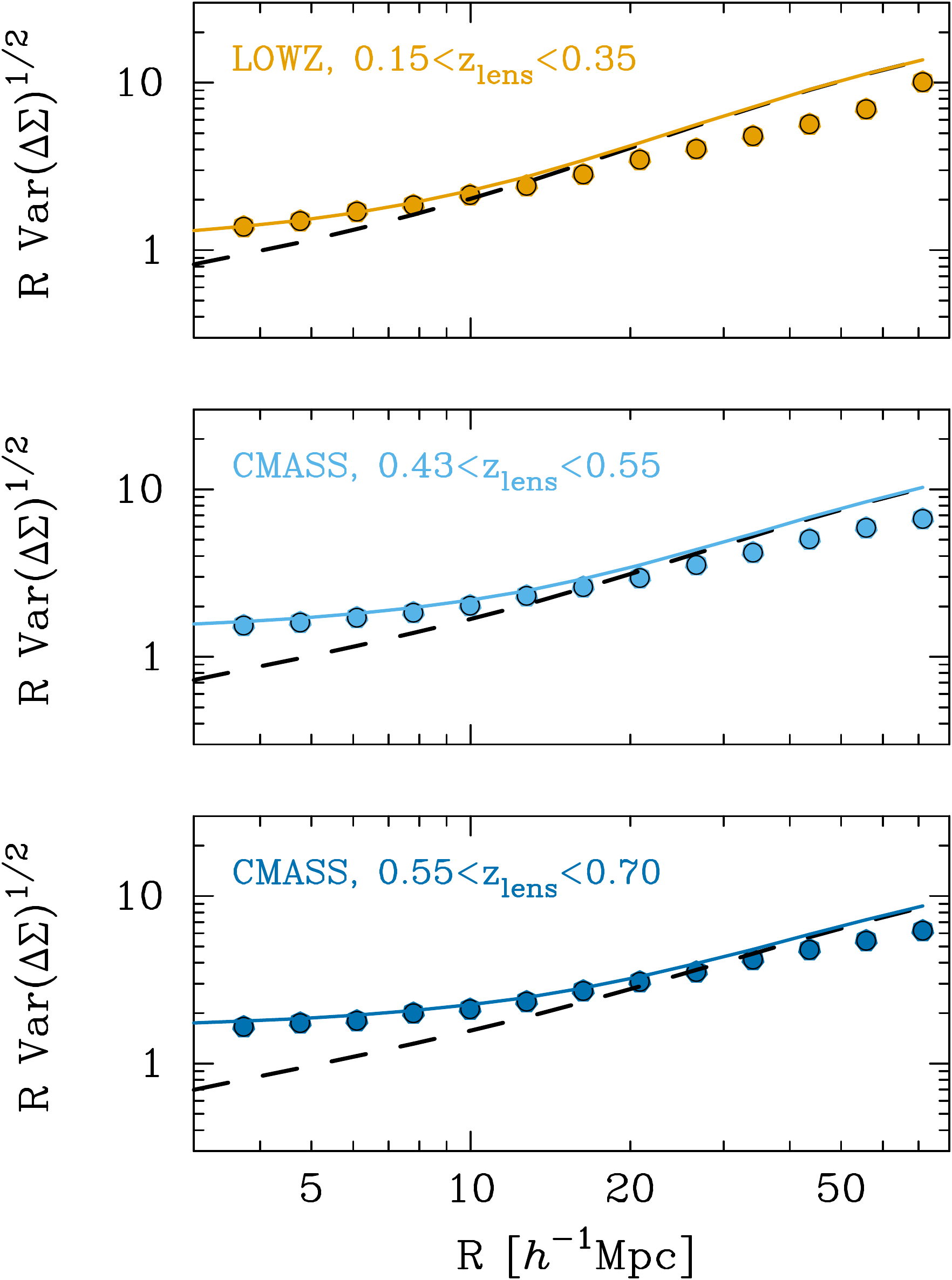}
     \caption{
     Comaprison of the diagonal components in the covariance matrices for the galaxy-galaxy lensing signals.
     From top to bottom, we show the variances of the lensing signals at three different redshift bins (the lower panel corresponds to the case at the higher redshift).
     The points represent the simulation results based on 2268 realizations, while the solid lines 
     stand for the Gaussian predictions developed in Ref.~\cite{2018MNRAS.478.4277S}.
     In each panel, the dashed line shows the variance in the absence of shape noises, highlighting
     the contribution from uncorrelated large-scale structures along a line of the sight.
     Note that we define $\Delta \Sigma$ in units of $h\, M_{\odot}\, \mathrm{pc}^{-2}$ in this figure. Hence, the unit in each vertical axis is given by $\mathrm{Mpc}\, (M_{\odot}\, \mathrm{pc}^{-2})$.
     \label{fig:var_ggl_mock_comp}
  } 
    \end{center}
\end{figure}

\begin{figure*}
\begin{center}
       \includegraphics[clip, width=0.9\columnwidth]{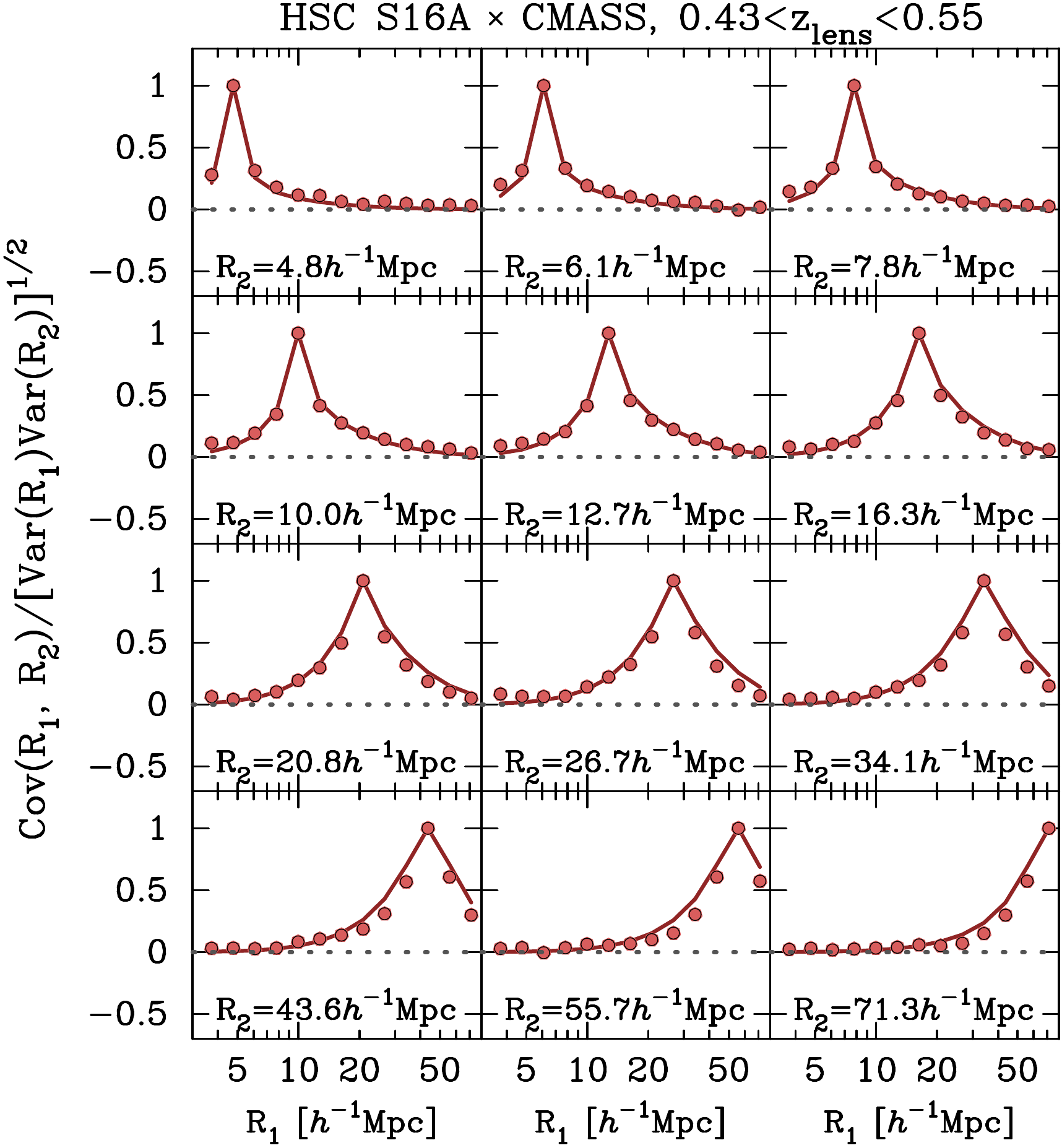}
       \includegraphics[clip, width=1.10\columnwidth]{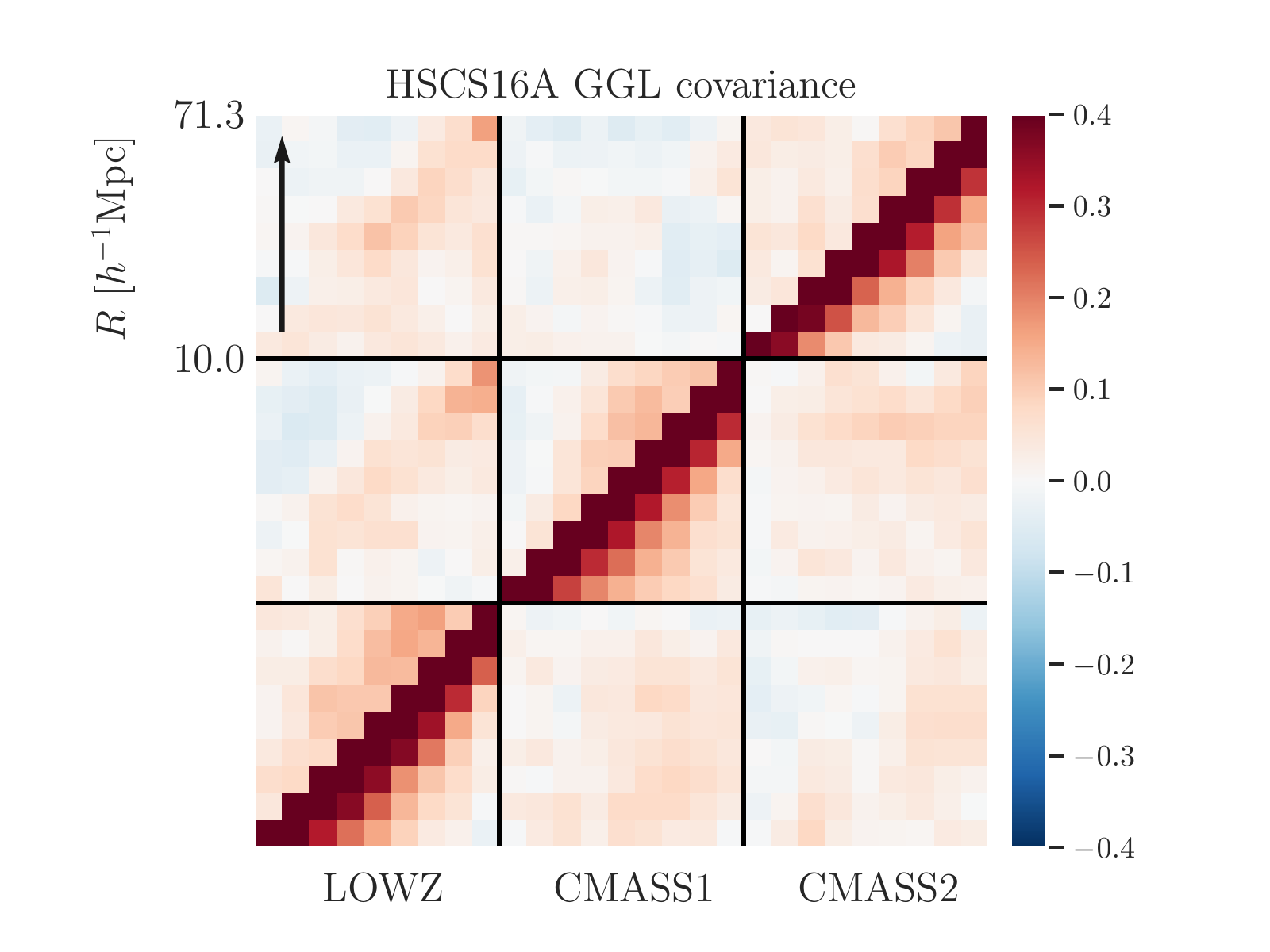}
     \caption{
     Comparisons of the off-diagonal elements in the covariance matrices with the simulation results and 
     their Gaussian predictions. The left panels show the cross correlation coefficient in the covariance 
     at the medium redshift bin (in the redshift range of $0.43<z<0.55$) as a function of radii.
     In the left, the points show the simulation results, while the lines are for the Gaussian predictions.
     Each small panel in the left represents the scale dependence of the cross correlation coefficient.
     In the right, we show the cross correlation coefficients in the full covariance 
     across three redshift bins. The lower triangular panels show the simulation results, 
     while the upper one displays the difference between the simulation results and the Gaussian predictions.
     In the right, we use the labels of  ``LOWZ", ''CMASS1" and ``CMASS2" from the lowest to the highest redshift bins. Note that we limit the length scales to be in the range of $R>10\, h^{-1}{\rm Mpc}$
     in the right panels to focus on the sample covariance.
     The diagonal elements in the right is set to 1 by construction.
     \label{fig:cov_ggl_mock_comp}
  } 
    \end{center}
\end{figure*}

In this section, we validate the performance of our covariance estimation 
with the mock measurements of galaxy-galaxy lensing signals.
For comparison, we predict the Gaussian covariances 
of the lensing signals by a halo-model framework developed in Ref.~\cite{2018MNRAS.478.4277S}.
In this framework, we can properly take into account the weight function related to 
the conversion between the lensing shear 
and the excess surface mass density in the covariance matrix.
It would be worth noting that the covariance model has been extensively validated with a set of numerical simulations for a variety of the weight functions in the lensing analysis 
(see Ref.~\cite{2018MNRAS.478.4277S} for details).

Fig.~\ref{fig:var_ggl_mock_comp} shows the comparison of the variance of the lensing signal in each of three redshift bins with the simulation results and their Gaussian predictions.
In each panel, the simulation results are shown in the colored points, while
the solid and dashed lines represent the Gaussian covariances with and without the shape noise, respectively.
The figure clearly shows that the sample variance caused by the line-of-sight large scale structures
dominates the mock variance, and the shot noise is dominant at the length scale less 
than $10\, h^{-1}{\rm Mpc}$. 

We then study the off-diagonal elements of the mock covariance matrices.
The left panels in Fig.~\ref{fig:cov_ggl_mock_comp} summarize the comparison of 
the cross correlation coefficient in the covariance for the lensing signals at the lens redshift $0.43<z<0.55$.
This figure shows that our simple Gaussian prediction provides a reasonable agreement with the off-diagonal 
elements of the mock covariance, indicating the super-sample covariance (SSC) \cite{2013PhRvD..87l3504T}
is not important to our galaxy-galaxy lensing analyses.
The SSC is expected to arise from the four-point correlation among super- and sub-survey modes \cite{2013PhRvD..87l3504T, 2014PhRvD..89h3519L}, and the recent simulations have shown that the SSC in the halo-matter cross correlation becomes important only at $1\, h^{-1}{\rm Mpc}$ \cite{2019MNRAS.482.4253T}.
At the scale of $R\sim1\, h^{-1}{\rm Mpc}$, the statistical uncertainties in our lensing analyses
are mostly determined by the shot noise terms. 

The right panel of Fig.~\ref{fig:cov_ggl_mock_comp} shows the cross correlation coefficient of the mock covariance matrix across 
three redshift bins. The lower triangular panel shows the simulation results,
while the upper panels represent the difference between the simulation results and the Gaussian predictions.
We find the large cross correlation coefficients
in the radius range of $R>10\, h^{-1}{\rm Mpc}$ 
with a level of $\sim0.5$ at most for single redshifts and 
the cross covariance between two different redshifts is less prominent.
Our Gaussian covariance is in good agreement with the simulation results, allowing to explain the mock covariance with a $20\%$-level accuracy.

\section{Posterior distribution of parameters in a full-dimension parameter space}
\label{sec:full_posterior}

\begin{figure*}
\begin{center}
    \includegraphics[width=2.\columnwidth]{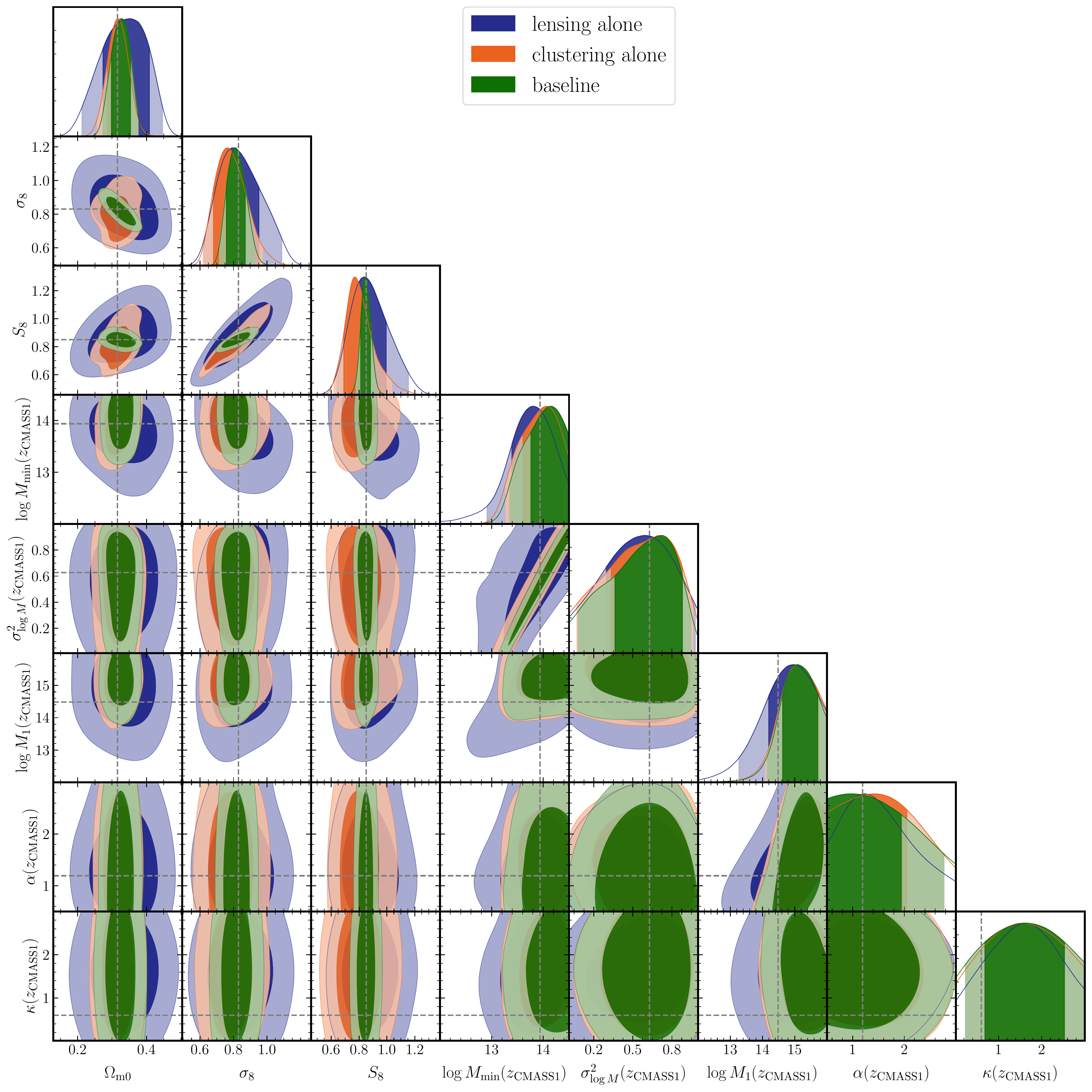}
    \caption{Posterior distributions of cosmological parameters and HOD parameters, as in Fig.~\ref{fig:2Dpost_baseline_cosmo_hod}. Here 
    we show the results for all the HOD parameters for the CMASS1-like galaxies (see Table~\ref{tab:sample_HODparameters}).}
    \label{fig:2Dpost_baseline_all}
\end{center}
\end{figure*}
For comprehensiveness of our discussion, in Fig.~\ref{fig:2Dpost_baseline_all} we show the posterior distribution for all the parameters including all the HOD parameters for
the CMASS1-like galaxies. The results for the LOWZ- and CMASS2-like galaxies at different redshifts are similar to this plot. 
Even if we use the joint information of $\wgg$ and $\dSigma$, each of the HOD parameters is not well constrained, but the cosmological parameters are 
recovered. 


\end{document}